\documentclass[eqsecnum,amsmath,amsfonts,amssymb,prd,a4paper,12pt]{revtex4}
\usepackage{graphicx}
\usepackage{color}
\usepackage{psfrag}
\usepackage{psfig}
\usepackage{epsfig}
\usepackage{ifthen}
\usepackage{calc}
\usepackage{graphpap}
\usepackage{eufrak}
\usepackage{supertabular}
\usepackage{longtable}
\usepackage{bm}
\usepackage{float}
\usepackage{mathrsfs}
\newcommand{\rot}{\circlearrowleft}
\newcommand{\manifold}{\mathscr}

\renewcommand{\vec}[1]{\boldsymbol{#1}}
\newcommand{\cosec}{\mathop{\mathrm{cosec}}}

\renewcommand{\d}{\mathrm{d}}
\newcommand{\diff}[2]{
  \ifthenelse{\equal{#1}{}}%
{\frac{\mathrm{d}\hphantom{#2}}{\mathrm{d}#2}}%
{\frac{\mathrm{d}#1}{\mathrm{d}#2}}}
\newcommand{\ddiff}[2]{
  \ifthenelse{\equal{#1}{}}%
{\frac{\mathrm{d}^2\hphantom{#2}}{\mathrm{d}#2^2}}
{\frac{\mathrm{d}^2#1}{\mathrm{d}#2^2}}}
\newcommand{\pardiff}[2]{\frac{\partial#1}{\partial#2}}

\newcommand{\expct}[3][]{\left\langle#3\left|#2\right|#3\right\rangle_{\mathrm{#1}}}
\newcommand{\bra}[1]{\left\langle\right.#1\left|\right.}
\newcommand{\ket}[1]{\left.\left|#1\right.\right\rangle}

\newcommand{\vac}[3][]{\left\langle#2\right\rangle^{#3}_{\mathrm{#1}}}   
\newcommand{\abs}[1]{\left|#1\right|}

\newcommand{\topbott}[2]{\left\{ #1 \atop #2 \right\}}
\newcommand{\indhel}{h}

\begin{document}

\title{\textbf{
Canonical Quantization of the Electromagnetic Field on the Kerr Background}}

\author{Marc Casals}
\email{marc.casals@ucd.ie}
\author{Adrian C. Ottewill}
\email{adrian.ottewill@ucd.ie}
 \affiliation{Department of Mathematical Physics,University
College Dublin, Belfield, Dublin 4, Ireland}

\begin{abstract}
We investigate the canonical quantization of the electromagnetic field on the Kerr background.
We give new expressions for the expectation value of the electromagnetic stress-energy tensor in various vacua states and give
a physical interpretation of the separate terms appearing in them. We numerically calculate the luminosity
in these states. We also study the form of the renormalized stress-energy tensor close to the horizon
when the electromagnetic field is in the past Boulware state.
\end{abstract}

\maketitle


\section{Introduction} \label{sec:Intro. in stress-energy tensor}

The properties of quantum states of physical interest in the Kerr background has been broadly investigated by studying the
scalar field. The study of the quantized electromagnetic field in this background has been studied
much less (notably, in ~\cite{ar:CCH}, ~\cite{ar:F&Z'85} and ~\cite{ar:Bolash&Frolov}). 
In this paper we perform the canonical quantization of the electromagnetic field in the Kerr background and calculate
the renormalized expectation value of the stress-energy tensor (abbreviated as RSET) when the electromagnetic field is in various states of physical interest.
We analytically calculate new expressions which are 
symmetric under parity for the expectation value of the stress-energy tensor in these physical states.
These expressions correct those given in Candelas, Chrzanowski and Howard ~\cite{ar:CCH}, hereafter referred to as CCH,
which are not
symmetric under the parity operation even when the state is
and even though the parity operation is a symmetry of both the Kerr metric and Maxwell's equations. 
We also numerically calculate differences between two states of the RSET.
We show that the difference in the RSET between a Hartle-Hawking-type state and the past Boulware state
is exactly (minus) thermal near the horizon, unlike a result in CCH.
We also study the rate of rotation of this difference, for which
there is no unanimous consensus in the literature,
and find that it is rigidly rotating with the horizon to at least second order in the distance from the horizon.

This paper is organized as follows.
We outline the classical theory of the electromagnetic field on the Kerr background in Section \ref{sec:classical}.
In Section \ref{sec:CCH} we lay out a canonical quantization of the electromagnetic potential and field on this background. 
In the following section we give a description of the main physical quantum states.
Section \ref{sec:theta->pi-theta Symmetry} is split into two subsections. 
In the first one we derive expressions for the expectation value of the electromagnetic stress-energy tensor when the field
is in these states. 
We also show that similar expressions given by CCH result in RSETs
which are not symmetric under the parity operation.
In the second subsection we give a physical interpretation of the various sets of terms appearing in 
these expectation values.
In Section \ref{sec:luminosity} we calculate the luminosity of the Kerr black hole in the past Boulware and past
Unruh states for the spin-1 case. 
In the last section we study the behaviour close to the horizon of the difference in the RSET between a Hartle-Hawking-type state defined in CCH and
the past Boulware state.

The method we followed to solve the radial Teukolsky equation is described in Appendix \ref{sec:radial func.},
whereas we solved the angular equation as described in ~\cite{ar:Casals&Ott'04}. 
Further details can be found in ~\cite{th:CasalsPhD}.
In Appendix \ref{sec:asympts. close to r_+} we extend to the spin-1 case an asymptotic analysis in ~\cite{ar:Candelas'80} for 
the radial solutions close to the horizon in the scalar case.

We use  geometrized Planck units: $c=G=\hbar=1$,
and follow the sign conventions of Misner, Thorne and Wheeler ~\cite{bk:M&T&W}.       
All figures have been obtained for the values: $a=0.95M$ and $M=1$;
the Boyer-Lindquist radius of the event horizon for such values of the black hole
parameters is $r_+\simeq 1.3122$.


\section{Classical theory}  \label{sec:classical}

The Kerr metric corresponding to a black hole of mass $M$ and intrinsic angular 
momentum $a$ per unit mass as viewed from infinity is, in the Boyer-Lindquist co-ordinate system  $\{t,r,\theta,\phi\}$, given by
\begin{equation} \label{eq:K-N metric}
\begin{aligned}
\d{s}^2
&=-\frac{\Delta}{\Sigma}\left(\d{t}-a\sin^2\theta\d{\phi}\right)^2+\frac{\sin^2\theta}{\Sigma}\left[(r^2+a^2)\d{\phi}-a\d{t}\right]^2+
\frac{\Sigma}{\Delta}\d{r}^2+\Sigma\d{\theta}^2
\end{aligned}
\end{equation}
where $\Sigma\equiv r^2+a^2\cos^2\theta$ and $\Delta \equiv r^2-2Mr+a^2$. 
The Kerr metric has co-ordinate singularities at $\Delta=0$, which has for solutions $r_{\pm}=M\pm\sqrt{M^2-a^2}$.
The hypersurface $r=r_+$ is the outer event horizon.
The surface gravity on the outer[inner] horizon is $\kappa_+[\kappa_-]$, given by
$\kappa_{\pm}\equiv (r_{\pm}-r_{\mp})/(2(r_{\pm}^2+a^2))$. 

The Kerr space-time is stationary and axially symmetric;
the two Killing vectors associated with these two symmetries are 
$\vec{k}_{\manifold{I}}\equiv \partial / \partial t$ and $\vec{k}_{\rot}\equiv \partial/\partial\phi$ 
respectively.
We can construct another Killing vector as
$\vec{k}_{\manifold{H}}\equiv \vec{k}_{\manifold{I}}+\Omega_+\vec{k}_{\rot}$
with $\Omega_+\equiv a/(r_+^2+a^2)$. 
The surface where the Killing vector $\vec{k}_{\manifold{I}}$ becomes a null vector 
is called the stationary limit surface. 
The vector $\vec{k}_{\manifold{I}}$ is timelike outside this surface and is spacelike in
the region between the event horizon and the stationary limit surface. 
This region is called the ergosphere.
The surface where the Killing vector $\vec{k}_{\manifold{H}}$ becomes null is called the speed-of-light surface.
The vector $\vec{k}_{\manifold{H}}$ is timelike between the event horizon and the speed-of-light surface, and it is spacelike 
outside it.

An observer who moves along a world line of constant $r$ and $\theta$ with an angular velocity 
$\Omega\equiv\d{\phi}/\d{t}$
relative to the asymptotic rest frame has a tetrad $\{\vec{e}_{(t)},\vec{e}_{(r)},\vec{e}_{(\theta)},\vec{e}_{(\phi)}\}$ 
associated with him. Such an observer sees no local change in the geometry and is therefore considered a stationary
observer relative to the local geometry. If his angular velocity is zero, and therefore he moves along a world line of 
constant $r$, $\phi$ and $\theta$, he is a static observer (SO) relative to radial infinity. A SO moves along the integral
curves of $\vec{k}_{\manifold{I}}$.         
If we require $\vec{e}_{(r)}$ and $\vec{e}_{(\theta)}$ to be parallel to 
$\partial/\partial r$ and $\partial/\partial \theta$ 
respectively, we then find that the two other
vectors in the tetrad of a stationary observer are given by        
\begin{equation} \label{eq:tetrad of stationary obs.}
\begin{aligned}
\vec{e}_{(t)}&=\frac{1}{\sqrt{\left|g_{tt}+2\Omega g_{t\phi}+\Omega^2g_{\phi\phi}\right|}}\left(\pardiff{}{t}+\Omega\pardiff{}{\phi}\right) \\
\vec{e}_{(\phi)}&=\frac{1}{\sqrt{\left|g_{tt}+2\Omega g_{t\phi}+\Omega^2g_{\phi\phi}\right|}}\frac{1}{\sqrt{g^2_{t\phi}-g_{tt}g_{\phi\phi}}}
\left[-(g_{t\phi}+\Omega g_{\phi\phi})\pardiff{}{t}+(g_{tt}+\Omega g_{t\phi})\pardiff{}{\phi}\right]
\end{aligned}
\end{equation}
The quantity $p_{\phi}\equiv \vec{p}\vec{k}_{\rot}=p_{\alpha}k_{\rot}^{\alpha}$, where $\vec{p}$ is the 4-momentum of a certain observer, is the component
of angular momentum of that observer along the black hole's spin axis.
This quantity is conserved for geodesic observers.
The only stationary observers for whom this quantity is zero are those with an angular velocity $\Omega_{\text{ZAMO}}=-g_{t\phi}/g_{\phi\phi}$. 
These observers are called zero angular momentum observers (ZAMO). 
These observers are the closest analogue to the static observers in Schwarzschild space-time on the Kerr-Newman space-time in the
sense that their 4-velocity is orthogonal to $\vec{k}_{\rot}$, the hypersurfaces of constant $t$. 
The angular velocity $\Omega_{\text{ZAMO}}$ of the ZAMOs as they approach the horizon tends to $\Omega_+$, which can therefore be
interpreted as the angular velocity of the horizon. The stationary observers whose angular velocity is constant and equal to $\Omega_+$ are observers
that follow integral curves of $\vec{k}_{\manifold{H}}$ and are called rigidly rotating observers (RRO).        
The last orthonormal tetrad that is of interest to us is the Carter orthonormal tetrad ~\cite{ar:Carter'68b}, 
which corresponds to a stationary observer (\ref{eq:tetrad of stationary obs.}) with angular velocity $\Omega_{\text{CARTER}}=a/(r^2+a^2)$. 

The Newman-Penrose (NP) formalism (~\cite{ar:N&P'62}) is based on four null basis vectors: $\{\vec{l},\vec{n},\vec{m},\vec{m^*}\}$.
In this paper we will use the Kinnersley tetrad:
\begin{subequations} \label{eq:def. null Kinnersley tetrad}
\begin{align}
\vec{l}&=\frac{1}{\Delta}\left[(r^2+a^2)\pardiff{}{t}+\Delta\pardiff{}{r}+a\pardiff{}{\phi}\right]\\
\vec{n}&=\frac{1}{2\Sigma}\left[(r^2+a^2)\pardiff{}{t}-\Delta\pardiff{}{r}+a\pardiff{}{\phi}\right]\\
\vec{m}&=\frac{-\rho^*}{\sqrt 2}\left[ia\sin\theta\pardiff{}{t}+\pardiff{}{\theta}+i\cosec\theta\pardiff{}{\phi}\right]
\end{align}
\end{subequations}
where $\rho=-1/(r-ia\cos\theta)$ is one of the spin coefficients of the Kerr metric in the Kinnersley tetrad.
The other spin coefficients and NP scalars can be found in ~\cite{bk:Chandr}. Of interest to us are the 
NP Maxwell scalars: $\phi_{-1} \equiv F_{\vec{l}\vec{m}}$, $\phi_{0} \equiv \left(F_{\vec{l}\vec{n}}+F_{\vec{m^*}\vec{m}}\right)/2$
and $\phi_{+1} \equiv F_{\vec{m^*}\vec{n}}$, where $F_{\alpha \beta}\equiv A_{\beta ;\alpha}-A_{\alpha;\beta}$ is the anti-symmetric Maxwell tensor
and $A_{\alpha}$ is the electromagnetic potential.
From these equations, one can obtain the operator $K_{\indhel}^{\mu}$ which maps the potential onto the NP scalars, i.e., 
\begin{equation}\label{eq:phi as func. of potential with K}
\phi_{\indhel}=K_{\indhel}^{\mu}A_{\mu}
\end{equation}
The explicit form of the operator $K_{\indhel}^{\mu}$ can be found in ~\cite{ar:Chrzan'75}.

It is easy to see that the effect of the parity operator $\mathcal{P}:(\theta,\phi)\to(\pi-\theta,\phi+\pi)$ on the NP Kinnersley tetrad 
is
\begin{equation} \label{eq:parity op. on NP objs.}
\mathcal{P}\vec{l}=\vec{l}, \qquad\qquad \mathcal{P}\vec{n}=\vec{n},  \qquad\qquad \mathcal{P}\vec{m}=-\vec{m}^*
\end{equation}
In the limit $r\rightarrow+\infty$, the Kinnersley tetrad becomes
\begin{equation} \label{eq:Kinnersley tetrad, r->inf}
\begin{aligned}\vec{l}&\rightarrow -\hat{\vec{e}}_t+\hat{\vec{e}}_r, & 
\qquad 
\vec{n}&\rightarrow -\frac{1}{2}\left(\hat{\vec{e}}_t+\hat{\vec{e}}_r\right), & 
\qquad 
\vec{m}&\rightarrow +\frac{1}{\sqrt{2}}\left(\hat{\vec{e}}_{\theta}+i\hat{\vec{e}}_{\phi}\right) & 
\qquad (r\rightarrow+\infty) 
\end{aligned}
\end{equation}
in terms of the usual unit polar vectors 
$\{\hat{\vec{e}}_t,\hat{\vec{e}}_r,\hat{\vec{e}}_{\theta},\hat{\vec{e}}_{\phi}\}$ 
in flat space-time.

By making use of the NP formalism, Teukolsky (~\cite{ar:Teuk'72}, ~\cite{ar:Teuk'73}) decoupled the equations describing the 
spin-0,-1/2,-1 and -2 linear perturbations in a general Type D vacuum background.
He further showed that, using the Kinnersley tetrad, some of the decoupled equations are separable in Boyer-Lindquist co-ordinates in the Kerr background.
We will refer to the second order ordinary differential equations for the variables $r$ and $\theta$ resulting from this separation as the 
radial and angular Teukolsky equations respectively.
The terminology we will use is the following.
A bullet $\bullet$ (and a primed bullet $\bullet'$) superscript indicates either `in', `up', `down' or `out' modes, which are standard but, for clarity, are
defined in Appendix \ref{sec:radial func.}.
Correspondingly, the symbol $\omega^{\bullet}$ is defined as being equal to $\omega$ when it is part of an expression
containing `in' modes and it is equal to $\tilde{\omega}$ when the expression contains `up' modes.
The separability of the equations describing the spin-1 perturbations implies that the electromagnetic potential
can be expressed as a Fourier mode sum as
\begin{subequations}
\begin{align}
A_{\mu}^{\bullet}&=\int_{-\infty}^{+\infty}\d{\omega}\sum_{l=|\indhel|}^{+\infty}\sum_{m=-l}^{+l}\sum_{P=\pm 1}a_{lm\omega P}^{\bullet}\ {}_{lm\omega P}A_{\mu}^{\bullet}
\label{eq:Fourier expansion for A} \\
{}_{lm\omega P}A_{\mu}^{\bullet}&\equiv {}_{lm\omega}A_{\mu}^{\bullet}+P\mathcal{P}{}_{lm\omega}A_{\mu}^{\bullet} \label{eq: def. of lmwPA_mu}
\end{align}
\end{subequations}
where $a_{lm\omega P}^{\bullet}$ are the coefficients of the Fourier series.
The parameter $l$ labels the eigenvalues of the angular Teukolsky equation.
The parameter $P$ is summed in (\ref{eq:Fourier expansion for A}) over the values $+1$ and $-1$, corresponding 
to two linearly independent polarization states for the potential, as we shall see in Section \ref{sec:theta->pi-theta Symmetry}.
The decomposition of the potential into eigenstates of the parity operator $\mathcal{P}$ is the natural choice because of the 
invariance of the Kerr metric under this operation. 

Chrzanowski ~\cite{ar:Chrzan'75} was the first author to give
analytic expressions, using Teukolsky's results,  for the linear electromagnetic and gravitational perturbation potentials in the Kerr background. 
He obtained, for the electromagnetic potential modes:
\begin{subequations} \label{eq:tableIChrzan.}
\begin{align}
\begin{split}
&{}_{l-m-\omega P}A{}^{\text{in}}_{\mu}
=
\\
&=\Big\{
\left[l_{\mu}(\delta^*+2\beta^*+\tau^*)-m^{*}_{\mu}(D+2\epsilon^*+\rho^*)\right]
{}_{-1}R_{lm\omega }^{\text{in}}(r){}_{+1}Z_{lm\omega }(\theta,\phi)e^{-i\omega t}
+\\&+P
\left[l_{\mu}(\delta+2\beta+\tau)-m_{\mu}(D+2\epsilon+\rho)\right]
{}_{-1}R_{lm\omega }^{\text{in}}(r){}_{-1}Z_{lm\omega }(\theta,\phi)e^{-i\omega t}
\Big\}
\end{split} \label{eq:A_1, tableIChrzan.} \\
\begin{split}
&{}_{l-m-\omega P}A{}^{\text{up}}_{\mu}
=
\\
&=\Big\{
\rho^{* -2}\left[-n_{\mu}(\delta-2\alpha^*+\pi^*)+m_{\mu}(\Delta-2\gamma^*+\mu^*)\right] 
{}_{+1}R_{lm\omega }^{\text{up}}(r){}_{-1}Z_{lm\omega }(\theta,\phi)e^{-i\omega t}
+\\&+P
\rho^{* -2}\left[-n_{\mu}(\delta^*-2\alpha+\pi)+m^{*}_{\mu}(\Delta-2\gamma+\mu)\right] 
{}_{+1}R_{lm\omega }^{\text{up}}(r){}_{+1}Z_{lm\omega }(\theta,\phi)e^{-i\omega t}
\Big\}
\end{split}
\label{eq:A1, tableIChrzan.}
\end{align}
\end{subequations}
where
\begin{equation}
{}_{\indhel}Z_{lm\omega}(\theta,\phi)\equiv \frac{(-1)^{m+1}}{\sqrt{2\pi}}{}_{\indhel}S_{lm\omega}(\theta)e^{+im\phi}.
\end{equation}
The only difference here with Chrzanowski's original expressions 
is an overall change in the sign of $m$ and $\omega$, 
justified by the sum over $m$ and integration over $\omega$.

It is immediate from the radial and angular Teukolsky equations that their respective solutions ${}_{\indhel}R_{lm\omega}(r)$
and ${}_{\indhel}S_{lm\omega}(\theta)$ satisfy the following symmetries:
\begin{subequations} \label{eq:R symms.}
\begin{align}
&{}_{\indhel}R_{lm\omega}(r)=\Delta^{-\indhel }{}_{-\indhel }R_{lm\omega}^*(r)   \label{eq:R symm.->cc,-s} \\
&{}_{\indhel}R_{lm\omega}(r)={}_{\indhel}R_{l-m-\omega}^*(r)  \label{eq:R symm.->cc,-m,-w}
\end{align}
\end{subequations}
and
\begin{subequations} \label{eq: S symms}
\begin{align}
&{}_{\indhel}S_{lm\omega}(\theta)=(-1)^{l+m}{}_{-\indhel}S_{lm\omega}(\pi-\theta)   \label{eq:S symm.->pi-t,-s} \\
&{}_{\indhel}S_{lm\omega}(\theta)=(-1)^{l+\indhel }{}_{\indhel}S_{l-m-\omega}(\pi-\theta)  \label{eq:S symm.->pi-t,-m,-w}\\
&{}_{\indhel}S_{lm\omega}(\theta)=(-1)^{\indhel+m}{}_{-\indhel}S_{l-m-\omega}(\theta)     \label{eq:S symm.->-s,-m,-w}
\end{align}
\end{subequations}

Note the symmetry property
\begin{equation}
\mathcal{P}{}_{lm\omega}A_{\mu}^{\bullet}=(-1)^{l+m}{}_{l-m-\omega}A_{\mu}^{\bullet *}
\end{equation}
which is easily obtained from (\ref{eq:R symm.->cc,-m,-w}), (\ref{eq:S symm.->-s,-m,-w}) and (\ref{eq:tableIChrzan.}).
It then follows that the Fourier sum coefficients must satisfy the following condition so that the potential remains real:
\begin{equation} \label{eq: c.c. of lmwPa}
a_{lm\omega P}^{\bullet *}=(-1)^{l+m}Pa_{l-m-\omega P}^{\bullet}
\end{equation}

The Maxwell scalars $\phi_{\indhel}^{\bullet}$ can be decomposed in a Fourier mode sum analogous to that in (\ref{eq:Fourier expansion for A}).
Following ~\cite{ar:CMSR} and ~\cite{bk:Chandr}, the Maxwell scalars modes can be expressed as
\begin{equation} \label{eq:phi_0/2(in/up)}
\begin{aligned}
{}_{l-m-\omega}\phi_{-1}^{\text{in}}&=-\frac{1}{2}{}_{+1}R^{\text{in}}_{lm\omega}{}_{+1}Z_{lm\omega}e^{-i\omega t} \\ 
{}_{l-m-\omega}\phi_{+1}^{\text{in}}&=-\frac{{}_{1}B_{lm\omega}}{2}\rho^2{}_{-1}R^{\text{in}}_{lm\omega}{}_{-1}Z_{lm\omega}e^{-i\omega t} \\
{}_{l-m-\omega}\phi_{-1}^{\text{up}}&=-\frac{{}_{1}B_{lm\omega}}{2}{}_{+1}R^{\text{up}}_{lm\omega}{}_{+1}Z_{lm\omega}e^{-i\omega t}      \\
{}_{l-m-\omega}\phi_{+1}^{\text{up}}&=-\frac{{}_{1}B_{lm\omega}^2}{2}\rho^2{}_{-1}R^{\text{up}}_{lm\omega}{}_{-1}Z_{lm\omega}e^{-i\omega t}
\end{aligned}
\end{equation}
and 
\begin{equation} \label{eq:phi0(ch)}
\begin{aligned}
{}_{l-m-\omega}\phi_{0}^{\text{in}}&=-\frac{\rho^2}{2^{3/2}{}_{1}B_{lm\omega}}\left[\left(\rho^{-1}\mathcal{D}_0^{\dagger}+1\right)
\mathcal{L}^{\dagger}_{1}+ia\sin\theta\mathcal{D}_0^{\dagger}\right]\Delta{}_{+1}R_{lm\omega}^{\text{in}}{}_{-1}Z_{lm\omega}e^{-i\omega t}= \\
&=\frac{\rho^2}{\sqrt{2}}\left[\left(\rho^{-1}\mathcal{D}_0+1\right)
\mathcal{L}_{1}+ia\sin\theta\mathcal{D}_0\right]{}_{-1}R_{lm\omega}^{\text{in}}{}_{+1}Z_{lm\omega}e^{-i\omega t} \\
{}_{l-m-\omega}\phi_{0}^{\text{up}}&=-\frac{\rho^2}{2^{3/2}}\left[\left(\rho^{-1}\mathcal{D}_0^{\dagger}+1\right)
\mathcal{L}^{\dagger}_{1}+ia\sin\theta\mathcal{D}_0^{\dagger}\right]\Delta{}_{+1}R_{lm\omega}^{\text{up}}{}_{-1}Z_{lm\omega}e^{-i\omega t}= \\
&=\frac{\rho^2{}_{1}B_{lm\omega}}{\sqrt{2}}\left[\left(\rho^{-1}\mathcal{D}_0+1\right)
\mathcal{L}_{1}+ia\sin\theta\mathcal{D}_0\right]{}_{-1}R_{lm\omega}^{\text{up}}{}_{+1}Z_{lm\omega}e^{-i\omega t}
\end{aligned}
\end{equation}
where ${}_1B_{lm\omega}^2\equiv {}_{-1}\lambda_{lm\omega}^2+4ma\omega-4a^2\omega^2$ and 
${}_{\indhel}\lambda_{lm\omega}$ is the eigenvalue of the angular Teukolsky equation.
We are using the definitions of the operators
\begin{equation} \label{eq:def. L_n}
\mathcal{L}_{n}^
{ \topbott{}{\dagger}} \equiv \partial_{\theta} \pm \mathcal{Q}+n\cot\theta  
\qquad \text{and} \qquad 
\mathcal{D}_{n}^
{ \topbott{}{\dagger}} \equiv \partial_{r} \mp \frac{iK}{\Delta}+2n\frac{r-M}{\Delta}    
\end{equation}
where $\mathcal{Q}\equiv -a\omega\sin\theta+m/\sin\theta$ and $K\equiv (r^2+a^2)\omega-am$.
We use the convention that the upper and lower symbols inside braces go 
with the upper and lower signs in the equation.

The asymptotic behaviour of the NP Maxwell scalars, separately for outgoing and ingoing waves, in the limit $r\to +\infty$ is 
\begin{equation} \label{eq:peeling th}
\begin{aligned}
{}_{lm\omega}\phi_{+1}&\sim r^{-1}e^{+i\omega r},     &\quad r^{-3}e^{-i\omega r}&\quad (r\rightarrow +\infty) \\
{}_{lm\omega}\phi_{0}&\sim r^{-2}e^{+i\omega r},     &\quad r^{-2}e^{-i\omega r}&\quad (r\rightarrow +\infty) \\
{}_{lm\omega}\phi_{-1}&\sim r^{-3}e^{+i\omega r},     &\quad r^{-1}e^{-i\omega r}&\quad (r\rightarrow +\infty)
\end{aligned}
\end{equation}
It is therefore the scalar $\phi_{+1[-1]}$ the one with the asymptotically dominant behaviour 
for the upgoing[ingoing] waves. The above asymptotic behaviour (\ref{eq:peeling th}) was originally obtained by
Newman and Penrose ~\cite{ar:N&P'62} and is commonly referred to as the peeling theorem. 

The Teukolsky-Starobinski\u{\i} identities relate solutions, radial or angular,
of opposite helicity $\indhel$. These identities are (~\cite{ar:Teuk&Press'74}):
\begin{subequations} \label{eq:Teuk-Starob. ids.}
\begin{align}
\mathcal{D}_0\mathcal{D}_0{}_{-1}R_{lm\omega}&=
\frac{1}{2}{}_{+1}R_{lm\omega} & \mathcal{L}_0\mathcal{L}_1{}_{+1}S_{lm\omega}&={}_1B_{lm\omega}{}_{-1}S_{lm\omega} \label{eq:a)Teuk-Starob. ids.} \\
\Delta\mathcal{D}^{\dagger}_0\mathcal{D}^{\dagger}_0\Delta{}_{+1}R_{lm\omega}&=2{}_1B_{lm\omega}^2{}_{-1}R_{lm\omega} & 
\mathcal{L}^{\dagger}_0\mathcal{L}^{\dagger}_1{}_{-1}S_{lm\omega}&={}_1B_{lm\omega}{}_{+1}S_{lm\omega} \label{eq:b)Teuk-Starob. ids.}
\end{align}
\end{subequations}

The angular Teukolsky equation has singular points at $x=\pm 1$, where $x\equiv \cos\theta$. 
By using the Frobenius method it can be found that the solution that is regular at both 
boundary points $x=+1$ and $-1$ is given by                             
\begin{equation} \label{eq:asympt. S for x->+/-1}
{}_{\indhel}S_{lm\omega}(x)=(1-x)^{|m+\indhel |/2}(1+x)^{|m-\indhel |/2}{}_{\indhel}y_{lm\omega}(x)
\end{equation}
and the function 
${}_{\indhel}y_{lm\omega}(x)$ behaves close to the boundary points as
\begin{equation} \label{eq:asympt. y for x->+/-1}
{}_{\indhel}y_{lm\omega}(x)=\sum_{n=0}^{\infty} {{}_{\indhel}\topbott{a}{b}_{n,lm\omega}(1\mp x)^n}    \quad   \text{for} \quad x\to \pm1            
\end{equation}

The solutions ${}_{\indhel}Z_{lm\omega}$ are called the spin-weighted spheroidal harmonics (SWSH).
In the spherical limit $a=0$, the SWSH reduce to the spin-weighted spherical harmonics ${}_{\indhel}Y_{lm}$.
We give here two properties (~\cite{ar:J&McL&Ott'91}, ~\cite{th:McL'90} and ~\cite{ar:J&McL&Ott'95}) 
satisfied by the spin-weighted spherical harmonics which we will need later on:
\begin{equation} \label{eq:eq.2aJ,McL,Ott'91}
\sum_{m=-l}^l {}_{-1}Y_{lm}(\theta,\phi){}_{+1}Y_{lm}^*(\theta,\phi)=0
\end{equation}
and the ``addition theorem''  
\begin{equation} \label{eq:eq.B6J,McL,Ott'95}
\sum_{m=-l}^l {}_{\indhel}Y_{lm}(\theta,\phi){}_{\indhel}Y^*_{lm}(\theta',\phi')=\frac{2l+1}{4\pi}P_l(\cos\gamma)
\end{equation}
where $\gamma$ is defined by $\cos\gamma\equiv \cos\theta\cos\theta'+\sin\theta\sin\theta'\cos(\phi-\phi')$.

The classical, electromagnetic stress-energy tensor can be expressed
in terms of the NP Maxwell scalars as
\begin{equation} \label{eq:stress tensor, spin 1}
\begin{aligned}
T_{\mu\nu}&=\Big\{\phi_{-1}\phi_{-1}^*n_{\mu}n_{\nu}+2\phi_0\phi_0^*\left[l_{(\mu}n_{\nu)}+
m_{(\mu}m_{\nu)}^*\right]+\phi_{+1}\phi_{+1}^*l_{\mu}l_{\nu}-
\\ 
&
-4\phi_0\phi_{-1}^*n_{(\mu}m_{\nu)}-4\phi_{+1}\phi_0^*l_{(\mu}m_{\nu)}+2\phi_{+1}\phi_{-1}^*m_{\mu}m_{\nu}\Big\}+c.c.
\end{aligned}
\end{equation}
in the case of absence of a charge current.
By virtue of the Maxwell field equations, the stress-energy tensor 
(\ref{eq:stress tensor, spin 1})
satisfies the conservation equation $T^{\mu\nu}{}_{;\mu}=0$. 

Note that it follows from (\ref{eq:parity op. on NP objs.}) that all the pairs of null tetrad vectors  
appearing in the different terms in (\ref{eq:stress tensor, spin 1}) remain invariant under
the parity operation, except for the ones that have a factor containing $\phi_0$ together with either $\phi_{-1}$ or $\phi_{+1}$, which
change sign. That is, a pair of null 
vectors $\vec{e}_{(a)}\vec{e}_{(b)}$ 
appearing in (\ref{eq:stress tensor, spin 1}) with a factor
$\phi_{\indhel}\phi_{\indhel'}^*$ changes under the parity operation as
\begin{equation}  \label{eq:pairs of vects. in stress tensor under parity}
\mathcal{P}\left(\vec{e}_{(a)}\vec{e}_{(b)}\right)=(-1)^{\indhel+\indhel'}\vec{e}_{(a)}\vec{e}_{(b)}
\end{equation}

The variable $\Gamma$ refers to either the potential components $A_{\mu}$ or the NP scalars $\phi_{\indhel}$.
The Fourier series expansion for either the potential components or the NP Maxwell scalars may then be expressed as
\begin{equation} \label{eq:Fourier series of Gamma field}
\Gamma^{\bullet}= \sum_{lmP} \int_{-\infty}^{+\infty}\d{\omega^{\bullet}}a_{lm\omega P}^{\bullet}{}_{lm\omega P}\Gamma^{\bullet}
\end{equation}
which may be re-arranged as
\begin{equation} \label{eq:Fourier series of Gamma field for pos.freq.}
\Gamma^{\bullet}=\sum_{lmP} \int_{0}^{+\infty}\d{\omega^{\bullet}}\left(a_{lm\omega P}^{\bullet}{}_{lm\omega P}\Gamma^{\bullet}+
(-1)^{l+m}Pa_{lm\omega P}^{\bullet *}{}_{l-m-\omega P}\Gamma^{\bullet}\right)
\end{equation}

We can now use the symmetry relations
\begin{subequations} \label{eq: symms. of lmwA_mu and lmwphi_i}
\begin{align}
\mathcal{P}{}_{lm\omega P}A_{\mu}^{\bullet}&=(-1)^{l+m}{}_{l-m-\omega P}A_{\mu}^{\bullet *}=
P{}_{lm\omega P}A_{\mu}^{\bullet} \label{eq: symm. of lmwA_mu}
\\
\mathcal{P}{}_{lm\omega}\phi_{\indhel}^{\bullet}&=(-1)^{l+m+1+\indhel}{}_{l-m-\omega}\phi_{\indhel}^{\bullet *} \label{eq: symm. of lmwphi_i}
\end{align}
\end{subequations}

The equations above for $\Gamma$ are equally valid for $A_{\mu}$ and $\phi_{\indhel}$. In particular, we only need to 
apply the operator $K^{\mu}_{\indhel}$ to an equation for $A_{\mu}$ in order to obtain the 
corresponding equation for $\phi_{\indhel}$. However the last step in (\ref{eq: symm. of lmwA_mu}) has no 
equivalent for  $\phi_{\indhel}$ in (\ref{eq: symm. of lmwphi_i}).   
The reason is that
\begin{equation} \label{eq:parity term is pure gauge}
K^{\mu}_{\indhel}\ \mathcal{P}{}_{lm\omega}A_{\mu}^{\bullet} \propto 
K^{\mu}_{\indhel}\ {}_{l-m-\omega}A_{\mu}^{\bullet *} \equiv 0
\end{equation}
as can be checked; that is, this term is pure gauge. 
Hence the fact that
${}_{lm\omega}\phi_{\indhel}^{\bullet}=K^{\mu}_{\indhel}\ {}_{lm\omega P}A_{\mu}^{\bullet}$ 
does not actually depend on $P$ which is why $P$ is not a subindex of the NP scalar modes. 
The potential is real whereas the field 
components are not, as seen in equations (\ref{eq:classical mode expansion for A(in)_mu}) 
and (\ref{eq:classical mode expansion for phi(in)_i}) below. 
From (\ref{eq:Fourier series of Gamma field for pos.freq.}) and using (\ref{eq: symm. of lmwA_mu}) and 
(\ref{eq: symm. of lmwphi_i}) for the potential and the field respectively, we have
\begin{equation} \label{eq:classical mode expansion for A(in)_mu}
\begin{aligned}
A_{\mu}^{\bullet}
&=\sum_{lmP} \int_{0}^{+\infty}\d{\omega^{\bullet}}\left(a_{lm\omega P}^{\bullet}{}_{lm\omega P}A_{\mu}^{\bullet}+
Pa_{lm\omega P}^{\bullet *}\mathcal{P}{}_{lm\omega P}A_{\mu}^{\bullet *}\right)=            \\    
&=\sum_{lmP} \int_{0}^{+\infty}\d{\omega^{\bullet}}\left(a_{lm\omega P}^{\bullet}{}_{lm\omega P}A_{\mu}^{\bullet}+
a_{lm\omega P}^{\bullet *}{}_{lm\omega P}A_{\mu}^{\bullet *}\right)
\end{aligned}
\end{equation}
and
\begin{equation} \label{eq:classical mode expansion for phi(in)_i}
\phi_{\indhel}^{\bullet}=\sum_{lmP} \int_{0}^{+\infty}\d{\omega^{\bullet}}\left(a_{lm\omega P}^{\bullet}{}_{lm\omega}\phi_{\indhel}^{\bullet}+
(-1)^{\indhel+1}Pa_{lm\omega P}^{\bullet *}\mathcal{P}{}_{lm\omega}\phi_{\indhel}^{\bullet *}\right)
\end{equation}


\section{Quantization of the electromagnetic potential/field}  \label{sec:CCH}

The abundance in the literature of the quantization of the scalar field in a curved background is in sharp contrast
with the scarce treatment of the quantization of the electromagnetic -or gravitational- field in such a background.
CCH did quantize both the electromagnetic and the
gravitational fields in the Kerr background. They used a canonical quantization method, which is the one
we have chosen to use in this paper.

We quantize the field by promoting $a_{lm\omega P}^{\bullet}$ and $a_{lm\omega P}^{\bullet *}$ to 
operators $\hat{a}_{lm\omega P}^{\bullet}$ and $\hat{a}_{lm\omega P}^{\bullet \dagger}$ respectively.
Cohen and Kegeles ~\cite{ar:Coh&Keg'74} and Wald ~\cite{ar:Wald'78} have shown
that the theory may be expressed in terms of one
single NP complex scalar, which represents the two radiative degrees of freedom of the electromagnetic perturbations. 
If we introduce expansion (\ref{eq:classical mode expansion for phi(in)_i}) 
for the NP scalars into $T^{00}$ given by (\ref{eq:stress tensor, spin 1}), we then obtain a hamiltonian which is a superposition of 
independent harmonic oscillator hamiltonians, one for each mode of the electromagnetic field. 
From the standard quantization of the harmonic oscillator, we know that the operators 
$\hat{a}_{lm\omega P}^{\bullet}$ and $\hat{a}_{lm\omega P}^{\bullet \dagger}$ must
satisfy the commutation relations:
\begin{equation} \label{eq:commut. rlns. for a,a_dagger}
\begin{aligned}
\left[\hat{a}_{lm\omega P}^{\bullet},\hat{a}_{l'm'\omega'P'}^{\bullet \dagger}\right]=\delta(\omega-\omega')\delta_{ll'}\delta_{mm'}\delta_{PP'} \\
\Big[\hat{a}_{lm\omega P}^{\bullet},\hat{a}_{l'm'\omega'P'}^{\bullet}\Big]=\left[\hat{a}_{lm\omega P}^{\bullet \dagger},\hat{a}_{l'm'\omega'P'}^{\bullet \dagger}\right]=0
\end{aligned}
\end{equation}

These commutation relations are satisfied provided that the orthonormality conditions
\begin{equation} \label{eq:orthonormality conds. for potential}
\begin{aligned}
\left\langle {}_{lm\omega P}A^{\bullet}_{\alpha},{}_{l'm'\omega'P'}A^{\bullet'}_{\alpha}\right\rangle_{\mathcal{S}}&=
\delta_{\bullet \bullet'}\delta_{ll'}\delta_{mm'}\delta(\omega-\omega')\delta_{PP'} \\
\left\langle {}_{lm\omega P}A^{\bullet *}_{\alpha},{}_{l'm'\omega'P'}A^{\bullet'}_{\alpha}\right\rangle_{\mathcal{S}}&=0
\end{aligned}
\end{equation}
are satisfied, where $\mathcal{S}$ is any complete Cauchy hypersurface for the outer region of the space-time 
and where the Klein-Gordon inner product is taken as  
\begin{equation} \label{eq:def. inner prod.}
\left\langle \psi_{\alpha},\varphi_{\alpha}\right\rangle_{\mathcal{S}}=
i\int_{\mathcal{S}}\d^3{\Sigma}^{\mu}\left(\psi^{\alpha *}\nabla_{\mu}\varphi_{\alpha}-\varphi^{\alpha}\nabla_{\mu}\psi^*_{\alpha}+
\varphi_{\mu}\nabla_{\alpha}\psi^{\alpha *}-\psi^*_{\mu}\nabla_{\alpha}\varphi^{\alpha}\right)
\end{equation}
The inner product (\ref{eq:def. inner prod.}) has the same form as the one taken by CCH. 
However, CCH give an expression for the stress-energy tensor which includes a factor $4\pi$ in (\ref{eq:stress tensor, spin 1}),
corresponding to unrationalized units. If unrationalized units are used, then a factor $4\pi$ should also be included in the
inner product (\ref{eq:def. inner prod.}). 

Constants of normalization are to be included in front of the radial functions so that the potential modes (\ref{eq:tableIChrzan.}) satisfy
the orthonormality conditions (\ref{eq:orthonormality conds. for potential}) given the asymptotic behaviour of the radial functions in (\ref{eq:R_in/up}).
We find that the constants of normalization are given by:
\begin{subequations}  \label{eq:normalization consts.}
\begin{align}
|N_{-1}^{\text{in}}|^2&=\frac{1}{2^5\omega^3\pi} \label{eq:N_1(in) normalization const.} \\
|N_{+1}^{\text{up}}|^2&=\frac{1}{2^3\pi|\EuFrak{N}|^2\tilde{\omega}(r_+^2+a^2)} \label{eq:N1(up) normalization const.}\\
|N_{-1}^{\text{up}}|^2&=|N_{+1}^{\text{up}}|^2\left|\frac{{}_{-1}R^{\text{up,inc}}_{lm\omega}}{{}_{+1}R^{\text{up,inc}}_{lm\omega}}\right|^2=
\frac{\tilde{\omega}(r_+^2+a^2)}{2\pi {}_1B_{lm\omega}^4} \label{eq:N_1(up) normalization const.}
\end{align}
\end{subequations}
where $\EuFrak{N} \equiv iK_++(r_+-r_-)/2$ and $K_+\equiv K(r_+)$.
We have chosen ${}_{-1}R^{\text{in,inc}}_{lm\omega}=1$ and ${}_{+1}R^{\text{up,inc}}_{lm\omega}=1$ in equations (\ref{eq:N_1(in) normalization const.})
and (\ref{eq:N1(up) normalization const.}) respectively.
The constant of normalization $|N_{-1}^{\text{up}}|$ is calculated as indicated with the use of (\ref{eq:R1 coeffs from R_1's}). It is 
therefore the constant of normalization that corresponds to using the radial function (\ref{eq:R_up}) when setting ${}_{-1}R^{\text{up,inc}}_{lm\omega}=1$,
which is the actual normalization we have used in the numerical calculation of the `up' solutions.
The NP scalars are therefore assumed to include the constants of normalization (\ref{eq:normalization consts.}). 
That is, the NP scalar modes ${}_{lm\omega}\phi_{\indhel}^{\bullet}$ are to be calculated from expressions 
(\ref{eq:phi_0/2(in/up)}) and (\ref{eq:phi0(ch)}) with the inclusion of the appropriate constant of normalization (\ref{eq:normalization consts.}),
while the radial functions remain unaltered.

It can be checked that the set of modes $\{{}_{lm\omega P}A_{\mu}^{\text{in}},{}_{lm\omega P}A_{\mu}^{\text{up}}\}$
forms a complete set of orthonormal solutions to the Maxwell equations in the outer region of the Kerr space-time. 
Similarly, it can be checked that $\{{}_{lm\omega P}A_{\mu}^{\text{out}},{}_{lm\omega P}A_{\mu}^{\text{down}}\}$ also
form a complete set.
We may expand the electromagnetic potential by using the
complete set of solutions $\{{}_{lm\omega P}A_{\mu}^{\text{in}},{}_{lm\omega P}A_{\mu}^{\text{up}}\}$ and then quantize it as:
\begin{equation} \label{eq:quantum mode expansion for A in in/up modes}
\begin{aligned}
\hat{A}_{\mu}&
=\sum_{lmP} \int_{0}^{+\infty}\d{\omega}
\left(\hat{a}_{lm\omega P}^{\text{in}}{}_{lm\omega P}A_{\mu}^{\text{in}}+
\hat{a}_{lm\omega P}^{\text{in} \dagger}{}_{lm\omega P}A_{\mu}^{\text{in} *}\right)+\\ 
&+\sum_{lmP} \int_{0}^{+\infty}\d{\tilde{\omega}}
\left(\hat{a}_{lm\omega P}^{\text{up}}{}_{lm\omega P}A_{\mu}^{\text{up}}+
\hat{a}_{lm\omega P}^{\text{up} \dagger}{}_{lm\omega P}A_{\mu}^{\text{up} *}\right)
\end{aligned}
\end{equation}

Alternatively, we could proceed exactly in the same manner but using the complete set of solutions
$\{{}_{lm\omega P}A_{\mu}^{\text{out}},{}_{lm\omega P}A_{\mu}^{\text{down}}\}$ instead.
The result is then:
\begin{equation} \label{eq:quantum mode expansion for A in out/dn modes}
\begin{aligned}
\hat{A}_{\mu}&=
\sum_{lmP} \int_{0}^{+\infty}\d{\omega}\left(\hat{a}_{lm\omega P}^{\text{out}}{}_{lm\omega P}A_{\mu}^{\text{out}}+
\hat{a}_{lm\omega P}^{\text{out} \dagger}{}_{lm\omega P}A_{\mu}^{\text{out} *}\right)+\\ 
&+\sum_{lmP} \int_{0}^{+\infty}\d{\tilde{\omega}}\left(\hat{a}_{lm\omega P}^{\text{down}}{}_{lm\omega P}A_{\mu}^{\text{down}}+
\hat{a}_{lm\omega P}^{\text{down} \dagger}{}_{lm\omega P}A_{\mu}^{\text{down} *}\right)
\end{aligned}
\end{equation}

The asymptotic behaviour in terms of the advanced $v$ and retarded $u$ time co-ordinates of the electromagnetic
potential and NP scalars for the `in' and `up' modes is the same as the one exhibited by the modes (\ref{eq:X_in/up as func. of u,v}).
The same applies to the asymptotic behaviour of the 
`out' and `down' modes exhibited in (\ref{eq:X_out/down as func. of u,v}).
Accordingly, the operators $\hat{a}_{lm\omega P}^{\text{in} \dagger}$, $\hat{a}_{lm\omega P}^{\text{up} \dagger}$, 
$\hat{a}_{lm\omega P}^{\text{out} \dagger}$ and $\hat{a}_{lm\omega P}^{\text{down} \dagger}$ are creation operators of particles incident
from past null infinity $\mathcal{I}^-$, past horizon $\mathcal{H}^-$, future null infinity $\mathcal{I}^+$ and 
future horizon $\mathcal{H}^+$ respectively.  

Since the `in' and `out' modes are only defined for $\omega$ non-negative, they have non-negative energy as measured by
an observer following the integral curve of $\vec{k}_{\manifold{I}}$, by virtue of (\ref{eq:hamiltonians on in/up modes}).
Similarly, the `up' and `down' modes, defined for $\tilde{\omega}$ non-negative, have non-negative energy with respect
to observers following the integral curve of $\vec{k}_{\manifold{H}}$. 

We may now construct the stress-energy tensor operator from either the potential operator 
(\ref{eq:quantum mode expansion for A in in/up modes}) or (\ref{eq:quantum mode expansion for A in out/dn modes}).
It is well-known that the stress-energy tensor as an operator does not have a well-defined meaning. It suffers from ultra-violet divergences 
and its expectation value when the field is in a certain state $\ket{\Psi}$ must be renormalized. 
There are several techniques for renormalization.
The point-splitting technique consists in starting from each quadratic term in the stress-energy tensor and
temporarily displacing one field point, thus forming the object $\vac{\hat{T}_{\alpha\beta}(x,x')}{\Psi}$, 
which is finite. 
Specific divergent terms, gathered in the bitensor $T^{\text{div}}_{\alpha\beta}(x,x')$, 
which are purely geometric and thus independent of the quantum state, are then subtracted from $\vac{\hat{T}_{\alpha\beta}(x,x')}{\Psi}$.
The end result is obtained by finally bringing the separated points together:
\begin{equation}
\vac[ren]{\hat{T}_{\alpha\beta}(x)}{\Psi}=\lim_{x'\to x}\left(\vac{\hat{T}_{\alpha\beta}(x,x')}{\Psi}-T^{\text{div}}_{\alpha\beta}(x,x')\right)
\end{equation}
It is this renormalized expectation value of the stress-energy tensor (RSET)
that is the source in Einstein's field equations in the semiclassical theory.
Christensen ~\cite{ar:Christ'78} has explicitly calculated the divergent terms $T^{\text{div}}_{\alpha\beta}$ by using covariant geodesic point separation. 
Jensen, McLaughlin and Ottewill ~\cite{ar:J&McL&Ott'88} calculated a linearly divergent term for the spin-1 case, 
which was not explicitly given by Christensen. The reason being that this term does not have to be included when an average 
is taken over the covariant derivative of the biscalar of geodetic interval $\sigma^{\mu}$ and $-\sigma^{\mu}$, 
as performed by Christensen.

Before we start a description of the various physical states of the field, we give an important result found by 
Unruh ~\cite{ar:Unruh'76} and further established by ~\cite{ar:Brown&Ott'83} and ~\cite{ar:Grove&Ott'83}. The result is that a `particle detector'
will react to states of the field which have positive frequency with respect to the detector's proper time.  
If a certain observer $\mathcal{A}$ makes measurements relative to a certain vacuum state $\ket{\Xi}$, then 
he or she measures a stress tensor 
\begin{equation}   \label{eq:stress for gral. obs.&vac.}
\vac[\mathcal{A}]{\hat{T}_{\alpha\beta}}{\Psi}=\vac{\hat{T}_{\alpha\beta}}{\Psi}-\vac{\hat{T}_{\alpha\beta}}{\Xi}
\end{equation}
when the field is in a certain state $\ket{\Psi}$.


\section{Vacua states} \label{sec:vac. states}

The properties of the various states described in this section have been obtained in the literature for the scalar case,
except where explicitly indicated otherwise. 

The Boulware vacuum state, denoted by $\ket{B}$, is defined in Schwarzschild space-time as the vacuum that corresponds 
to quantizing the field with normal modes that have all positive frequency with respect to the space-time's hypersurface-orthogonal
timelike killing vector $\vec{k}_{\manifold{I}}$. 
This state respects the isometries of Schwarzschild space-time.
Since it is the static observers SO the ones that move along integral curves
of $\vec{k}_{\manifold{I}}$, from Unruh's result stated in the previous section it follows that these observers will make
measurements relative to the Boulware vacuum $\ket{B}$. 
Candelas ~\cite{ar:Candelas'80}, based on conjectures made previously by Christensen and Fulling ~\cite{ar:Christ&Fulling'77},
has found that the RSET when the scalar field is in the Boulware vacuum is zero at both $\mathcal{I}^-$ and $\mathcal{I}^+$
in the Schwarzschild space-time. 
Candelas also found that the RSET, close to the horizon, when the field is in the Boulware vacuum diverges and corresponds to the
absence from the vacuum of black-body radiation at the black hole temperature appropriately red-shifted. 
The Boulware vacuum is therefore irregular at 
$\mathcal{H}^-$ and $\mathcal{H}^+$.

In Schwarzschild space-time the Boulware vacuum may be associated with the field expansion in terms of either the
`in' and `up' modes or the `out' and `down', both pairs of sets of complete modes defining the same vacuum $\ket{B}$.
We can perform a similar expansion for the electromagnetic potential in the Kerr space-time.
The past Boulware state is defined by
\begin{equation} \label{eq:a^in/up on B}
\begin{aligned}
\hat{a}_{lm\omega P}^{\text{in}}\ket{B^-}&=0  \\
\hat{a}_{lm\omega P}^{\text{up}}\ket{B^-}&=0
\end{aligned}
\end{equation}
corresponding to an absence of particles at $\mathcal{H}^-$ and $\mathcal{I}^-$. 
We can also define the future Boulware state, as that state which is empty at $\mathcal{I}^+$ and $\mathcal{H}^+$:  
\begin{equation} \label{eq:a^out/down on B}
\begin{aligned}
\hat{a}_{lm\omega P}^{\text{out}}\ket{B^+}&=0  \\
\hat{a}_{lm\omega P}^{\text{down}}\ket{B^+}&=0
\end{aligned}
\end{equation}
The Bogolubov transformation between the pair of operators $\hat{a}_{lm\omega P}^{\text{in}}$ and $\hat{a}_{lm\omega P}^{\text{up}}$
and the pair $\hat{a}_{lm\omega P}^{\text{down}}$ and $\hat{a}_{lm\omega P}^{\text{out}}$ is non-trivial: 
the expression for $\hat{a}_{lm\omega P}^{\text{in} \dagger}\left[\hat{a}_{lm\omega P}^{\text{up} \dagger}\right]$ 
in terms of `out' and `down' operators contains $\hat{a}_{l,-m,-\omega,P}^{\text{down} \dagger}\left[\hat{a}_{l,-m,-\omega,P}^{\text{out} \dagger}\right]$
for modes in the superradiant regime.    
This implies that the past Boulware state contains both outgoing and downgoing superradiant particles, 
and is therefore not empty at $\mathcal{I}^+$ and $\mathcal{H}^+$.
This flux of particles out to $\mathcal{I}^+$ corresponds to the Starobinski\u{\i}-Unruh effect.  
Similarly, the future Boulware state contains ingoing and upgoing superradiant particles, and is therefore not empty 
at $\mathcal{I}^-$ and $\mathcal{H}^-$. 
As $\hat{a}_{lm\omega P}^{\text{up}}$, when expressed in terms of `out' and `down' operators, contains 
the creator operator $\hat{a}_{l,-m,-\omega,P}^{\text{out} \dagger}$, it is not possible to construct a state which is empty at both 
$\mathcal{I}^-$ and $\mathcal{I}^+$, unlike the situation in the Schwarzschild space-time.

From the definitions (\ref{eq:a^in/up on B}) and (\ref{eq:a^out/down on B}) together with the relations (\ref{NP scalars in/up->out/down}),
the past and future Boulware states are obtainable
one from the other under the transformation $(t,\phi)\to (-t,-\phi)$. 
Because the two states are not equivalent, it follows that neither is invariant under this symmetry of the Kerr space-time. 

The defining features of a Hartle-Hawking state (~\cite{ar:Hartle&Hawk'76})  is that it possesses the symmetries
of the space-time and that it is regular everywhere, including on both the past and the future event horizons. 
Kay and Wald ~\cite{ar:Kay&Wald'91} have proven that for any globally hyperbolic space-time which has a Killing field with a bifurcate Killing 
horizon there can be at most one state with the above features. 
Kay and Wald have further shown that for the Kerr space-time this state does not exist. 
Rindler and Schwarzschild space-times are covered by Kay and Wald's theorem;
in Rindler space-time this state is clearly the Minkowski vacuum.  

In Schwarzschild space-time the state $\ket{H}$ corresponds to quantizing the
field with upgoing normal modes which on $\mathcal{H}^-$ have positive frequency with respect to the Kruskal co-ordinate $U\equiv -e^{-\kappa_+u}$
and with ingoing normal modes which on $\mathcal{H}^+$ have positive frequency with respect to the Kruskal co-ordinate $V\equiv e^{\kappa_+v}$. 
Candelas ~\cite{ar:Candelas'80} showed that the state defined in this manner is regular on both the past and 
future horizons. He also found that the RSET at infinity when the field is in the $\ket{H}$ state corresponds 
to that of a bath of black body radiation at the black hole temperature $T_H=\kappa_+/(2\pi)$. The Hartle-Hawking state models a black
hole in (unstable) thermal equilibrium with an infinite distribution at the Hawking temperature. 

From the above results and from the previous section we know that $\vac{\hat{T}_{\alpha\beta}}{H-B}$ is thermal both for
$r\rightarrow r_+$ and for $r\rightarrow +\infty$. Christensen and Fulling conjectured that this is the case everywhere.
However, Jensen, McLaughlin and Ottewill \cite{ar:J&McL&Ott'92} numerically showed that $\vac{\hat{T}_{\alpha\beta}}{H-B}$
deviates from isotropic, thermal form as one moves away from the horizon.

Candelas showed that the RSET close to the horizon when the field is in the $\ket{B}$ state diverges
like minus the stress tensor of black body radiation at the black hole temperature, and that it must tend  
to $-\vac[SO]{\hat{T}_{\alpha\beta}}{H}$, due to (\ref{eq:stress for gral. obs.&vac.}) and to the regularity of $\ket{H}$. 
Analogously, Unruh ~\cite{ar:Unruh'76} showed that in flat space-time and when the field is in the Minkowski vacuum, 
a Rindler observer RO will also see a bath of black body radiation at the Hawking temperature of a black hole 
with surface gravity $\kappa_+=a\alpha$, where $a$ is the RO's acceleration and $\alpha$ is the lapse function in Rindler space.   

Frolov and Thorne ~\cite{ar:F&T'89} defined a new ``Hartle-Hawking'' state $\ket{FT}$ invariant under the symmetries of the Kerr space-time
by using a variant of the $\eta$ formalism (which employs non-standard commutation relations for the creation and annihilation operators). They
proved that the RSET when the field is in the $\ket{FT}$ state is finite at the horizon but that, at least
for arbitrarily slow rotation, it is equal to the stress tensor of a thermal distribution at the Hawking
temperature rigidly rotating with the horizon. This suggests that it becomes irregular wherever $\vec{k}_{\manifold{H}}$ is not timelike,
that is on and outside the speed-of-light surface.          
Ottewill and Winstanley ~\cite{ar:Ott&Winst'00}, however, proved that although $\ket{FT}$ has a Feynman propagator
with the correct properties for regularity on the horizons, its two-point function is actually pathological almost everywhere, not just outside the
speed-of-light surface. Only at the axis of symmetry, where all the modes in the two-point function for the scalar field 
are evaluated for $\tilde{\omega}=\omega$ (i.e., $m=0$), it does not suffer from this pathology. 

Frolov and Thorne claim that close to the horizon ZAMOs make measurements relative to an unspecified Boulware vacuum.
They also claim that, when the field is in the state $\ket{FT}$, ZAMOs measure close to the horizon 
a stress tensor equal to that of a thermal distribution at the Hawking temperature rigidly rotating with the horizon. 

Duffy ~\cite{th:GavPhD} modified the Kerr space-time by introducing a mirror and constructed a state $\ket{H_{\mathcal{M}}}$
for the scalar field that is invariant under the isometries of the modified space-time.
He then showed that $\ket{H_{\mathcal{M}}}$ is regular everywhere in the modified space-time if, and only if, the mirror
removes the region outside the speed-of-light surface.  
He constructed another state, $\ket{B_{\mathcal{M}}}$, invariant under the isometries of the modified space-time and empty on both the past
and future horizons. 
This is the state that RROs make measurements relative to in the modified space-time. 
He also numerically showed that when the field is in the $\ket{H_{\mathcal{M}}}$ state the stress tensor measured by a RRO
is, close to the horizon, that of a thermal distribution at the Hawking temperature rigidly rotating with the horizon. 

CCH defined a new Hartle-Hawking-type state, which we will hereafter denote by $\ket{CCH^-}$. 
This state is obtained by thermalizing the `in' and `up' modes with respect to their natural energy. 
Ottewill and Winstanley ~\cite{ar:Ott&Winst'00} showed that this state is, however, not invariant under the symmetry transformation 
$(t,\phi)\to (-t,-\phi)$ of the space-time.
They further argued that the RSET when the scalar field is in the $\ket{CCH^-}$ state is regular on the future horizon but irregular on the past horizon.
We must note that these results were derived in ~\cite{ar:Ott&Winst'00} based on a stress-energy tensor for which the $t\theta$- and $\phi\theta$-components
are identically zero. We shall see in Section \ref{sec:luminosity} that although this is indeed the case for the scalar field, which is the case
they considered, this is most probably not true for the electromagnetic field.
A similar state could be constructed by applying the transformation $(t,\phi)\to (-t,-\phi)$ to the state $\ket{CCH^-}$. 
This state, suitably named $\ket{CCH^+}$, would then be irregular on the future horizon and regular on the past horizon.
In Section \ref{sec:RRO} we will investigate
the form close to the horizon of $\vac[ren]{\hat{T}^{\mu}{}_{\nu}}{CCH^--B^-}$ for electromagnetism.

Finally, Unruh ~\cite{ar:Unruh'76} constructed a state in the Schwarzschild space-time 
by expanding the
scalar field in modes that are positive frequency with respect to the proper time $t$ of 
inertial observers in $\mathcal{I^-}$ and modes that are positive frequency with respect to
the proper time of inertial observers close to $\mathcal{H^-}$.            
It is possible to construct a state in the Kerr space-time with the same positive-frequency mode definitions.
We call this state the past Unruh state, $\ket{U^-}$.


\section{Expectation value of the stress tensor} \label{sec:theta->pi-theta Symmetry}


\subsection{Analytic expressions} \label{subsec:new expressions}

In this subsection we derive new, analytic expressions for the expectation value of the stress-energy tensor
when the electromagnetic field is in quantum states of interest in the Kerr space-time.
It is well-known that there is an operator-ordering ambiguity in the transition from classical to quantum theory:
the classical term $\phi_{\indhel}\phi_{\indhel'}^*$ should be quantized to
the symmetrized form 
$\left(\hat{\phi}_{\indhel}\hat{\phi}_{\indhel'}^{\dagger}+\hat{\phi}_{\indhel'}^{\dagger}\hat{\phi}_{\indhel}\right)/2$.
We will calculate separately the expectation value of the two quadratic terms in the symmetrized form.

It is straight-forward to check that in the past Boulware state we have
\begin{equation} \label{eq:phi(in)_iphi(in,dagger)_j on vac.0}
\begin{aligned}
\bra{B^-}
\hat{\phi}_{\indhel} \hat{\phi}_{\indhel'}^{\dagger}
\ket{B^-}
=&
\sum_{lmP}\Bigg(\int_{0}^{+\infty}\d{\tilde{\omega}}\ {}_{lm\omega}\phi^{\text{up}}_{\indhel}\ {}_{lm\omega}\phi_{\indhel'}^{\text{up} *}
+\int_{0}^{+\infty}\d{\omega}\ {}_{lm\omega}\phi^{\text{in}}_{\indhel}\ {}_{lm\omega}\phi_{\indhel'}^{\text{in} *}\Bigg)
=
\\=
&(-1)^{\indhel+\indhel'}\mathcal{P}\left(\bra{B^-}
\hat{\phi}_{\indhel'}^{\dagger} \hat{\phi}_{\indhel}
\ket{B^-}\right)
^*
\end{aligned}
\end{equation}

The sign $(-1)^{\indhel+\indhel'}$ is precisely the same sign appearing in (\ref{eq:pairs of vects. in stress tensor under parity}). 
This implies that if the quadratic terms in the expression (\ref{eq:stress tensor, spin 1})
are quantum-mechanically symmetrized when promoting the NP scalars to operators, 
then the expectation value in the state $\ket{B^-}$ of the stress-energy tensor will be invariant under parity.

In order to calculate the expectation value of the quadratic terms 
$\hat{\phi}_{\indhel}\hat{\phi}_{\indhel'}^{\dagger}$ and
$\hat{\phi}_{\indhel'}^{\dagger}\hat{\phi}_{\indhel}$ in the past Unruh state we are going to
make use of the expression calculated in ~\cite{ar:F&T'89} which gives the past Unruh state in terms of the past Boulware state:
\begin{equation} \label{eq:U in terms of B}
\ket{U^-}=\prod_{lm\tilde{\omega}P}C_{lm\omega P}\exp\left(e^{-\pi\tilde{\omega}/\kappa_+}\hat{a}_{lm\omega P}^{\text{up} \dagger}
\hat{a}_{lm\omega P}^{\text{up'} \dagger}\right)\ket{B^-}
\end{equation}
where $C_{lm\omega P}$ are normalization constants and $\hat{a}_{lm\omega P}^{\text{up'} \dagger}$ are
creation operators in the left hand region of the extended Kerr space-time. 
We will also make use of the following expression in ~\cite{ar:Schum&Caves'85}:
\begin{equation} \label{eq:eq3.66Schum&Caves'85}
\begin{aligned}
&S(r,\phi)=\\
&=(\cosh r)^{-1}\exp\left(-\hat{a}^{\dagger}_+\hat{a}^{\dagger}_-e^{2i\phi}\tanh r\right)
\exp\left(-(\hat{a}^{\dagger}_+\hat{a}_{+}+\hat{a}^{\dagger}_-\hat{a}_-)\ln(\cosh r)\right)\exp\left(\hat{a}_+\hat{a}_{-}e^{-2i\phi}\tanh r\right)
\end{aligned}
\end{equation}
where $S(r,\phi)$ is the two-mode squeeze operator
\begin{equation} \label{eq:S op.} 
S(r,\phi)=\exp\left(r(\hat{a}_+\hat{a}_{-}e^{-2i\phi}-\hat{a}^{\dagger}_+\hat{a}^{\dagger}_-e^{2i\phi})\right)
\end{equation}
and the independent operators $\hat{a}_+$ and $\hat{a}_{-}$ satisfy the standard commutation relations.
By using (\ref{eq:eq3.66Schum&Caves'85}) and (\ref{eq:S op.}) we can re-express (\ref{eq:U in terms of B}) as
\begin{equation}
\ket{U^-}=\exp{\left\{\sum_{lmP}\int_0^{\infty}\d{\tilde{\omega}}\Big[\ln C_{lm\omega P}+\ln(\cosh r_{\tilde{\omega}})\Big]\right\}}e^{-\hat{A}}\ket{B^-}
\end{equation}
with
\begin{equation}
\hat{A}\equiv\sum_{lmP}\int_0^{\infty}\d{\tilde{\omega}}\  
r_{\tilde{\omega}}\left(\hat{a}_{lm\omega P}^{\text{up} \dagger}\hat{a}_{lm\omega P}^{\text{up'} \dagger}-
\hat{a}_{lm\omega P}^{\text{up}}\hat{a}_{lm\omega P}^{\text{up'}}\right)
\qquad \text{and}\qquad
r_{\tilde{\omega}}\equiv -\tanh^{-1} \left(e^{-\pi\tilde{\omega}/\kappa_+}\right).
\end{equation}
Since $\left(e^{\hat{A}}\right)^{\dagger}=e^{-\hat{A}}$, 
the normalization $\left\langle\right.U^- \left|\right.U^-\left.\right\rangle=1$ implies 
\begin{equation}
\exp{\left\{\sum_{lmP}\int_0^{\infty}\d{\tilde{\omega}}\Big[\ln C_{lm\omega P}+\ln C^*_{lm\omega P}+2\ln(\cosh r_{\tilde{\omega}})\Big]\right\}}=1
\end{equation}

Using now the Baker-Campbell-Hausdorff equation (~\cite{bk:Louisell}) we find that
\begin{equation} \label{}
e^{\hat{A}}\hat{a}_{lm\omega P}^{\text{up}}e^{-\hat{A}}=
\hat{a}_{lm\omega P}^{\text{up}}\cosh r_{\tilde{\omega}}+\hat{a}_{lm\omega P}^{\text{up'} \dagger}\sinh r_{\tilde{\omega}}
\end{equation}
and finally
\begin{subequations} \label{eq:a_dagger*a on unruh}
\begin{align}
\bra{U^-}\hat{a}_{lm\omega P}^{\text{up} \dagger}\hat{a}_{l'm'\omega'P'}^{\text{up}}\ket{U^-}&=
\frac{1}{2}\left[\coth \left(\frac{\pi\tilde{\omega}}{\kappa_+}\right)-1\right]\delta(\omega-\omega')\delta_{ll'}\delta_{mm'}\delta_{PP'} \\
\bra{U^-}\hat{a}_{lm\omega P}^{\text{up} \dagger}\hat{a}_{l'm'\omega'P'}^{\text{up} \dagger}\ket{U^-}&=0
=\bra{U^-}\hat{a}_{lm\omega P}^{\text{up}}\hat{a}_{l'm'\omega'P'}^{\text{up}}\ket{U^-}
\end{align}
\end{subequations}
It immediately follows that
\begin{equation}  \label{eq:phi*phi^dagger `up' on U-}
\begin{aligned}
&\bra{U^-}\hat{\phi}_{\indhel}\hat{\phi}_{\indhel'}^{\dagger}\ket{U^-}=  
\\ &=
\sum_{lmP}\Bigg(\frac{1}{2}\int_0^{\infty}\d{\tilde{\omega}}
\begin{aligned}[t]
\bigg\{
&
\left[{}_{lm\omega}\phi_{\indhel}^{\text{up}}{}_{lm\omega}\phi_{\indhel'}^{\text{up} *}
+(-1)^{\indhel+\indhel'}\mathcal{P}({}_{lm\omega}\phi_{\indhel}^{\text{up} *}{}_{lm\omega}\phi_{\indhel'}^{\text{up}})\right]
\coth\left(\frac{\pi\tilde{\omega}}{\kappa_+}\right)+
\\ +&
\left[{}_{lm\omega}\phi_{\indhel}^{\text{up}}{}_{lm\omega}\phi_{\indhel'}^{\text{up} *}-
(-1)^{\indhel+\indhel'}\mathcal{P}({}_{lm\omega}\phi_{\indhel}^{\text{up} *}{}_{lm\omega}\phi_{\indhel'}^{\text{up}})\right]
\bigg\}
+ 
\end{aligned}
\\ &\qquad\qquad+
\int_0^{\infty}\d{\omega}\ {}_{lm\omega}\phi_{\indhel}^{\text{in}}\ {}_{lm\omega}\phi_{\indhel'}^{\text{in} *}
\Bigg)=
\\ & =
(-1)^{\indhel+\indhel'}\mathcal{P}\left(\bra{U^-}\hat{\phi}_{\indhel'}^{\dagger}\hat{\phi}_{\indhel}\ket{U^-}\right)^*
\end{aligned}
\end{equation}
Note the minus sign in the second term in (\ref{eq:phi*phi^dagger `up' on U-}). Its presence may seem a bit surprising at first but,
as we shall now see, it is precisely this sign that causes the expectation value of the stress-energy tensor in the
past Unruh state to adopt a more familiar form by having all `up' terms multiplied by a $\coth$ factor. 
This is already clear from looking at (\ref{eq:phi*phi^dagger `up' on U-}) and realizing that
when quantum-symmetrizing the classical expression $\phi_{\indhel}^{\text{up}}\phi_{\indhel'}^{\text{up} *}$ 
the terms without a $\coth$ factor will cancel out.

The following identities are therefore immediately satisfied
\begin{equation}
\begin{aligned}
\vac{\left[\hat{\phi}_{\indhel}^{\bullet},\hat{\phi}_{\indhel'}^{\bullet \dagger}\right]}{U^-}&=
\vac{\left[\hat{\phi}_{\indhel}^{\bullet},\hat{\phi}_{\indhel'}^{\bullet \dagger}\right]}{B^-}=\\
&=\sum_{lmP}\int_0^{\infty}\d{\omega^{\bullet}}
\left[{}_{lm\omega}\phi_{\indhel}^{\bullet}{}_{lm\omega}\phi_{\indhel'}^{\bullet *}
-(-1)^{\indhel+\indhel'}\mathcal{P}\left({}_{lm\omega}\phi_{\indhel}^{\bullet *}{}_{lm\omega}\phi_{\indhel'}^{\bullet}\right)\right]
\end{aligned}
\end{equation}
and hence
\begin{equation}
\vac{\left[\hat{\phi}_{\indhel}^{\bullet},\hat{\phi}_{\indhel'}^{\bullet \dagger}\right]}{U^--B^-}=0
\end{equation}
as it should be.

When the classical term $\left(\phi_{\indhel}\phi_{\indhel'}^*+c.c.\right)$ is quantized to
the symmetrized term $\left(\hat{\phi}_{\indhel}\hat{\phi}_{\indhel'}^{\dagger}+\hat{\phi}_{\indhel'}^{\dagger}\hat{\phi}_{\indhel}\right)/2+h.c.$,
where the symbol $`h.c.'$ stands for hermitian conjugate,
we obtain the following real, parity-invariant expressions in the past Boulware and past Unruh states:
\begin{subequations}
\begin{align}
\begin{split}
&\vac{\frac{\hat{\phi}_{\indhel}\hat{\phi}_{\indhel'}^{\dagger}+\hat{\phi}_{\indhel'}^{\dagger}\hat{\phi}_{\indhel}}{2}+h.c.}{B^-}= 
\\ &=
\frac{1}{2}\sum_{lmP}\bigg(\int_0^{\infty}\d{\tilde{\omega}}
\left[
%
{}_{lm\omega}\phi_{\indhel}^{\text{up}}{}_{lm\omega}\phi_{\indhel'}^{\text{up} *}+
(-1)^{\indhel+\indhel'}\mathcal{P}({}_{lm\omega}\phi_{\indhel}^{\text{up}}{}_{lm\omega}\phi_{\indhel'}^{\text{up} *})\right]+
\\ &
\qquad \quad\ \
+
\int_0^{\infty}\d{\omega}
\left[{}_{lm\omega}\phi_{\indhel}^{\text{in}}{}_{lm\omega}\phi_{\indhel'}^{\text{in} *}+
(-1)^{\indhel+\indhel'}\mathcal{P}({}_{lm\omega}\phi_{\indhel}^{\text{in}}{}_{lm\omega}\phi_{\indhel'}^{\text{in} *})\right]
\bigg)+c.c.
\end{split}\\
\begin{split}
&\vac{\frac{\hat{\phi}_{\indhel}\hat{\phi}_{\indhel'}^{\dagger}+\hat{\phi}_{\indhel'}^{\dagger}\hat{\phi}_{\indhel}}{2}+h.c.}{U^-}= 
\\ &=
\frac{1}{2}\sum_{lmP}\bigg(\int_0^{\infty}\d{\tilde{\omega}}
\left[{}_{lm\omega}\phi_{\indhel}^{\text{up}}{}_{lm\omega}\phi_{\indhel'}^{\text{up} *}
+(-1)^{\indhel+\indhel'}\mathcal{P}({}_{lm\omega}\phi_{\indhel}^{\text{up}}{}_{lm\omega}\phi_{\indhel'}^{\text{up} *})\right]
\coth\left(\frac{\pi\tilde{\omega}}{\kappa}\right)+
\\ &
\qquad \quad\ \
+
\int_0^{\infty}\d{\omega}
\left[{}_{lm\omega}\phi_{\indhel}^{\text{in}}{}_{lm\omega}\phi_{\indhel'}^{\text{in} *}+
(-1)^{\indhel+\indhel'}\mathcal{P}({}_{lm\omega}\phi_{\indhel}^{\text{in}}{}_{lm\omega}\phi_{\indhel'}^{\text{in} *})\right]
\bigg)+c.c.
\end{split}
\end{align}
\end{subequations}

Note that in the above expressions we have been able to complex conjugate the mode functions that are operated on by $\mathcal{P}$
because of the existence of the $+c.c.$ terms.
This immediately leads to the following real, parity-invariant expressions for the stress-energy tensor in
the past Boulware and past Unruh states:
\begin{subequations} \label{eq:corrected stress tensor for s=1 on B-,U-}
\begin{align}
\begin{split}
&\expct{\hat{T}_{\mu\nu}}{B^-}= \\
&=\frac{1}{2}\sum_{lmP}\left(
\int_0^{\infty}\d{\tilde{\omega}}\, 
\Big\{T_{\mu\nu}
\left[{}_{lm\omega}\phi_{\indhel}^{\text{up}},{}_{lm\omega}\phi_{\indhel}^{\text{up} *}\right]
+(-1)^{\vartheta}\mathcal{P}\left(T_{\mu\nu}
\left[{}_{lm\omega}\phi_{\indhel}^{\text{up}},{}_{lm\omega}\phi_{\indhel}^{\text{up} *}\right]
\right)\Big\}
+  \right. \\
& 
\qquad \quad \ \
\left. +\int_0^{\infty}\d{\omega}\, 
\Big\{T_{\mu\nu}
\left[{}_{lm\omega}\phi_{\indhel}^{\text{in}},{}_{lm\omega}\phi_{\indhel}^{\text{in} *}\right]
+(-1)^{\vartheta}\mathcal{P}\left(T_{\mu\nu}
\left[{}_{lm\omega}\phi_{\indhel}^{\text{in}},{}_{lm\omega}\phi_{\indhel}^{\text{in} *}\right]
\right)\Big\}
\right)   
\end{split} \label{eq:corrected stress tensor for s=1 on B-} \\ \begin{split}
&\expct{\hat{T}_{\mu\nu}}{U^-}= 
\frac{1}{2}\sum_{lmP}
\\&
\left(
\int_0^{\infty}\d{\tilde{\omega}}\, 
\coth\left(\frac{\pi\tilde{\omega}}{\kappa_+}\right)
\Big\{T_{\mu\nu}
\left[{}_{lm\omega}\phi_{\indhel}^{\text{up}},{}_{lm\omega}\phi_{\indhel}^{\text{up} *}\right]
+(-1)^{\vartheta}\mathcal{P}\left(T_{\mu\nu}
\left[{}_{lm\omega}\phi_{\indhel}^{\text{up}},{}_{lm\omega}\phi_{\indhel}^{\text{up} *}\right]
\right)\Big\}
+  \right. \\
& 
\qquad \qquad \quad
\left. +\int_0^{\infty}\d{\omega}\, 
\Big\{T_{\mu\nu}
\left[{}_{lm\omega}\phi_{\indhel}^{\text{in}},{}_{lm\omega}\phi_{\indhel}^{\text{in} *}\right]
+(-1)^{\vartheta}\mathcal{P}\left(T_{\mu\nu}
\left[{}_{lm\omega}\phi_{\indhel}^{\text{in}},{}_{lm\omega}\phi_{\indhel}^{\text{in} *}\right]
\right)\Big\}
\right) \end{split} \label{eq:corrected stress tensor for s=1 on U-}
\end{align}
\end{subequations}
We use the obvious notation that $T_{\mu\nu}\left[{}_{lm\omega}\phi_{\indhel}^{\bullet},{}_{lm\omega}\phi_{\indhel}^{\bullet *}\right]$ 
denotes the general expression for the stress-energy tensor (\ref{eq:stress tensor, spin 1}) where the scalars 
$\phi_{\indhel}$ have been replaced by the modes ${}_{lm\omega}\phi_{\indhel}^{\bullet}$. 
We will also use the symbol ${}_{lm\omega}T^{\bullet}_{\mu\nu}$ to refer to 
$T_{\mu\nu}\left[{}_{lm\omega}\phi_{\indhel}^{\bullet},{}_{lm\omega}\phi_{\indhel}^{\bullet *}\right]$.
The variable $\vartheta$ is defined so that $(-1)^{\vartheta}$ is equal to -1 if one
index of the component of the stress tensor is $\theta$ and the other one is not, and it is equal to +1 otherwise. 
Note that the sign $(-1)^{\vartheta}$ appears in the above expressions instead of $(-1)^{\indhel+\indhel'}$ by virtue of the change under the parity operation 
of the coefficients of the quadratic field operators that appear in the expression for the stress-energy tensor, as seen in
(\ref{eq:pairs of vects. in stress tensor under parity}).

To our knowledge, these important expressions for the electromagnetic field in the Kerr space-time have only been given so far by CCH.
Their expressions, however, differ from ours in that they do not include the terms explicitly containing the operator $\mathcal{P}$.
As a result, the expressions for the expectation value of the stress tensor that CCH give are not symmetric under the parity operation 
$\mathcal{P}:(\theta,\phi)\to (\pi-\theta,\phi+\pi)$,
even though this is a symmetry of both the Kerr metric and the Maxwell field equations.

We initially used CCH's expressions in our numerical calculations and the numerical results we
obtained were clearly not symmetric under the parity operation.
We also showed analytically that at least for one particular instance the RSET is not symmetric under $\mathcal{P}$.
We outline here the derivation.

From equation (\ref{eq:asympt. y for x->+/-1}) 
and the two equivalent expressions for ${}_{lm\omega}\phi_{0}$ given in (\ref{eq:phi0(ch)}) it follows that
 $|{}_{lm\omega}\phi_{0}|^2$ is only non-zero at $\theta=0,\pi$ if $m=0$ and 
its value, for the `up' modes, in that case is
\begin{equation} \label{eq: phi1^2 0->pi}  
\begin{aligned}
&\abs{{}_{l,m=0,\omega}\phi_{0}^{\text{up}}(r,\theta=0)}^2-\abs{{}_{l,m=0,\omega}\phi_{0}^{\text{up}}(r,\theta=\pi)}^2= \\
&=\frac{-4ia|N^{\text{up}}_{+1}|^2W[{}_{+1}R,{}_{-1}R^{*}]^{\text{up}}_{l,m=0,\omega}({}_{-1}a_{n=0,l,m=0,\omega}\ {}_{+1}a_{n=0,l,m=0,\omega})}
{(r^2+a^2)^2{}_1B_{l,m=0,\omega}^2}
\end{aligned}
\end{equation}
By virtue of the property (\ref{eq: wronskian in=-up}),
the corresponding value for the `in' modes is obtained by merely changing the sign.

Since $T^{\text{div}}_{\mu\nu}$ is a purely geometrical object and the metric is invariant under $\mathcal{P}$, this
divergent stress tensor must also be invariant under $\mathcal{P}$.  
From this property together with equations (\ref{eq:stress tensor, spin 1}) and (\ref{eq: phi1^2 0->pi})
we find that CCH's expressions for the expectation value of the stress tensor yield:
\begin{equation} \label{eq:T_thetatheta 0->pi,past Unruh}
\begin{aligned}
&\vac[ren]{\hat{T}_{\theta\theta}(r,\theta=0)}{U^-}-\vac[ren]{\hat{T}_{\theta\theta}(r,\theta=\pi)}{U^-}= \\
&=\sum_{l=0}^{\infty}
\int_{0}^{\infty}\d{\omega}\left[\coth\left(\frac{\pi \omega}{\kappa}\right)-1\right]
\times \\ & \qquad \quad \times
\frac{2^3ia|N^{\text{up}}_{+1}|^2W[{}_{+1}R,{}_{-1}R^{*}]^{\text{up}}_{l,m=0,\omega}\ {}_{+1}a_{n=0,l,m=0,\omega}^2}
{(r^2+a^2)^2{}_1B_{l,m=0,\omega}^2}
\sqrt{\frac{{}_{-1}\lambda_{l,m=0,\omega}+2a\omega}{{}_{-1}\lambda_{l,m=0,\omega}-2a\omega}}
\end{aligned}
\end{equation}
where we have made use of the Teukolsky-Starobinski\u{\i} identities (\ref{eq:Teuk-Starob. ids.})
to relate the SWSH for $\indhel=+1$ with the one for $\indhel=-1$ in the limit $x\to +1$.

It is easy to check that 
$iW[{}_{+1}R,{}_{-1}R^{*}]^{\text{up}}_{l,m=0,\omega}=
{}_1B^2_{l,m=0,\omega}\abs{{}_{-1}R^{\text{up,tra}}_{l,m=0,\omega}}^2/\omega \ge 0$ 
as long as 
$\omega \ge 0$, and therefore
the integrand in (\ref{eq:T_thetatheta 0->pi,past Unruh}) is non-negative for $\omega \ge 0$.
It follows that
$\vac[ren]{\hat{T}_{\theta\theta}}{U^-}$ is not symmetric under $\mathcal{P}$ at the axis 
when CCH's expressions are used.

On the Schwarzschild background, CCH's expressions for the expectation value of the electromagnetic field are straight-forwardly invariant under parity:
when $a=0$, the angular mode function does not depend on $\omega$ 
and the radial mode function does not depend on $m$. Therefore, applying the 
transformation $(\theta \to \pi -\theta)$ on any mode in their expressions
is equivalent to performing $(m \to -m)$ only, by virtue of the relations (\ref{eq:R symm.->cc,-m,-w})
and (\ref{eq:S symm.->pi-t,-m,-w}) and the reality of the stress tensor.
The presence of the sum $\sum_{m=-l}^{l}$ in their expressions
guarantees their invariance under $\mathcal{P}$.
Such a straight-forward reasoning does not follow in the Kerr background because applying the parity operation on 
CCH's expressions implies a change in the sign of $\omega$ as well as in the sign of $m$.
Whereas the sum over $m$ is symmetric with respect to $m=0$, the integration over the frequency is not symmetric
with respect to $\omega=0$. 

CCH's expressions are, however, valid when applied to the scalar field case in the Kerr background.
By comparing the scalar field's expectation values and CCH's expectation values with the symmetrized 
versions (\ref{eq:corrected stress tensor for s=1 on B-,U-}) for the states $\ket{B^-}$ and $\ket{U^-}$, 
we can give analogous symmetrized versions for the states $\ket{FT}$ (not included in CCH as this state was only defined later) and $\ket{CCH^-}$:
\begin{subequations} \label{eq:stress tensor for s=1 on FT and CCH^-}
\begin{align}
\begin{split}
&\expct{\hat{T}_{\mu\nu}}{FT}= 
\frac{1}{2}\sum_{lmP}
\\ &
\left(
\int_0^{\infty}\d{\tilde{\omega}}\, 
\coth\left(\frac{\pi\tilde{\omega}}{\kappa_+}\right)
\Big\{T_{\mu\nu}
\left[{}_{lm\omega}\phi_{\indhel}^{\text{up}},{}_{lm\omega}\phi_{\indhel}^{\text{up} *}\right]
+(-1)^{\vartheta}\mathcal{P}\left(T_{\mu\nu}
\left[{}_{lm\omega}\phi_{\indhel}^{\text{up}},{}_{lm\omega}\phi_{\indhel}^{\text{up} *}\right]
\right)\Big\}
+ \right.  \\
& \left. +\int_0^{\infty}\d{\omega}\, 
\coth\left(\frac{\pi\tilde{\omega}}{\kappa_+}\right)
\Big\{T_{\mu\nu}
\left[{}_{lm\omega}\phi_{\indhel}^{\text{in}},{}_{lm\omega}\phi_{\indhel}^{\text{in} *}\right]
+(-1)^{\vartheta}\mathcal{P}\left(T_{\mu\nu}
\left[{}_{lm\omega}\phi_{\indhel}^{\text{in}},{}_{lm\omega}\phi_{\indhel}^{\text{in} *}\right]
\right)\Big\}
\right) 
\end{split} \label{eq:corrected stress tensor for s=1 on CCH^-}
\\
\begin{split}
&\expct{\hat{T}_{\mu\nu}}{CCH^-}= 
\frac{1}{2}\sum_{lmP}
\\ &
\left(
\int_0^{\infty}\d{\tilde{\omega}}\, 
\coth\left(\frac{\pi\tilde{\omega}}{\kappa_+}\right)
\Big\{T_{\mu\nu}
\left[{}_{lm\omega}\phi_{\indhel}^{\text{up}},{}_{lm\omega}\phi_{\indhel}^{\text{up} *}\right]
+(-1)^{\vartheta}\mathcal{P}\left(T_{\mu\nu}
\left[{}_{lm\omega}\phi_{\indhel}^{\text{up}},{}_{lm\omega}\phi_{\indhel}^{\text{up} *}\right]
\right)\Big\}
+ \right.  \\
& \left. +\int_0^{\infty}\d{\omega}\, 
\coth\left(\frac{\pi\omega}{\kappa_+}\right)
\Big\{T_{\mu\nu}
\left[{}_{lm\omega}\phi_{\indhel}^{\text{in}},{}_{lm\omega}\phi_{\indhel}^{\text{in} *}\right]
+(-1)^{\vartheta}\mathcal{P}\left(T_{\mu\nu}
\left[{}_{lm\omega}\phi_{\indhel}^{\text{in}},{}_{lm\omega}\phi_{\indhel}^{\text{in} *}\right]
\right)\Big\}
\right) 
\end{split} \label{eq:stress tensor for s=1 on FT}
\end{align}
\end{subequations}


\subsection{Polarization} \label{sec:polarization}


In this subsection we will give a physical interpretation of the non-parity term and the parity term appearing in the
expressions (\ref{eq:corrected stress tensor for s=1 on B-,U-}) and (\ref{eq:stress tensor for s=1 on FT and CCH^-})
for the expectation value of the stress-energy tensor in different states.
We denote by parity term in a certain expression a term that explicitly contains the parity operator $\mathcal{P}$, and by
non-parity term one in the same expression that does not explicitly contain this operator.

It is clear from the classical expression (\ref{eq:A1, tableIChrzan.}) for the `upgoing gauge' potential ${}_{lm\omega}A^{\text{up} \mu}$ 
that this potential only contains the null vectors $\vec{n}$ and $\vec{m}$.
The parity term $P\mathcal{P}{}_{lm\omega}A^{\text{up} \mu}$ in (\ref{eq: def. of lmwPA_mu}), 
because of the transformations (\ref{eq:parity op. on NP objs.}) of the null base under parity, 
contains the vectors $\vec{n}$ and $\vec{m}^*$. 
The potential ${}_{lm\omega P}A^{\text{up} \mu}$ therefore contains the vectors $\vec{n}$, $\vec{m}$ and $\vec{m}^*$.
However, it is well-known (~\cite{bk:Weinberg}) that only two of them are physically significant at radial infinity, where the space-time is flat. 

It is in the limit for large $r$ that the physical meaning of the various vectors becomes clear.
We know that in flat space-time an electric field mode of positive frequency 
with radial and timelike exponential behaviour $e^{-i\omega(t-r)}$ (which we shall see is the asymptotic behaviour in our case)
that is proportional to the vector 
$\left(\hat{\vec{e}}_{\theta}+i\hat{\vec{e}}_{\phi}\right)[\left(\hat{\vec{e}}_{\theta}-i\hat{\vec{e}}_{\phi}\right)]$
possesses a negative[positive] angular momentum and we thus say that it is negatively[positively] polarized.
If the mode is instead of negative frequency, the sign of the angular momentum changes and then an electric field mode
with radial and timelike exponential behaviour $e^{-i\omega(t-r)}$
proportional to $\left(\hat{\vec{e}}_{\theta}+i\hat{\vec{e}}_{\phi}\right)[\left(\hat{\vec{e}}_{\theta}-i\hat{\vec{e}}_{\phi}\right)]$
is said to be positively[negatively] polarized.
Therefore, according to (\ref{eq:Kinnersley tetrad, r->inf}), 
an electric and a magnetic field modes of positive frequency that are proportional to the vector $\vec{m}[\vec{m}^*]$ correspond, 
in the flat space limit, to a negative[positive] polarization, whereas the vectors $\vec{l}$ and $\vec{n}$ correspond both to neutral polarization.

We know from (\ref{eq:parity term is pure gauge}) that the parity term is
pure gauge and therefore the contribution to the NP scalars from the term with $\vec{m}^*$ in the potential 
is zero. Only the terms with $\vec{n}$ and $\vec{m}$ in the `upgoing' potential contribute to the NP scalars. 
In particular, it is immediate from expressions (\ref{eq:phi as func. of potential with K}) 
for the Maxwell scalars that only the term in the potential ${}_{lm\omega P}A^{\text{up}}_{\mu}$ that contains the vector
$\vec{m}$ contributes to ${}_{lm\omega}\phi^{\text{up}}_{+1}$ whereas only the term with $\vec{n}$ contributes
to ${}_{lm\omega}\phi^{\text{up}}_{-1}$. Both, terms with $\vec{n}$ and terms with $\vec{m}$, contribute to ${}_{lm\omega}\phi^{\text{up}}_{0}$.
This implies that the positive-frequency modes ${}_{lm\omega}\phi^{\text{up}}_{-1}$, ${}_{lm\omega}\phi^{\text{up}}_{+1}$ and 
${}_{lm\omega}\phi^{\text{up}}_{0}$ are obtained from terms in the potential that, in the flat space limit, are
neutrally-, negatively- and both neutrally- and negatively- polarized respectively. 
The parity term in the potential in (\ref{eq: def. of lmwPA_mu}), 
which is of the opposite polarization to that of the non-parity term, 
does not contribute to any NP scalar mode because it is pure gauge. 
The opposite polarization to that of the non-parity term in the potential contributes to the NP scalars
through the negative-frequency
modes when the integration is over all frequencies, as in (\ref{eq:Fourier series of Gamma field}). 
In the expression (\ref{eq:classical mode expansion for A(in)_mu}) for the potential
or (\ref{eq:classical mode expansion for phi(in)_i}) for the NP scalars, 
in which we have rid of the negative-frequency modes, the opposite polarization appears
via the complex-conjugate term or the parity-term respectively.

Even though all three Maxwell scalars appear in the classical expression for the electromagnetic stress tensor, 
due to their different asymptotic behaviour (\ref{eq:peeling th}) for large $r$, 
the terms in the stress tensor (\ref{eq:stress tensor, spin 1}) with ${}_{lm\omega}\phi^{\text{up}}_{+1}$ predominate in this limit.
That is, the radiation field components of the stress tensor $T^{\text{up}}_{\mu\nu}$ are calculated in the flat space limit 
from modes in the potential (\ref{eq:Fourier series of Gamma field}) which for positive[negative] frequency 
correspond to a negative[positive] polarization. 
Note that the complex-conjugation of NP scalars in the stress tensor does not change the polarization 
of the field since it is merely a consequence of the fact that the null tetrad contains complex vectors, and does
not imply the complex-conjugation of the tensor field components $F_{\mu\nu}$.

We give here expressions for the `upgoing' field modes in the limit for large $r$, where the space-time is flat.
We wish, however, to obtain expressions for the fields that are real mode by mode. 
We will therefore not calculate them from the potential modes ${}_{lm\omega P}A^{\text{up}}_{\mu}$
in (\ref{eq:tableIChrzan.}), since the transformation $(m,\omega)\to (-m,-\omega)$ has been applied to
the parity term in (\ref{eq: def. of lmwPA_mu}). 
Even though the potential $A^{\text{up}}_{\mu}$ is obviously real, the potential modes ${}_{lm\omega P}A^{\text{up}}_{\mu}$ are not.
Instead, we will calculate field modes from the potential modes: ${}_{lm\omega}A^{\text{up}}_{\mu}+{}_{lm\omega}A^{\text{up} *}_{\mu}$.
The result is:
\begin{equation}  \label{eq:E_up,B_up, r->inf}
\begin{aligned}
&{}_{lm\omega P}\vec{E}^{\text{up}}\rightarrow
\frac{\omega^2 |N^{\text{up}}_{+1}|}{\sqrt{2}r}
\times \\&\times
\left[{}_{-1}Y_{l-m-\omega}\ {}_{+1}R^{\text{up,tra}}_{l-m-\omega}e^{+i\omega (t-r)}\left(\hat{\vec{e}}_{\theta}+i\hat{\vec{e}}_{\phi}\right)+
{}_{-1}Y_{l-m-\omega}^*\ {}_{+1}R^{\text{up,tra} *}_{l-m-\omega}e^{-i\omega (t-r)}\left(\hat{\vec{e}}_{\theta}-i\hat{\vec{e}}_{\phi}\right)\right]
,
\\
&{}_{lm\omega P}\vec{B}^{\text{up}}\rightarrow
\frac{-\omega^2 i|N^{\text{up}}_{+1}|}{\sqrt{2}r}
\times \\&\times
\left[{}_{-1}Y_{l-m-\omega}\ {}_{+1}R^{\text{up,tra}}_{l-m-\omega}e^{+i\omega (t-r)}\left(\hat{\vec{e}}_{\theta}+i\hat{\vec{e}}_{\phi}\right)-
{}_{-1}Y_{l-m-\omega}^*\ {}_{+1}R^{\text{up,tra} *}_{l-m-\omega}e^{-i\omega (t-r)}\left(\hat{\vec{e}}_{\theta}-i\hat{\vec{e}}_{\phi}\right)\right]
, \\
&\qquad \qquad \qquad \qquad \qquad \qquad \qquad \qquad \qquad \qquad \qquad \qquad \qquad \qquad \qquad \qquad 
\quad (r\rightarrow+\infty)
\end{aligned}
\end{equation}
where we have used the fact that in flat space we can replace ${}_{\indhel}Z_{lm\omega}$ by ${}_{\indhel}Y_{lm\omega}$.
It is worth noting that even though the second term in either $({}_{lm\omega}A^{\text{up}}_{\mu}+{}_{lm\omega}A^{\text{up} *}_{\mu})$ 
or (\ref{eq: def. of lmwPA_mu})
does not contribute to the NP scalars it does contribute to the fields. 
The reason is that the NP scalars must be calculated from real fields. 
It does not make physical sense to consider the contribution to the NP scalars from a non-real field, such as the second
term in the above expressions for the field. We say that this term is `pure gauge' in the sense that it does not contribute
to the NP scalars even if it does contribute to the physical fields so as to make them real. 
The large-$r$ asymptotics for the NP Maxwell scalars in terms of the electric and magnetic fields are easily obtained:
\begin{equation}  \label{eq:phi_i, r->inf}
\begin{aligned}
\phi_{-1}&\rightarrow -\frac{1}{\sqrt{2}}\left(\vec{E}+\vec{B}i\right)\left(\hat{\vec{e}}_{\theta}+i\hat{\vec{e}}_{\phi}\right)  & (r\rightarrow+\infty)\\
\phi_{0}&\rightarrow \frac{1}{2}\left(\vec{E}+\vec{B}i\right)\hat{\vec{e}}_{r}  & (r\rightarrow+\infty) \\
\phi_{+1}&\rightarrow \frac{1}{2\sqrt{2}}\left(\vec{E}+\vec{B}i\right)\left(\hat{\vec{e}}_{\theta}-i\hat{\vec{e}}_{\phi}\right)  & (r\rightarrow+\infty)
\end{aligned}
\end{equation}
It is clear that the parity and non-parity terms correspond to opposite polarizations for
the fields (\ref{eq:E_up,B_up, r->inf}) (and also for the potential). 
It is also clear that the only contribution to $\phi^{\text{up}}_{+1}$ 
from the `upgoing' electric and magnetic fields comes from the non-parity term. 
To leading order in $r$ for the electric and magnetic fields both $\phi^{\text{up}}_{-1}$ and $\phi^{\text{up}}_{0}$ vanish, 
in agreement with (\ref{eq:peeling th}).
To next order in $r$, expressions (\ref{eq:E_up,B_up, r->inf}) and (\ref{eq:phi_i, r->inf}) must be calculated to include lower
order terms and it is therefore not valid to conclude from them that the only contribution to $\phi^{\text{up}}_{-1}$ and $\phi^{\text{up}}_{0}$ 
comes from positively- and neutrally- polarized terms respectively. 
We have indeed seen in the beginning of this subsection that this is not the case. 

The reasoning used so far for the `upgoing gauge' potential can be applied in the same manner to the `ingoing gauge' potential ${}_{lm\omega}A^{\text{in}}_{\mu}$.
In this case, the potential contains one term with the vector $\vec{l}$, which is the only one that contributes to
${}_{lm\omega}\phi^{\text{in}}_{+1}$, and one term with the vector $\vec{m}^*$, which is the only one that contributes to
${}_{lm\omega}\phi^{\text{in}}_{-1}$. Like in the `upgoing gauge' case, both terms contribute to ${}_{lm\omega}\phi^{\text{up}}_{0}$,
and the parity term (containing $\vec{l}$ and $\vec{m}$) does not contribute to any of the NP scalars.
The scalar ${}_{lm\omega}\phi^{\text{in}}_{-1}$ is the one that diminishes more slowly in the limit for large $r$.
The radiative components of the classical stress tensor $T^{\text{in}}_{\mu\nu}$ is thus calculated in the flat space limit from modes in the potential 
(\ref{eq:Fourier series of Gamma field}) that for positive[negative] frequency correspond to positive[negative] polarization.

So far in this subsection we have looked at the physical meaning of the different terms in classical expressions only.
We are now in a position to understand the physical meaning of the terms in the quantum field theory expressions.
The positive frequency modes in (\ref{eq:Fourier series of Gamma field}) correspond to the non-parity term
in the expression (\ref{eq:classical mode expansion for phi(in)_i}) for the NP scalar and, ultimately, 
give rise to the non-parity term in the expectation value of the stress tensor (\ref{eq:corrected stress tensor for s=1 on B-,U-}).
Similarly, the negative frequency modes in (\ref{eq:Fourier series of Gamma field}) give rise to the 
parity term in the NP scalars and the parity term in the expectation value of the stress tensor.
We therefore reach the conclusion that the non-parity terms in expressions (\ref{eq:corrected stress tensor for s=1 on B-,U-})
for the expectation value of the stress tensor 
correspond in the flat space limit to one specific polarization (negative in the `up' case and positive in the `in' case)
and that the corresponding parity terms in the same expressions correspond to the opposite polarization.
Both the contribution from the positive-polarization terms and from the negative-polarization terms are
separately real, as it should be.
We also know, from the end of Subsection \ref{subsec:new expressions}, that in the spherically-symmetrical case $a=0$, 
the contribution to the expectation value of the stress tensor from the positive-polarization terms 
is identical to the one from the negative-polarization terms, as one would expect.

The notable exception to this picture are the `up' superradiant modes. 
Indeed, these modes have a sign of $\omega$ opposite to the non-superradiant modes in the same term in the expectation
value, whether the parity term or the non-parity term. The polarization of the `up' superradiant modes
is therefore the opposite to the non-superradiant modes in the same term, that is, it is positive if part of the non-parity
term and negative if part of the parity term. Note, however, that the `in' superradiant modes
have the same sign of $\omega$ (positive), and therefore the same polarization, as the non-superradiant modes
in the same term in the expectation value.

When CCH only include non-parity terms in their expressions for the expectation value of the stress tensor
they are only including one polarization and leaving out the other one for the `in' modes. For the `up' modes, 
they are only including one polarization for the non-superradiant modes and the opposite polarization for the superradiant modes.
In particular, when subtracting the expectation value of the stress tensor 
in the past Boulware state from the one in the past Unruh state, only `up' modes are needed.
Neglecting the parity terms is in this case equivalent to neglecting positive polarization non-superradiant modes
as well as negative polarization superradiant modes. 
That is the case in the calculation of $\vac[ren]{\hat{T}^{\mu}{}_{\nu}}{B^-}$ close to the horizon in Section \ref{sec:RRO} 
but,
as explained in that section, in this limit
the non-parity and the parity terms coincide. 

\begin{figure}[H]
\rotatebox{90}
\centering
\includegraphics*[width=70mm,angle=270]{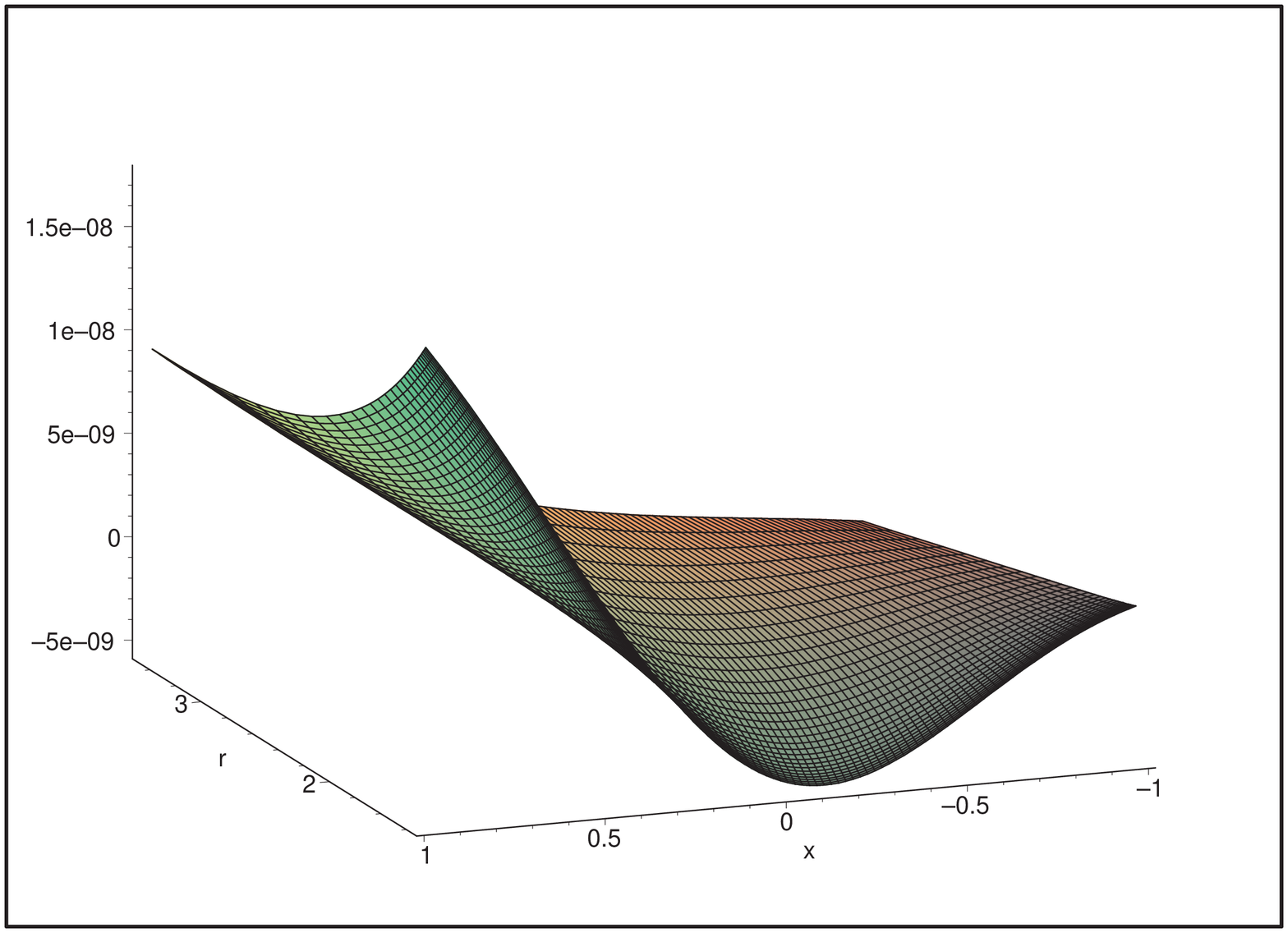}
\caption{Positive polarization terms of $\frac{1}{4\pi}\Delta\vac{\hat{T}_{tr}}{CCH^--U^-}$}
\label{fig:deltaTtr_cch_u_past}
\includegraphics*[width=70mm,angle=270]{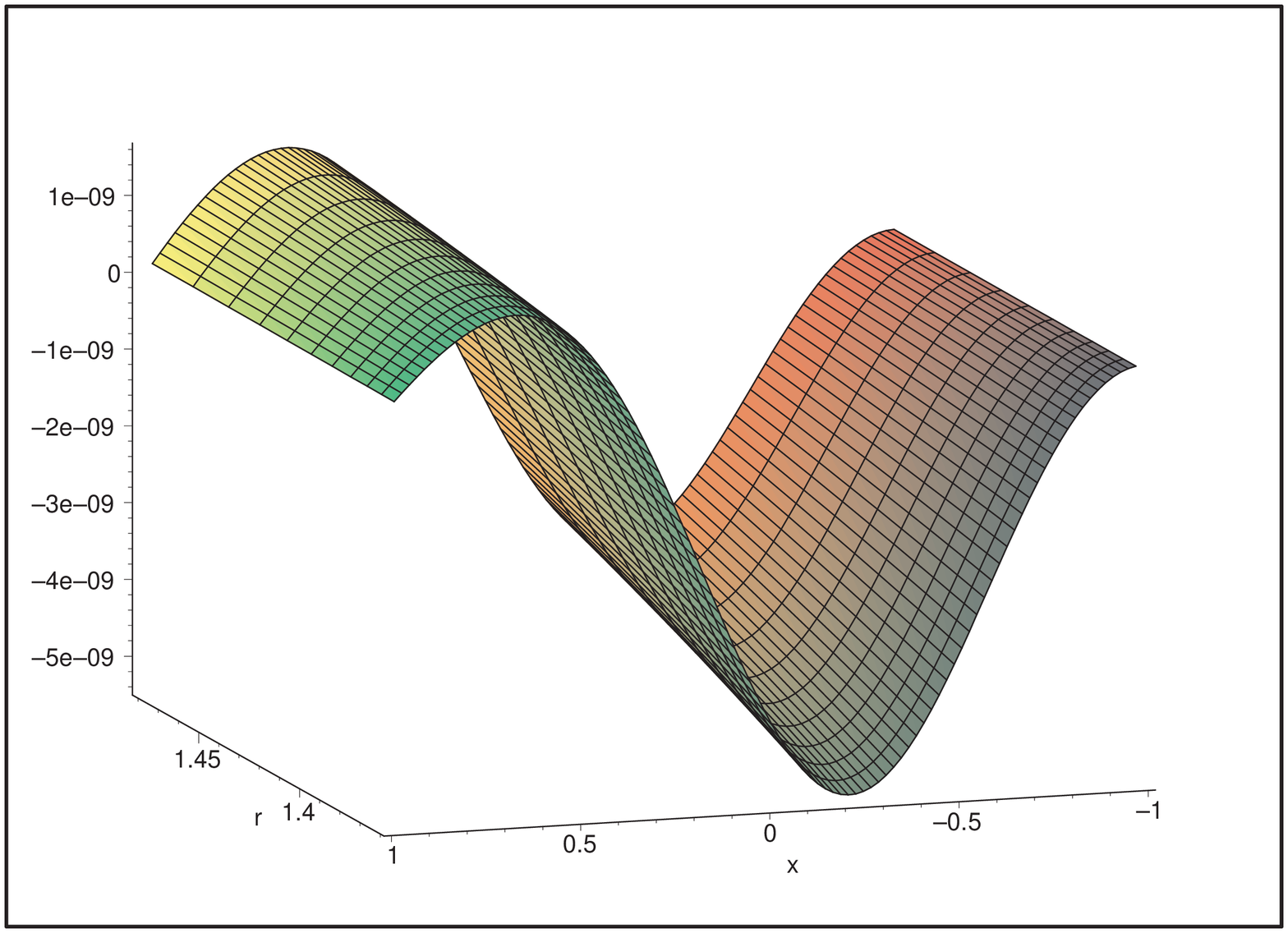}
\caption{Positive polarization terms of $\frac{1}{4\pi}\vac{\hat{T}_{t\phi}}{CCH^--U^-}$}
\label{fig:Ttphi_cch_u_past}
\end{figure}

\begin{figure}[H]
\rotatebox{90}
\centering
\includegraphics*[width=70mm,angle=270]{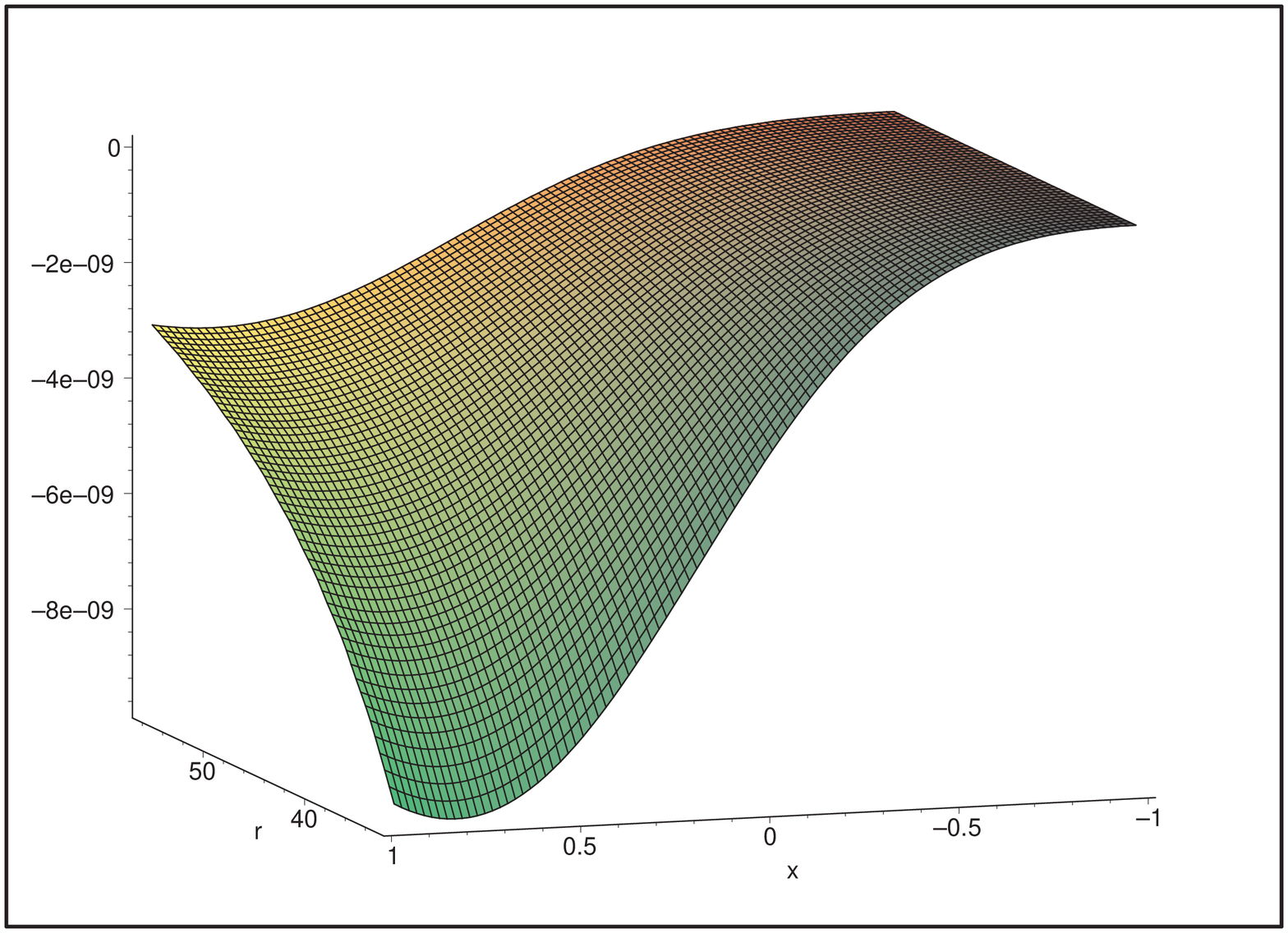}
\caption{Positive polarization terms of $\frac{1}{4\pi}\vac{\hat{T}_{tr}}{U^--B^-}$}
\label{fig:Ttr_polpos_u_b}
\includegraphics*[width=70mm,angle=270]{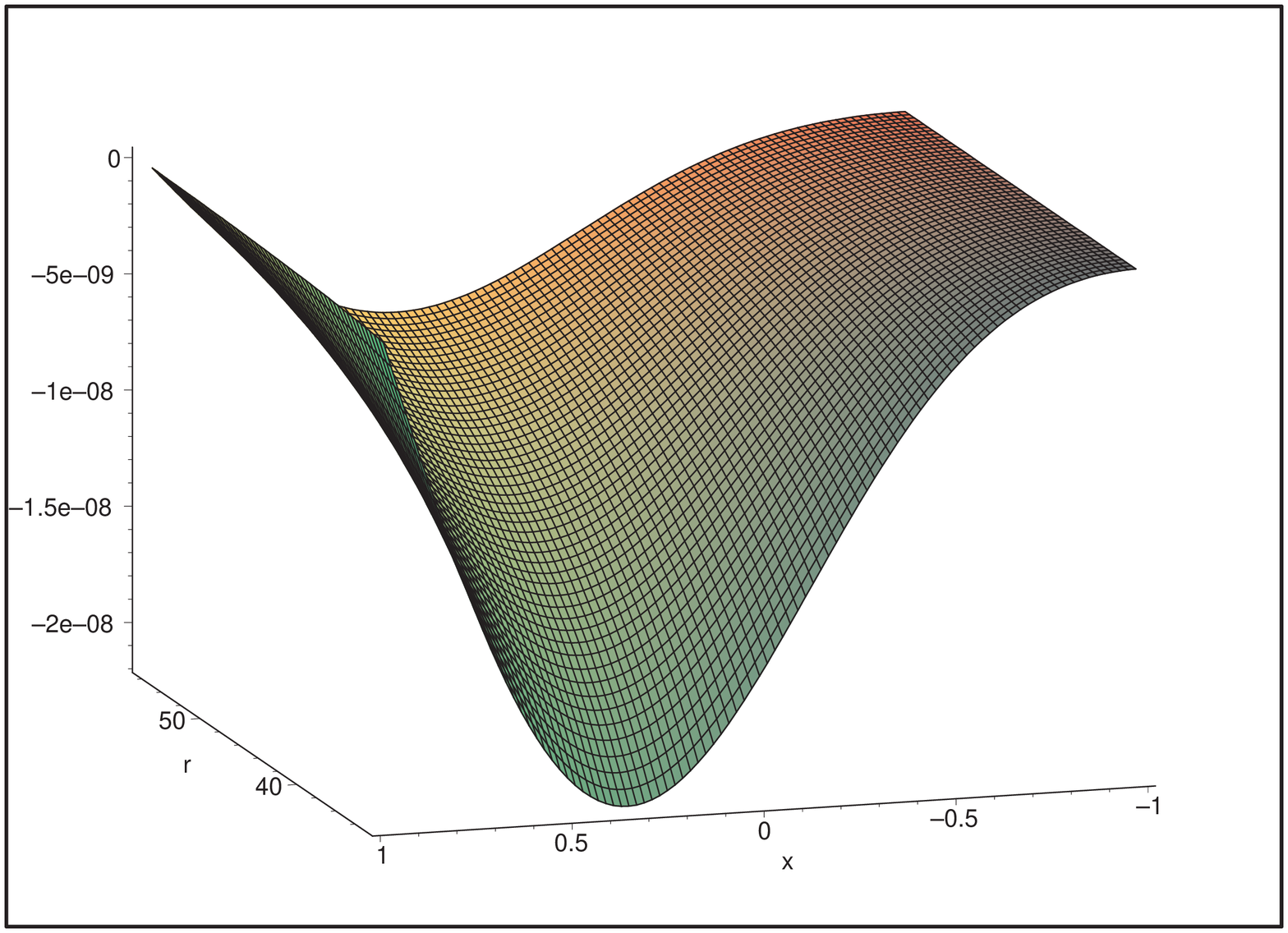}
\caption{Positive polarization terms of $\frac{1}{4\pi}\vac{\hat{T}_{t\phi}}{U^--B^-}$}
\label{fig:Ttphi_polpos_u_b}
\end{figure}

It is interesting to group the terms with the same polarization in the expectation value of the stress energy tensor.
Of course, in the case of the difference between the states $\ket{CCH^-}$ and $\ket{U^-}$, which only has contribution from the `in' modes,
the sum of the positive polarization terms coincides with the direct evaluation of CCH's expressions.
The negative polarization contribution can be obtained by applying the transformation $x\to -x$.
We include the plots of the tensor components corresponding to the fluxes of energy and angular momentum 
from the positive polarization terms in Figures \ref{fig:deltaTtr_cch_u_past}--\ref{fig:Ttphi_cch_u_past}.
The evaluation of the positive polarization contribution to the difference in the expectation value of the stress energy tensor 
between the states $\ket{U^-}$ and $\ket{B^-}$ requires carefully adding the contribution of the superradiant modes to the appropriate
polarization. We calculated the positive polarization contribution to these differences of 
expectation values and plot them in Figures \ref{fig:Ttr_polpos_u_b}--\ref{fig:Ttphi_polpos_u_b}.
The corresponding negative polarization contribution is, again, obtained by applying the transformation $x\to -x$.
The interest of these graphs lies in the region far from the horizon. Close to the horizon 
the irregularity of the state $\ket{B^-}$ dominates and thermality guarantees symmetry with respect to the equator.

\section{Luminosity} \label{sec:luminosity}

Let $E^{\text{(inc)}}$ and $E^{\text{(ref)}}$ denote, respectively, the energy incident and the energy reflected
by the black hole at infinity. 
Let $E^{\text{(tra)}}$ denote the energy going down across the event-horizon of the black hole. 
The reflection coefficient $\mathbb{R}_{lm\omega}$  and the transmission coefficient $\mathbb{T}_{lm\omega}$ 
of an incoming wave mode are then defined as the following flux ratios:
\begin{equation}
\begin{aligned}
\mathbb{R}_{lm\omega}&\equiv \frac{\d E^{\text{(ref)}}_{lm\omega}/\d{t}}{\d E^{\text{(inc)}}_{lm\omega}/\d{t}}, \qquad
\mathbb{T}_{lm\omega}&\equiv \frac{\d E^{\text{(tra)}}_{lm\omega}/\d{t}}{\d E^{\text{(inc)}}_{lm\omega}/\d{t}}
\end{aligned}
\end{equation}

The wronskian relations for the solutions of the radial Teukolsky equation 
correspond to the conservation of energy law that equates the net flux of energy coming in
from infinity to the net flux of energy going down into the black hole:
\begin{equation} \label{eq:conserv. of energy law} 
1-\mathbb{R}_{lm\omega}=\mathbb{T}_{lm\omega}=1-\left|A^{\text{in}}_{lm\omega}\right|^2=\frac{-i}{2^4\omega^3}W[{}_{+1}R,{}_{-1}R^{*}]^{\text{in}}_{lm\omega}
\end{equation}
where we have made use of the relations (\ref{eq:R1 coeffs from R_1's}) and (\ref{eq:R_1 coeffs from X's}).
For superradiant wave modes the fractional gain or loss of energy of an incoming wave mode,
i.e., $\mathbb{R}_{lm\omega}=|A^{\text{in}}_{lm\omega}|^2$, is greater than one while
the transmission coefficient $\mathbb{T}_{lm\omega}$ is negative.

The conservation equations $\nabla_{\nu}T_{\mu}{}^{\nu}=0$ can alternatively be written ~\cite{bk:Dirac} as
\begin{equation} \label{eq:compact conserv.eqs.}
\partial_{\nu}\left(T_{\mu}{}^{\nu}\sqrt{-g}\right)=\frac{1}{2}\sqrt{-g}\left(\partial_{\mu}g_{\alpha\beta}\right)T^{\alpha\beta}
\end{equation}
Assuming that the stress-energy tensor is independent of $t$ and $\phi$, like the Kerr metric, the $\mu=t$ and 
$\mu=\phi$ components of equations (\ref{eq:compact conserv.eqs.}) can be integrated over $r$ to yield:
\begin{subequations} \label{eq:integrated conserv.eqs., T_tr,Ttphi}
\begin{align}
T_{tr}&=\frac{K(\theta)}{\Delta}-\frac{1}{\Delta\sin\theta}\partial_{\theta}\left(\sin\theta\int_{r_+}^r\d{r'}T_{t\theta}\right) 
\label{eq:integrated conserv.eqs., T_tr} \\
T_{\phi r}&=\frac{L(\theta)}{\Delta}-\frac{1}{\Delta\sin\theta}\partial_{\theta}\left(\sin\theta\int_{r_+}^r\d{r'}T_{\phi\theta}\right) 
\label{eq:integrated conserv.eqs., Ttphi}
\end{align}
\end{subequations}
where $K(\theta)$ and $L(\theta)$ are arbitrary functions.
The luminosity when the field is in the state $\ket{\Psi}$ is given by
\begin{equation} \label{eq:def. dM/dt}
\left. \diff{M}{t}\right|_{\Psi}=\Delta\int_S\d{\Omega}\vac[ren]{\hat{T}_{tr}}{\Psi}
\end{equation}
where the surface $S$ can be any surface of constant $t$ and $r$. 
In the forthcoming the subindex $A$ refers to either $t$ or $\phi$ and the subindex $X$ to either $r$ or $\theta$.

We will now compare some spin-1 results with the corresponding spin-0 results. 
The following results for spin-0 are proven in part by Frolov and Thorne ~\cite{ar:F&T'89} and extended by Ottewill and Winstanley ~\cite{ar:Ott&Winst'00}:
\begin{equation} \label{eq:T_Ar for s=0}
\begin{aligned}
\vac[ren]{\hat{T}_{A\theta}}{\Psi} &=0, \quad &
\Delta \vac[ren]{\hat{T}_{tr}}{\Psi}&=K(\theta), \quad  &
\Delta \vac[ren]{\hat{T}_{\phi r}}{\Psi}&=L(\theta) 
\quad \text{for} \quad s=0
\end{aligned}
\end{equation}
where $\ket{\Psi}$ is any state among $\ket{B^\pm}$, $\ket{U^-}$, $\ket{CCH^-}$ or $\ket{FT}$. 
The last two equations in (\ref{eq:T_Ar for s=0}) are a direct consequence of the first one and of (\ref{eq:integrated conserv.eqs., T_tr,Ttphi}).
It may indeed be calculated directly from the expression for the spin-0 stress-energy tensor 
that all the radial dependence of $\Delta {}_{lm\omega}T_{tr}$ can be expressed as a radial wronskian. 
It can also be checked that ${}_{lm\omega}T_{tr}^{\text{in}}=-{}_{lm\omega}T_{tr}^{\text{up}}$ for spin-0 so that 
the only contribution to the luminosity in the past Boulware vacuum comes from the superradiant modes:
\begin{equation}
\Delta \vac[ren]{\hat{T}_{tr}}{B^-}=-2\sum_{l=1}^{\infty}\sum_{m=1}^{l}\int_0^{m\Omega_+}\d{\omega}\Delta {}_{lm\omega}T_{tr}^{\text{up}} \qquad \text{for} \quad s=0
\end{equation}  

It is immediately apparent that for the spin-1 case the task of proving analytically 
whether $\Delta T_{tr}$ is constant in $r$ or not is not as straight-forward as for spin-0. 
When using expressions (\ref{eq:corrected stress tensor for s=1 on B-,U-}) for the expectation value of the stress-energy tensor for a spin-1 field,
we are interested in the following calculation: 
\begin{equation} \label{eq:Delta(Ttr+Ttr(t->pi-t))}
\begin{aligned}
&\Delta \left({}_{lm\omega}T_{tr}^{\text{up}}+\mathcal{P}{}_{lm\omega}T_{tr}^{\text{up}}\right)=
\\&=
\frac{\mathbb{T}_{lm\omega}}{4\pi^2\Sigma}\Big\{-\omega\Sigma\left({}_{-1}S_{lm\omega}^2+{}_{+1}S_{lm\omega}^2\right)+ 
\frac{a^3\cos\theta\sin^2\theta}{\Sigma}\left({}_{-1}S_{lm\omega}^2-{}_{+1}S_{lm\omega}^2\right)+
\\ &+
a\sin\theta\left({}_{-1}S_{lm\omega}\partial_{\theta}{}_{-1}S_{lm\omega}-{}_{+1}S_{lm\omega}\partial_{\theta}{}_{+1}S_{lm\omega}\right)\Big\}
\end{aligned}
\end{equation}
where we have made use of the symmetry (\ref{eq:S symm.->pi-t,-s}) in order to relate terms that contain
$({}_{-1}S_{lm\omega}^2+\mathcal{P}{}_{-1}S_{lm\omega}^2)$ to terms that contain $({}_{+1}S_{lm\omega}^2+\mathcal{P}{}_{+1}S_{lm\omega}^2)$.
The corresponding result for the `in' modes is equal to (\ref{eq:Delta(Ttr+Ttr(t->pi-t))}) with a change of sign, by virtue of
(\ref{eq:conserv. of energy law}) and the property (\ref{eq: wronskian in=-up}).
We can get rid of the derivatives in (\ref{eq:Delta(Ttr+Ttr(t->pi-t))}) by using the Teukolsky-Starobinski\u{\i} identities
and the angular Teukolsky equation. But even when doing that, it is not possible to express the last term in (\ref{eq:Delta(Ttr+Ttr(t->pi-t))})
in terms of ${}_{-1}S_{lm\omega}^2$ and ${}_{+1}S_{lm\omega}^2$ only. 
As a matter of fact, when evaluated at the axis of symmetry $\theta=0$ or $\pi$, only one term containing ${}_{-1}S_{lm\omega}^2$ and one
term containing ${}_{+1}S_{lm\omega}^2$ appear, neither of which is constant in $r$.
It is therefore apparent that if we wish to prove that $\Delta \left({}_{lm\omega}T_{tr}^{\text{up}}+\mathcal{P}{}_{lm\omega}T_{tr}^{\text{up}}\right)$
is constant in $r$, or otherwise, we must then somehow relate ${}_{-1}S_{lm\omega}^2$ to ${}_{+1}S_{lm\omega}^2$.
It follows from the symmetries (\ref{eq: S symms}) that we can only relate at the same 
point a term whose spherical dependence is ${}_{-1}S_{lm\omega}^2$ to a term whose spherical dependence is ${}_{+1}S_{lm\omega}^2$ 
by applying the transformation $(m,\omega)\to (-m,-\omega)$ to one
of them. The change in sign of $m$ can be overturned due to the symmetric sum in $m$ in the Fourier sums 
in (\ref{eq:corrected stress tensor for s=1 on B-,U-}) and (\ref{eq:stress tensor for s=1 on FT and CCH^-}).
The change in sign of $\omega$, however, is a problem when trying to relate a term with 
${}_{-1}S_{lm\omega}^2$ to a term with ${}_{+1}S_{lm\omega}^2$
due to the non-symmetric nature under $(m,\omega)\to (-m,-\omega)$ of the integrals over $\omega$ or $\tilde{\omega}$ 
for all states involved in (\ref{eq:corrected stress tensor for s=1 on B-,U-}) and (\ref{eq:stress tensor for s=1 on FT and CCH^-}). 
It follows that $\Delta \left({}_{lm\omega}T_{tr}+\mathcal{P}{}_{lm\omega}T_{tr}\right)$ is not
constant in $r$ and therefore neither is $\Delta\vac[ren]{\hat{T}_{tr}}{B^-}$ nor $\Delta\vac[ren]{\hat{T}_{tr}}{U^-}$. 
Similarly, a calculation of $\left({}_{lm\omega}T_{t\theta}+\mathcal{P}{}_{lm\omega}T_{t\theta}\right)$
shows that it is not zero and therefore neither $\vac[ren]{\hat{T}_{t\theta}}{B^-}$ nor $\vac[ren]{\hat{T}_{t\theta}}{U^-}$
are zero. 

\begin{figure}[H]
\rotatebox{90}
\centering
\includegraphics*[width=70mm,angle=270]{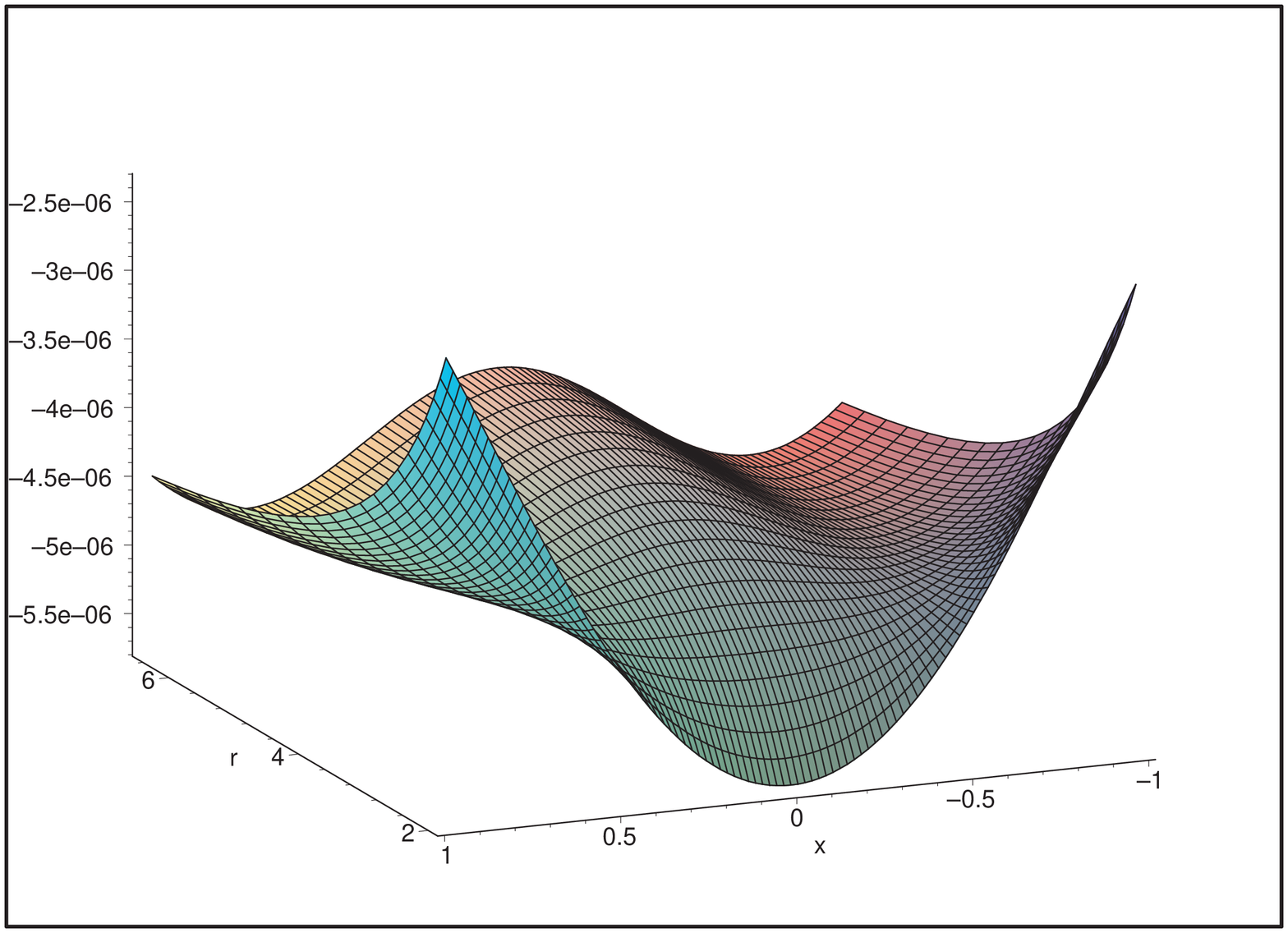} 
\caption{$\frac{1}{4\pi}\Delta\vac{\hat{T}_{tr}}{U^--B^-}$}    \label{fig:deltaSymmTtr_u_b}
\includegraphics*[width=70mm,angle=270]{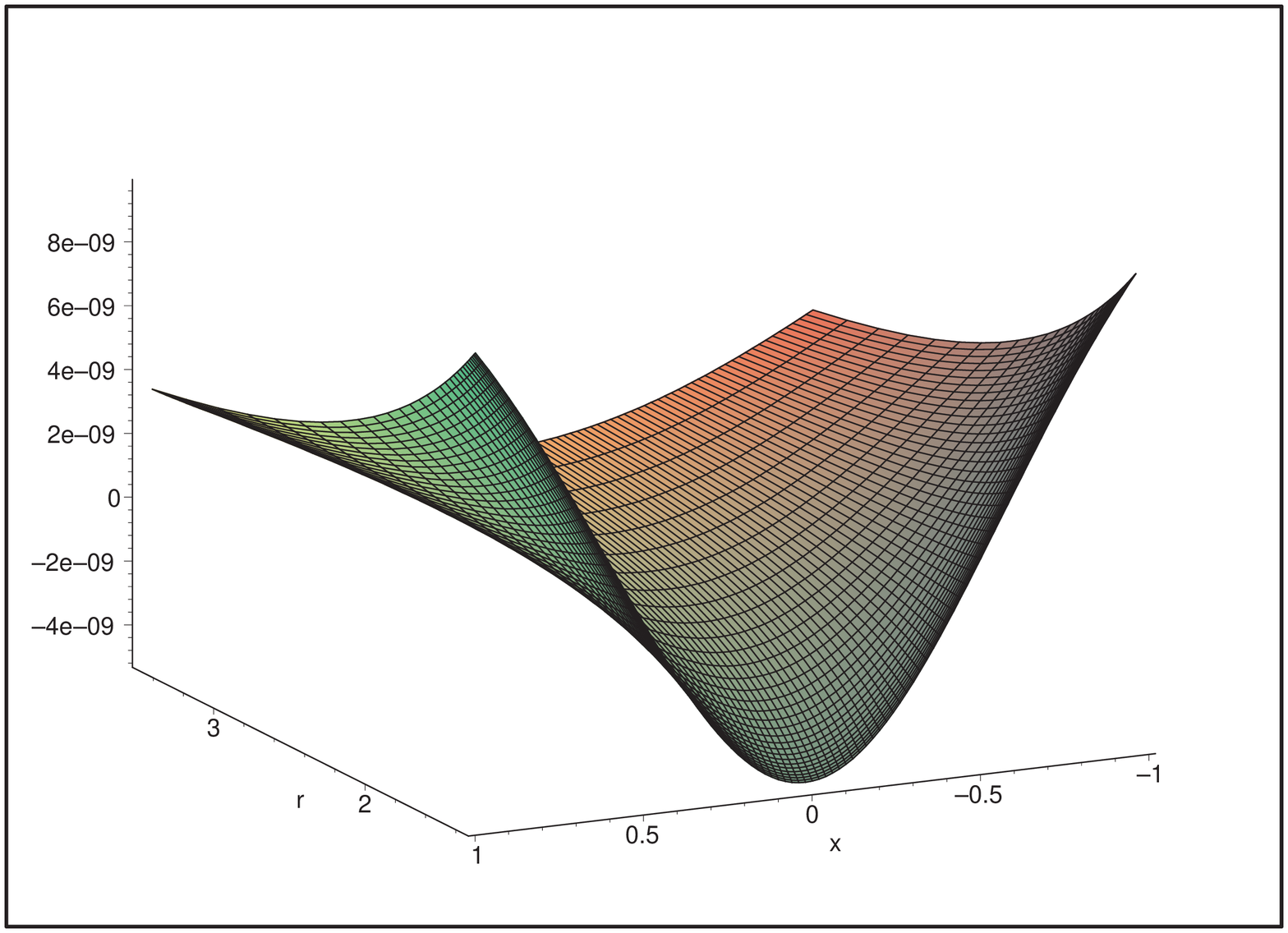} 
\caption{$\frac{1}{4\pi}\Delta\vac{\hat{T}_{tr}}{CCH^--U^-}$}    \label{fig:deltaSymmTtr_cch_u_past}
\end{figure}

\begin{figure}[H]
\rotatebox{90}
\centering
\includegraphics*[width=70mm,angle=270]{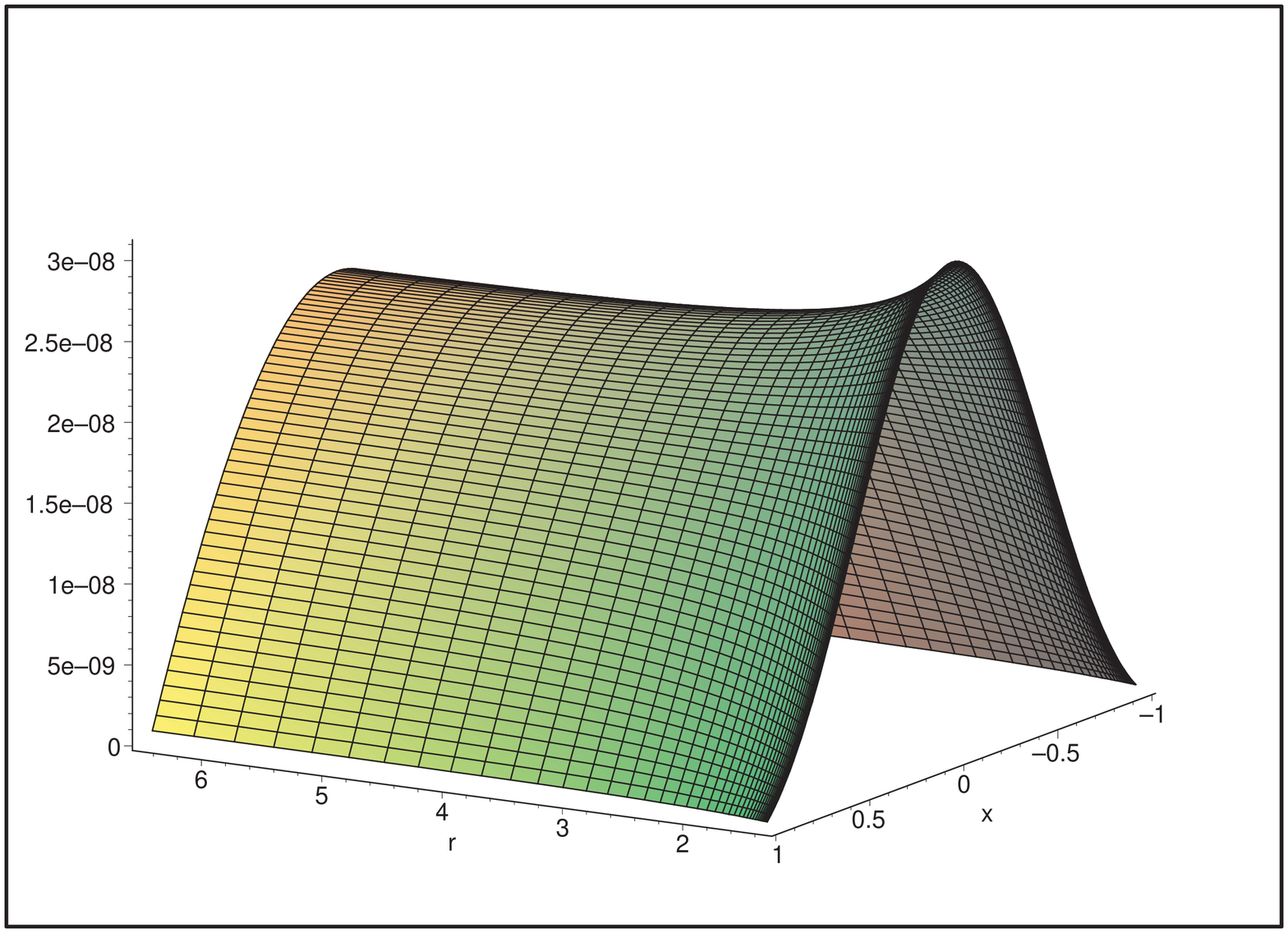} 
\caption{$\frac{1}{4\pi}\Delta\vac{\hat{T}_{r\phi}}{CCH^--U^-}$}    \label{fig:deltaSymmTrphi_cch_u_past}
\includegraphics*[width=70mm,angle=270]{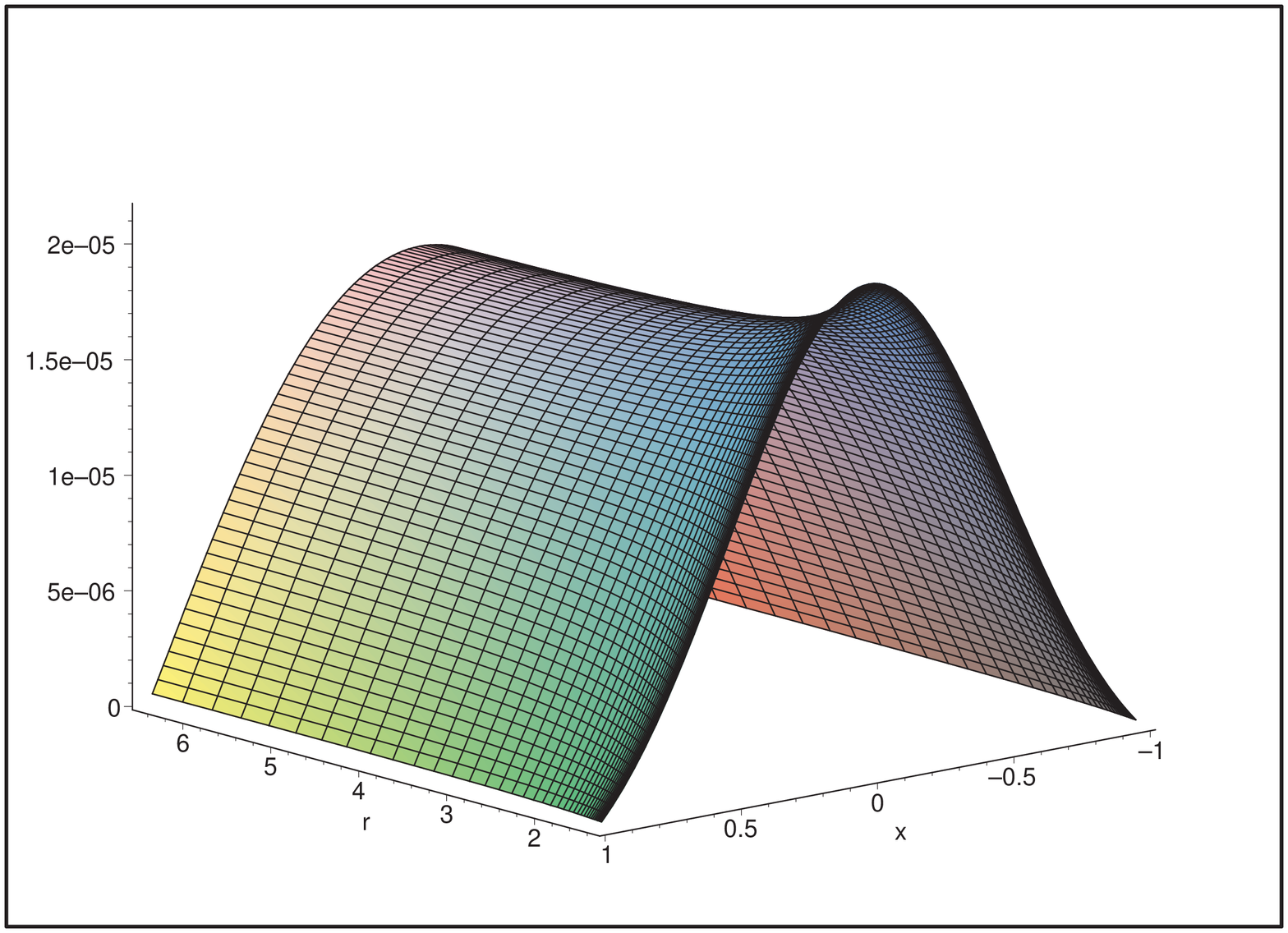} 
\caption{$\frac{1}{4\pi}\Delta\vac{\hat{T}_{r\phi}}{U^--B^-}$}    \label{fig:deltaSymmTrphi_u_b}
\end{figure}

\begin{figure}[H]
\rotatebox{90}
\centering
\includegraphics*[width=70mm,angle=270]{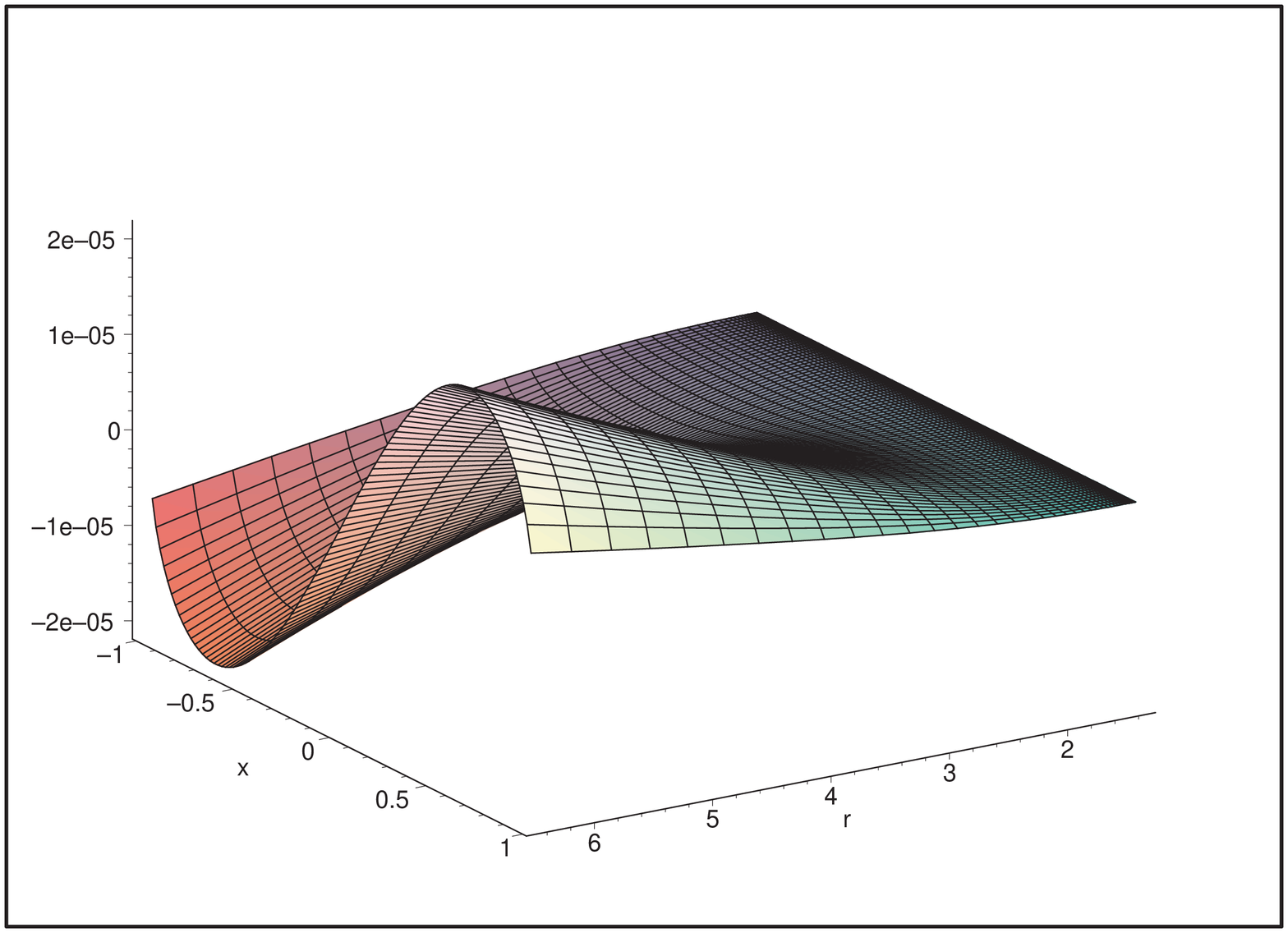} 
\caption{$\frac{1}{4\pi}\Delta^2\vac{\hat{T}_{t\theta}}{U^--B^-}$.
Note that the viewing angle in this and the following graphs is different from that of the previous graphs 
in order to make more visible the region far from the horizon.}    \label{fig:delta2SymmTttheta_u_b}
\includegraphics*[width=70mm,angle=270]{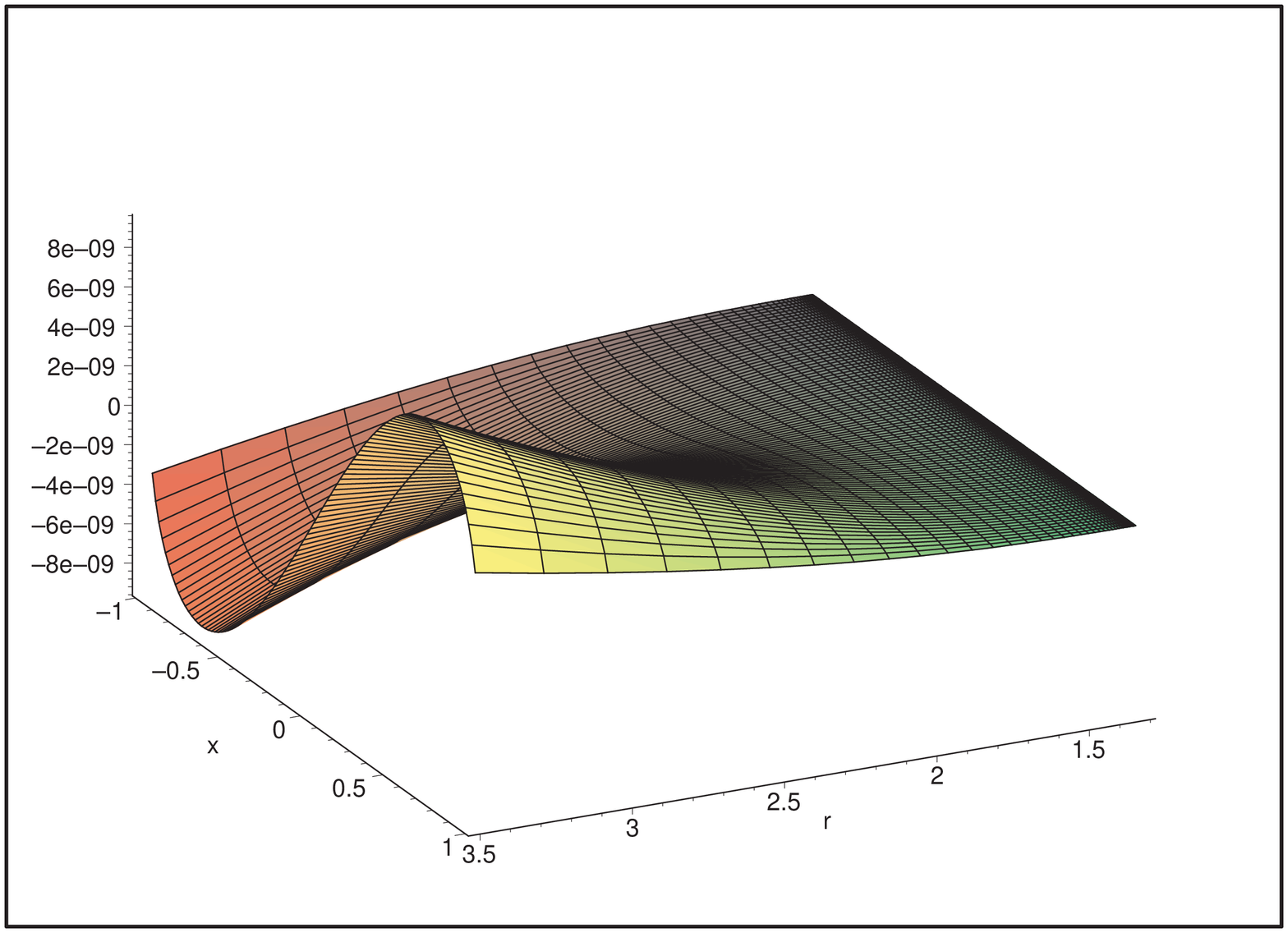} 
\caption{$\frac{1}{4\pi}\Delta^2\vac{\hat{T}_{t\theta}}{CCH^--U^-}$}    \label{fig:delta2SymmTttheta_cch_u_past}
\end{figure}

\begin{figure}[H]
\rotatebox{90}
\centering
\includegraphics*[width=70mm,angle=270]{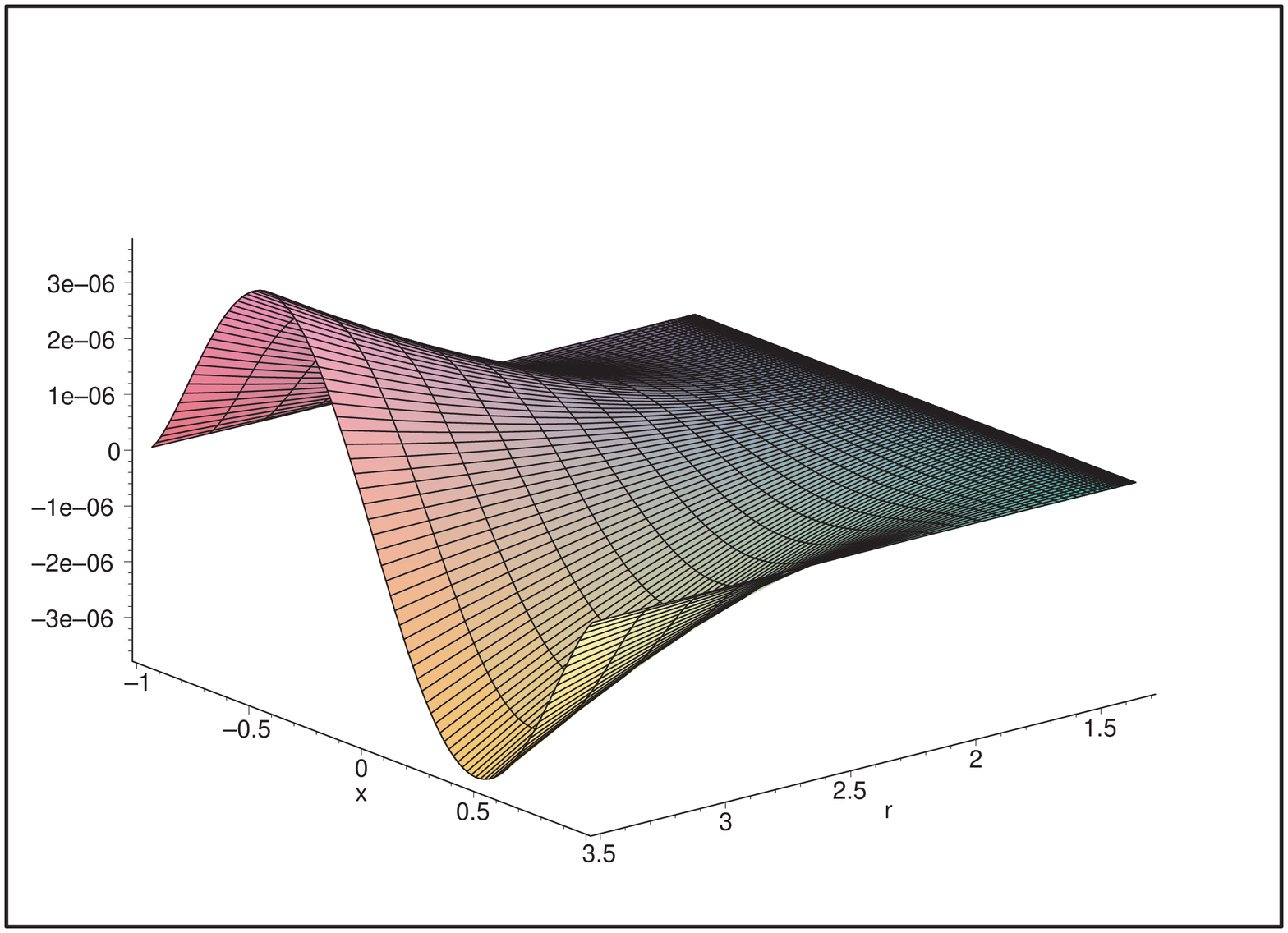} 
\caption{$\frac{1}{4\pi}\Delta^2\vac{\hat{T}_{\theta\phi}}{U^--B^-}$}    \label{fig:delta2SymmTthetaphi_u_b}
\includegraphics*[width=70mm,angle=270]{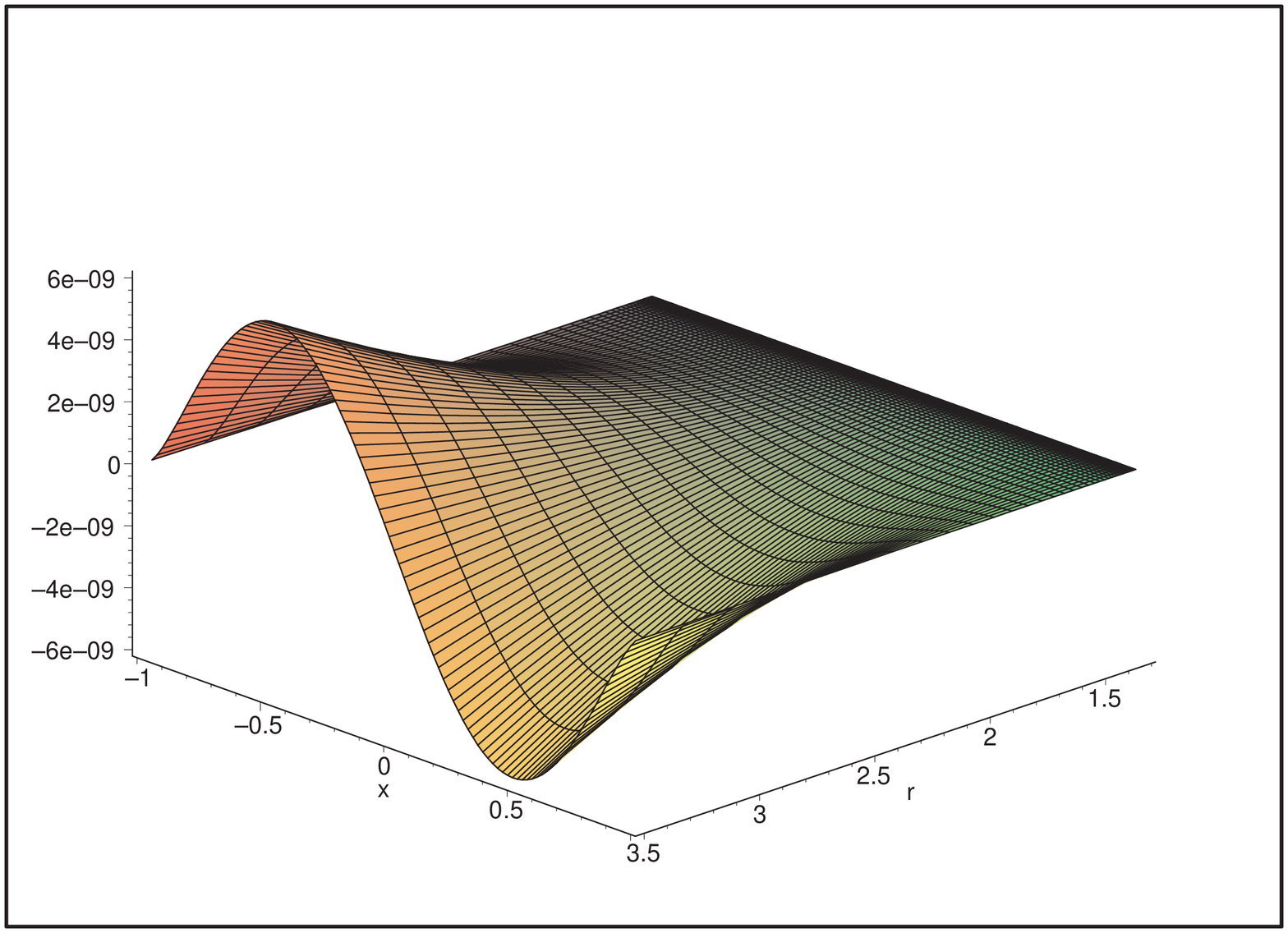} 
\caption{$\frac{1}{4\pi}\Delta^2\vac{\hat{T}_{\theta\phi}}{CCH^--U^-}$}    \label{fig:delta2SymmTthetaphi_cch_u_past}
\end{figure}

Indeed, Graphs \ref{fig:deltaSymmTtr_u_b}--\ref{fig:delta2SymmTthetaphi_cch_u_past} numerically corroborate the above 
conclusions.
Graphs \ref{fig:deltaSymmTtr_u_b} and \ref{fig:deltaSymmTtr_cch_u_past} show that neither 
$\Delta\vac{\hat{T}_{tr}}{U^--B^-}$ nor $\Delta\vac{\hat{T}_{tr}}{CCH^--U^-}$ are constant in $r$.
Graphs \ref{fig:delta2SymmTttheta_u_b}--\ref{fig:delta2SymmTthetaphi_cch_u_past} show that neither 
$\vac[ren]{\hat{T}_{A\theta}}{U^--B^-}$ nor $\vac[ren]{\hat{T}_{A\theta}}{CCH^--U^-}$ are zero.
Graphs \ref{fig:deltaSymmTrphi_cch_u_past}--\ref{fig:deltaSymmTrphi_u_b}, however, seem to indicate that
both $\Delta\vac{\hat{T}_{r\phi}}{CCH^--U^-}$ and $\Delta\vac{\hat{T}_{r\phi}}{U^--B^-}$ might actually be constant in $r$.

Note that if instead of using expressions (\ref{eq:corrected stress tensor for s=1 on B-,U-}) we use
CCH's expressions for the expectation value of the stress tensor, we encounter the same difficulty when trying to prove
whether $T_{t\theta}$ is zero and whether $\Delta T_{tr}$ is constant in $r$. 

In the case $a=0$, since the spin-weighted spherical harmonics ${}_{\indhel}Y_{lm}$ do not depend on $\omega$, we only need
a change in the sign of $m$ to relate terms with $|{}_{+1}Y_{lm}|^2$ to terms with $|{}_{-1}Y_{lm}|^2$ in $\Delta {}_{lm\omega}T_{tr}$. 
Indeed, use of (\ref{eq:eq.B6J,McL,Ott'95}) allows us
to prove that $\sum_m \Delta {}_{lm\omega}T_{tr}$ is constant in $r$ and that 
$\sum_m {}_{lm\omega}T_{tr}^{\text{up}}=-\sum_m {}_{lm\omega}T_{tr}^{\text{in}}$ in the Schwarzschild background.   
A numerical investigation of the luminosity and components of the RSET for spin-1 in the past Unruh state is presented in ~\cite{ar:J&McL&Ott'91}.

The solution to this deadlock for the spin-1 case in the Kerr background consists in integrating over the solid angle. 
This allows us to relate a term with
$\int\d{\Omega}{}_{-1}S_{lm\omega}^2$ to a term with $\int\d{\Omega}{}_{+1}S_{lm\omega}^2$, 
when both types of terms appear in $\int\d{\Omega}\Delta {}_{lm\omega}T_{tr}$ .
This is in accord with the fact that if we integrate the conservation
equation (\ref{eq:integrated conserv.eqs., T_tr}) over the solid angle we immediately obtain that
\begin{equation}
\int\d{\Omega}\Delta T_{tr}=\int\d{\Omega} K(\theta)=const.
\end{equation}
Indeed, we found that
\begin{equation} \label{eq:DeltaTtr integrated over solid angle}
\int\d{\Omega} \Delta {}_{lm\omega}T_{tr}^{\text{up}}=-\int\d{\Omega} \Delta {}_{lm\omega}T_{tr}^{\text{in}}=\frac{-1}{4\pi}\omega\mathbb{T}_{lm\omega}
\end{equation}
where we have included the constants of normalization (\ref{eq:normalization consts.}).
We can now give simple expressions for the luminosity when the electromagnetic field is in the past Boulware state and in the past Unruh state:
\begin{subequations} \label{eq:dM/dt for B-,U-}
\begin{align}
\left.\diff{M}{t}\right|_{B^-}&=
\frac{1}{2\pi}\sum_{l=1}^{\infty}\sum_{m=1}^{+l}\sum_{P=\pm1}\int_0^{m\Omega_+}\d{\omega}\omega\mathbb{T}_{lm\omega} \label{eq:dM/dt for B-}
\\
\left.\diff{M}{t}\right|_{U^-}&=
\frac{1}{2\pi}\sum_{l=1}^{\infty}\sum_{m=-l}^{+l}\sum_{P=\pm1}\int_0^{\infty}\d{\omega}\frac{\omega\mathbb{T}_{lm\omega}}{e^{2\pi\tilde{\omega}/\kappa}-1}
\label{eq:dM/dt for U-}
\end{align}
\end{subequations}
in agreement with Page's ~\cite{ar:PageII'76}. 
The former is a manifestation of the Starobinski\u{\i}-Unruh radiation and the latter of the Hawking radiation.
Since only superradiant modes are being included in the Starobinski\u{\i}-Unruh radiation (\ref{eq:dM/dt for B-}) and 
the transmission coefficient $\mathbb{T}_{lm\omega}$ is negative for these modes, there is a constant outflow of energy from the black hole
when the field is in the past Boulware state. 

We numerically evaluated (\ref{eq:dM/dt for B-,U-}) for the case $a=0.95M$.
The results, compared against values in the literature are:
\begin{subequations} \label{eq:dM/dt for B- for a=0.95,M=1}
\begin{align}
M^2\left.\diff{M}{t}\right|_{B^-}&=-4.750*10^{-4}    && (\text{spin-1})          \label{eq:num. dM/dt in B-, s=1}\\
M^2\left.\diff{M}{t}\right|_{B^-}&=-5.01*10^{-5}      && (\text{spin-0, Duffy})   \label{eq:Gav. num. dM/dt in B-, s=0} 
\end{align}
\end{subequations}
in the past Boulware state, and
\begin{subequations} \label{eq:dM/dt for U- for a=0.95,M=1}
\begin{align}
M^2\left.\diff{M}{t}\right|_{U^-}&=-1.1714*10^{-3}    && (\text{spin-1})                \label{eq:num. dM/dt in U-, s=1} \\
M^2\left.\diff{M}{t}\right|_{U^-}&=-1.18*10^{-3}      && (\text{spin-1, Page})   \label{eq:Page num. dM/dt in U-, s=1} 
\end{align}
\end{subequations}
in the past Unruh state.
The value (\ref{eq:Gav. num. dM/dt in B-, s=0}) for the scalar field is calculated by Duffy ~\cite{th:GavPhD} and 
we have calculated (\ref{eq:Page num. dM/dt in U-, s=1}) from splining Page's ~\cite{ar:PageII'76} numerical results. Both
of them have also been calculated for $a=0.95M$.


\section{RSET close to the horizon in the Boulware vacuum} \label{sec:RRO}

Candelas and Deutsch ~\cite{ar:Cand&Deutsch'77} consider flat space-time in the presence of an accelerating barrier
with acceleration $a_B^{-1}$. They then calculate the spin-1 RSET in the tetrad of an accelerating
observer RO with local acceleration $\xi^{-1}$. In the limit $\xi/a_B\rightarrow \infty$ 
the vacuum state above the accelerating mirror approximates the Fulling vacuum $\ket{F}$. 
The result is
\begin{equation} \label{eq:stress tensor in Fulling in RO tetrad}
\begin{aligned}
\vac[ren]{T^{\bar{\mu}}{}_{\bar{\nu}}}{F}
\sim& -\frac{1}{\pi^2\xi^4}\int_0^{\infty}\d{x}
\frac{x^3+x}{e^{2\pi x}-1}\text{diag}\left(-1,\frac{1}{3},\frac{1}{3},\frac{1}{3}\right)= \\
&=\frac{-11}{240\pi^2\xi^4}\text{diag}\left(-1,\frac{1}{3},\frac{1}{3},\frac{1}{3}\right)  \qquad (\xi/a_B\rightarrow \infty)
\end{aligned}
\end{equation}
where the bars on the indices indicate RO tetrad.  
Expression (\ref{eq:stress tensor in Fulling in RO tetrad}) is equivalent to minus the stress-energy tensor for thermal radiation
at a temperature of $(2\pi\xi)^{-1}$.
As mentioned in Section \ref{sec:vac. states}, in the Schwarzschild space-time, analogously to (\ref{eq:stress tensor in Fulling in RO tetrad}) in flat space,
the RSET close to the horizon when the field is in the Boulware vacuum diverges like minus the stress tensor of black body radiation at the black hole temperature. 
It is therefore reasonable to expect that if there existed a
state in Kerr with the defining features that the Boulware vacuum possesses in Schwarzschild, then
the RSET close to the horizon
when the field were in this vacuum, would 
diverge like minus the stress tensor of black body radiation at the black hole temperature rotating with the horizon.
The past Boulware vacuum is not invariant under $(t,\phi)$ reversal because of the existence of the Starobinski\u{\i}-Unruh radiation.
However, the stress tensor components $tr$ and $r\phi$, which correspond to the Starobinski\u{\i}-Unruh radiation, are expected 
(from Section \ref{sec:luminosity}) to have a divergence of one lower leading order than that of the diagonal components as the horizon is approached.
It is with this understanding that we say that a state is isotropic at the horizon and that, in particular, the past Boulware vacuum might be isotropic.
It is obvious that to next order in $\Delta$ the past Boulware vacuum cannot be isotropic, 
but $\vac[ren]{\hat{T}^{\mu}{}_{\nu}}{CCH^--B^-}$
might be since $\ket{CCH^-}$ is not invariant under $(t,\phi)$ reversal either. 

CCH claim that the RSET of the electromagnetic field in the past Boulware vacuum close to the horizon 
differs from that of minus the stress-energy tensor of a thermal distribution rotating at the angular 
velocity of a Carter observer by a factor which is a function of $\theta$.
In the present section, we will show that CCH's result is due to a flawed assumption in the asymptotic behaviour of the SWSH. 
We will show this by re-calculating their result using the assumptions we believe they used. 
The numerical results back up the fact that the mentioned RSET is (minus) thermal at the horizon.

We also mentioned in Section \ref{sec:vac. states} that Frolov and Thorne claim that close to the horizon ZAMOs measure a thermal 
stress tensor which is rigidly rotating with the horizon when the field is in the $\ket{FT}$ state.
That is, they argue that $\vac[ren]{\hat{T}^{\mu}{}_{\nu}}{FT-B}$ (where $\ket{B}$ is an unspecified Boulware-type state)
is thermal close to the horizon, and isotropic in the frame of a RRO.
Duffy, in turn, shows that close to the horizon RROs measure a thermal state which is rigidly
rotating with the horizon when the field is in the $\ket{H_{\mathcal{M}}}$ state in the Kerr space-time 
modified with the introduction of a mirror. 
That is to say, close to the horizon $\vac[ren]{\hat{T}^{\mu}{}_{\nu}}{H_{\mathcal{M}}-B_{\mathcal{M}}}$ is 
thermal and isotropic in the frame of a RRO.
Finally, as mentioned above, CCH claim that $\vac[ren]{\hat{T}^{\mu}{}_{\nu}}{CCH^--B^-}$ at the horizon differs by a factor from a thermal 
distribution isotropic in the Carter tetrad. 
Of course, the angular velocity at the horizon of a Carter observer, a RRO and a ZAMO is $\Omega=\Omega_+$ for them all,
so that CCH's result
does not actually distinguish between these observers.

Ottewill and Winstanley ~\cite{ar:Ott&Winst'00Lett} have proved that if a certain stress-energy tensor is thermal and rigidly-rotating with
the horizon everywhere, then it is divergent on the speed-of-light surface in the Boyer-Lindquist co-ordinates,
which are regular on this surface. 
This implies that if $\vac[ren]{\hat{T}^{\mu}{}_{\nu}}{CCH^--B^-}$ were thermal and
rigidly-rotating with the horizon everywhere then the state $\ket{CCH^-}$ would have to be irregular on the speed-of-light surface.
In the present section we will numerically investigate the rate of rotation of the thermal distribution in question.

The stress-energy tensor of a spin-1 thermal distribution at the Hawking temperature rigidly rotating with the horizon is given by
\begin{equation} \label{eq:thermal RR stress tensor}
T^{\text{(th,RR)} \mu}{}_{\nu}=\frac{11T^4\pi^2}{45}\left[\delta^{\mu}_{\nu}-4\frac{\chi^{\mu}\chi_{\nu}}
{\chi^{\rho}\chi_{\rho}}\right]
\end{equation}
where
\begin{equation} \label{eq:def. T}
T \equiv \frac{T_H}{\sqrt{-\chi^{\rho}\chi_{\rho}}}
\end{equation}
is the local temperature. Note that this stress-energy tensor is obviously isotropic in the frame of a RRO, but it is not
in the rigidly-rotating co-ordinate system $\{t_+,r,\theta,\phi_+\}$, where $\phi_+\equiv \phi-\Omega_+t$ and $t_+\equiv t$,
which is not adapted to a RRO.

In primed co-ordinates, which are adapted to a RRO, the rigidly-rotating thermal stress tensor becomes
\begin{equation} \label{eq:thermal RR stress tensor in RRO coords.}
T^{\text{(th,RR)} \mu'}{}_{\nu'}=\frac{11(r_+-r_-)^4}{2^8\cdot 3^2\cdot 5\pi^2}\frac{1}{\Delta^2\Sigma^2}\text{diag}\left(-1,\frac{1}{3},\frac{1}{3},\frac{1}{3}\right)
\end{equation}
in the Kerr space-time.

CCH calculate an expression for the RSET close to the horizon when the electromagnetic field is in the 
past Boulware state. They make the assumption that the RSET close to the horizon when the field 
is in this state is more irregular than when it is in the $\ket{CCH^-}$ state and therefore approximate
\begin{equation} \label{eq:RSET in B^- at horizon}
\vac[ren]{\hat{T}^{\mu}{}_{\nu}}{B^-}\sim\vac[ren]{\hat{T}^{\mu}{}_{\nu}}{B^-}-\vac[ren]{\hat{T}^{\mu}{}_{\nu}}{CCH^-}=
\vac{\hat{T}^{\mu}{}_{\nu}}{B^-}-\vac{\hat{T}^{\mu}{}_{\nu}}{CCH^-} \qquad (r\rightarrow r_+)
\end{equation}
One can then use the expressions for the expectation value of the stress tensor, and it is clear that only the
`up' modes are involved in the calculation.
Their result, when the components of the stress tensor are put in the Carter orthonormal
tetrad is:
\begin{equation}   \label{eq:CCH's RSET in B^- in Carter, for r->r_+}
\begin{aligned}
\vac[ren]{\hat{T}^{\hat{\mu}}{}_{\hat{\nu}}}{B^-} &\sim
-\frac{8M^3r_+}{\pi^2\Delta^2\Sigma}\int_0^{\infty}\d{\tilde{\omega}}
\frac{\tilde{\omega}\left(\tilde{\omega}^2+\kappa^2\right)}{e^{2\pi\tilde{\omega}/\kappa}-1}\text{diag}\left(-1,\frac{1}{3},\frac{1}{3},\frac{1}{3}\right)=\\
&=-\frac{1}{r_+}\frac{11(r_+-r_-)^4}{2^8\cdot 3^2\cdot 5\pi^2}\frac{1}{\Delta^2\Sigma (2Mr_+)}\text{diag}\left(-1,\frac{1}{3},\frac{1}{3},\frac{1}{3}\right)
\end{aligned}
\end{equation}
where the hats on the indices indicate adaptation to the Carter orthonormal tetrad.
This expression and the expected result, minus 
(\ref{eq:thermal RR stress tensor in RRO coords.}), differ in a factor of $r_+(2Mr_+)/\Sigma$.
We proceed to reproduce CCH's expression to explain
this disagreement.

We believe that 
CCH followed Candelas ~\cite{ar:Candelas'80} method for spin-0 to obtain asymptotic expansions 
for the radial solutions for spin-1 close to the horizon. This is the method that we develop for spin-1 in Appendix
\ref{sec:asympts. close to r_+}. Armed with the asymptotics of that appendix, we can proceed to calculate 
the different components of the stress-energy tensor. In order to do that,
we are first going to separately calculate the asymptotic expressions for the various terms that occur in 
the classical stress-energy tensor (\ref{eq:stress tensor, spin 1}).

As mentioned in Appendix \ref{sec:asympts. close to r_+}, for the asymptotic behaviour we are seeking here 
we can replace the spin-weighted spheroidal harmonics ${}_{\indhel}Z_{lm\omega}$ by the spin-weighted spherical 
harmonics ${}_{\indhel}Y_{lm}$. 
Note that this implies that the result of this asymptotic analysis is the same whether it is 
equations (\ref{eq:corrected stress tensor for s=1 on B-,U-}) and (\ref{eq:stress tensor for s=1 on FT and CCH^-})
or CCH's expressions for the expectation value of the stress tensor that are used.
We can make use of
(\ref{eq:eq.2aJ,McL,Ott'91}), which immediately leads to
\begin{equation} \label{eq:phi_iphi_j,i<>j,approx l->inf,r->rplus}
\sum_{m=-l}^l {}_{lm\omega}\phi_{\indhel}^{\text{up}}{}_{lm\omega}\phi_{h'}^{\text{up} *} 
\rightarrow 0 
\qquad (l\rightarrow +\infty, r \rightarrow r_+) 
\qquad \text{when } h\neq h'
\end{equation}

The asymptotic calculation of the term $\abs{{}_{lm\omega}\phi_{0}^{\text{up}}}^2$ requires a more careful treatment.
We observe in Appendix \ref{sec:asympts. close to r_+} that the large-$l$ modes dominate the Fourier series for the
`up' radial solution close to the horizon.
We can therefore replace $\sum_{l=0}^{\infty}$ with $\int_{0}^{\infty} \d{l}$  
and use the fact that of the two independent variables $\tilde{\omega}$ and $m$,
${}_{\indhel}R^{\text{up}}_{lm\omega}$ depends only on $\tilde{\omega}$ in the limit $(l\rightarrow +\infty, r \rightarrow r_+)$,
whereas ${}_{\indhel}Y_{lm}$ depends only on $m$.
From (\ref{eq:phi0(ch)}) and using
equations (\ref{eq:R1 `up' approx l->inf,r->rplus;compact version}),
(\ref{eq:Ddagger Delta R1 `up' approx l->inf,r->rplus}), (\ref{eq:eq.B6J,McL,Ott'95}),  (\ref{eq:eq.2aJ,McL,Ott'91}) 
and other properties of the spin-weighted spheroidal harmonics that can be found in ~\cite{ar:J&McL&Ott'95}
we then obtain that
\begin{equation} \label{eq:phi1^2 approx l->inf,r->rplus;2nd step}
\sum_{l,m,P}\abs{{}_{lm\omega}\phi_{0}^{\text{up}}}^2 \sim
\frac{1}{2^3\pi\Sigma}
\int_{0}^{\infty} \d{l}
l^3
|N^{\text{up}}_{+1}|^2 \abs{\mathcal{D}_0^{\dagger}(\Delta{}_{+1}R^{\text{up}}_{lm\omega})}^2
\qquad \quad  (r \rightarrow r_+) 
\end{equation}
We substitute (\ref{eq:def.A_s}), (\ref{eq:def.D}), and 
(\ref{eq:Ddagger Delta R+1 `up' approx l->inf,r->rplus})
in the above equation and approximate ${}_1B_{lm\omega} \sim l^2$, which is valid in the limit of large $l$.
The next integral, found in ~\cite{bk:GR}, is needed:
\begin{equation} \label{eq:eq.6.576(4)G&R,lambda=-3}
\int_{0}^{\infty} \d{l} l^3 K_{iq}^2(2lx^{1/2})=\frac{q^2(1+q^2)\abs{\Gamma(iq)}^2}{3\cdot 2^4x^2}
\end{equation}
We finally obtain 
\begin{equation} \label{eq:phi1^2 approx l->inf,r->rplus}
\begin{aligned}
\sum_{l,m,P}\abs{{}_{lm\omega}\phi_{0}^{\text{up}}}^2 \sim
\frac{Mr_+\tilde{\omega}\abs{\EuFrak{N}}^2}{6\pi^2\Sigma \Delta^2}
\qquad \quad  \text{($r \rightarrow r_+$)} 
\end{aligned}
\end{equation}

The other terms in the expression for the stress-energy tensor can be obtained in a similar manner,
but they are easier to calculate. We will therefore only give the final results:
\begin{subequations} \label{eq:phi0^2,phi2^2 approx l->inf,r->rplus}
\begin{align}
\sum_{l,m,P}\abs{{}_{lm\omega}\phi_{-1}^{\text{up}}}^2 & \sim
\frac{2Mr_+\tilde{\omega}\abs{\EuFrak{N}}^2}{3\pi^2 \Delta^3}
&
(r \rightarrow r_+)
\label{eq:phi0^2 approx l->inf,r->rplus} \\ 
\sum_{l,m,P}\abs{{}_{lm\omega}\phi_{+1}^{\text{up}}}^2 & \sim
\frac{Mr_+\tilde{\omega}\abs{\EuFrak{N}}^2}{6\pi^2 \Sigma^2 \Delta}
&
(r \rightarrow r_+)  \label{eq:phi2^2 approx l->inf,r->rplus}
\end{align}
\end{subequations}

We can now use equations (\ref{eq:phi_iphi_j,i<>j,approx l->inf,r->rplus}), (\ref{eq:phi1^2 approx l->inf,r->rplus}) 
and (\ref{eq:phi0^2,phi2^2 approx l->inf,r->rplus}) together with the quantum expressions (\ref{eq:RSET in B^- at horizon})
and the expectation value of the stress tensor to reproduce equation 3.7 in CCH. We obtain
\begin{equation} \label{eq:eq.3.7CCH;mine}
\begin{aligned}
&\vac[ren]{\hat{T}^{\mu}{}_{\nu}}{B^-} \sim \vac{\hat{T}^{\mu}{}_{\nu}}{B^--CCH^-} \sim
\frac{-8M^3r_+^3}{3\pi^2\Delta^2\Sigma^2}\int_0^\infty 
\frac{\d{\tilde{\omega}}\tilde{\omega}(\tilde{\omega}^2+\kappa^2)}{e^{2\pi\tilde{\omega}/\kappa}-1}
\times
\\
&
\times
\left(
\begin{array}{cccc}
-3(r_+^2+a^2)-a^2\sin^2\theta & 0 & 0 & 4a\sin^2\theta(r_+^2+a^2)  \\
0 & \Sigma & 0 & 0 \\
0 & 0 & \Sigma & 0 \\
-4a & 0 & 0 & (r_+^2+a^2)+3a^2\sin^2\theta
\end{array}
\right)
\qquad \quad 
(r \rightarrow r_+)
\end{aligned}
\end{equation}
in Boyer-Lindquist co-ordinates.
This is exactly equation 3.7 in CCH except for the fact that (\ref{eq:eq.3.7CCH;mine}) contains a factor $r_+^3$ instead
of a $r_+$ in CCH. We believe that the discrepancy is due to a typographical error in CCH 
since otherwise the stress-energy tensor would not have the correct units.
We have also checked that equation (\ref{eq:eq.3.7CCH;mine}), when the tensor indices are adapted
to the Carter orthonormal tetrad, produces the result (\ref{eq:CCH's RSET in B^- in Carter, for r->r_+}) above.
Again, the discrepancy with respect to (\ref{eq:CCH's RSET in B^- in Carter, for r->r_+}) is only 
in the power of $r_+$.
It seems that, despite the dicrepancy in the power of $r_+$, this is the method that CCH used to calculate
their expression (\ref{eq:CCH's RSET in B^- in Carter, for r->r_+}).
However, as we point out in Appendix \ref{sec:asympts. close to r_+}, this asymptotic analysis is only valid when both $\omega$ and $m$ are kept bounded
since otherwise we would not be able to replace the spin-weighted spheroidal harmonics by the spin-weighted
spherical harmonics. In the analysis we have just carried out $\omega$ and $m$ do not both remain bounded in general.
The only points in the Kerr space-time where both remain bounded are the points along the axis $\theta=0$ or $\pi$ since, 
there, the Newman-Penrose scalars ${}_{lm\omega}\phi_{\indhel}$ are only non-zero for $m=\pm1,0$ and thus $m$ is bounded. 
The frequency $\omega$ is then also kept bounded because the factor in the integrand diminishes
exponentially with $\tilde{\omega}$ and thus the contribution is only important when $\tilde{\omega}$ is bounded.
Equations (\ref{eq:eq.3.7CCH;mine}) and (\ref{eq:CCH's RSET in B^- in Carter, for r->r_+}) are therefore
only valid at the axis. An asymptotic behaviour of the `up' radial solutions uniform both in $l$
and $\tilde{\omega}$ is required.

Another issue is the fact that the state $\ket{CCH^-}$ has been used in (\ref{eq:RSET in B^- at horizon}) as a Hartle-Hawking
state, regular on both the past and future horizons. 
We know from Kay and Wald's work that there exists no such state on the Kerr space-time satisfying its isommetries.
Since $\ket{CCH^-}$ is not invariant under $(t,\phi)$ reversal it is not covered by Kay and Wald's result and thus it might
be regular on both $\mathcal{H^-}$ and $\mathcal{H^+}$.
We saw that Ottewill and Winstanley ~\cite{ar:Ott&Winst'00} argued that in the scalar case this state is 
 irregular on $\mathcal{H^-}$ and regular on $\mathcal{H^+}$.
Even if that were also the case for spin-1, using $\ket{CCH^-}$ in the preceding calculation could still be acceptable if the divergence
of $\ket{CCH^-}$ close to $r_+$ is of a smaller order than that of $\ket{B^-}$. 
In the Schwarzschild background Candelas has shown that the Unruh state is irregular on $\mathcal{H^-}$, regular on $\mathcal{H^+}$ and
that the order of its divergence close to $r_+$ is smaller than that of the Boulware state. 
It is therefore reasonable to expect that the order of the divergence of $\ket{CCH^-}$ close to $r_+$ in the Kerr background
is smaller than that of the past Boulware state.
Indeed, our numerical data indicate that the approximation in (\ref{eq:RSET in B^- at horizon}) is correct.

\begin{figure}[H]
\rotatebox{90}
\centering
\includegraphics*[width=70mm,angle=270]{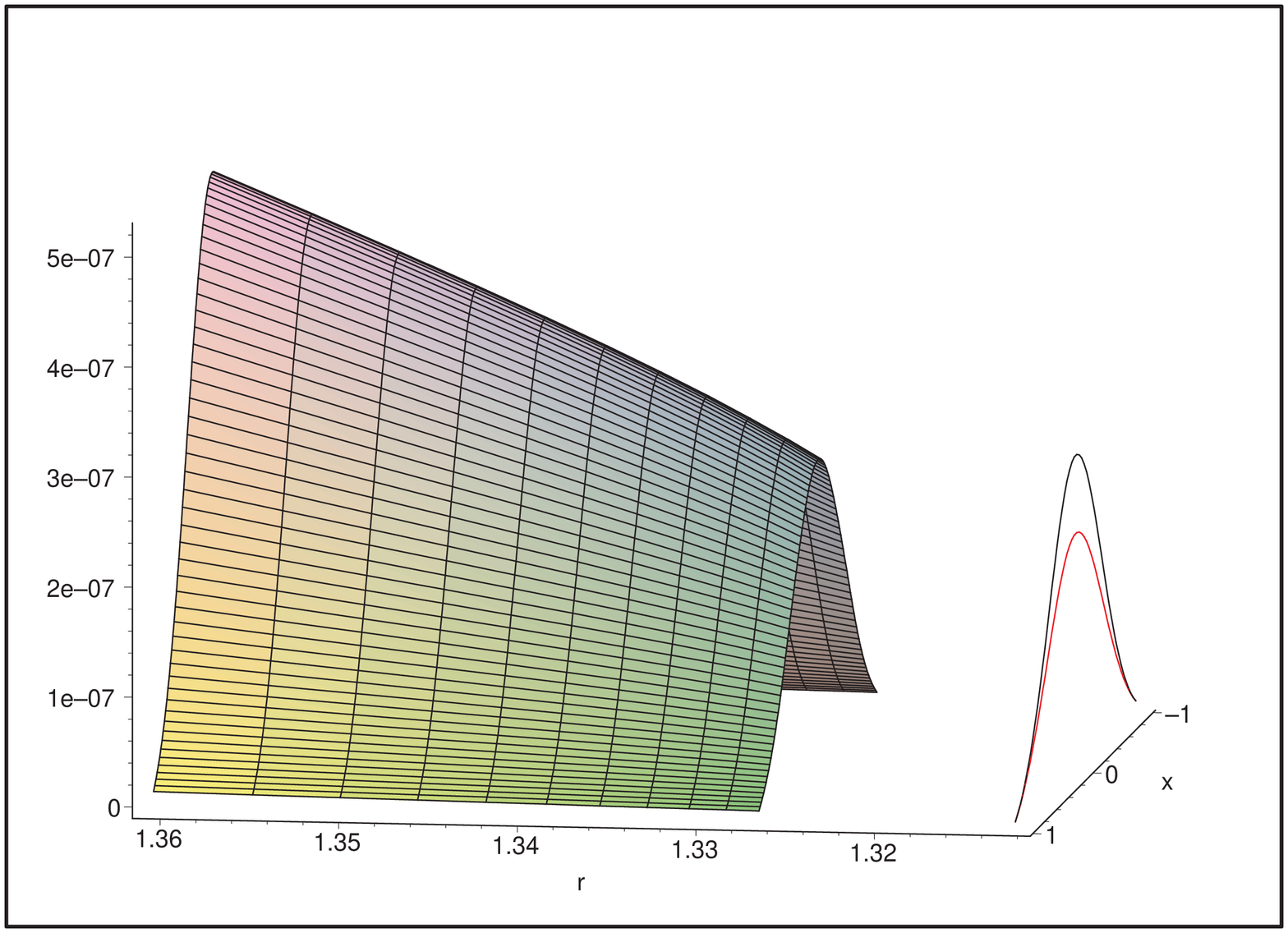} 
\caption{$\frac{1}{4\pi}\Delta^2\vac{\hat{T}_{tt}}{CCH^--B^-}$}    \label{fig:delta2Ttt_cch_b_past}
\includegraphics*[width=70mm,angle=270]{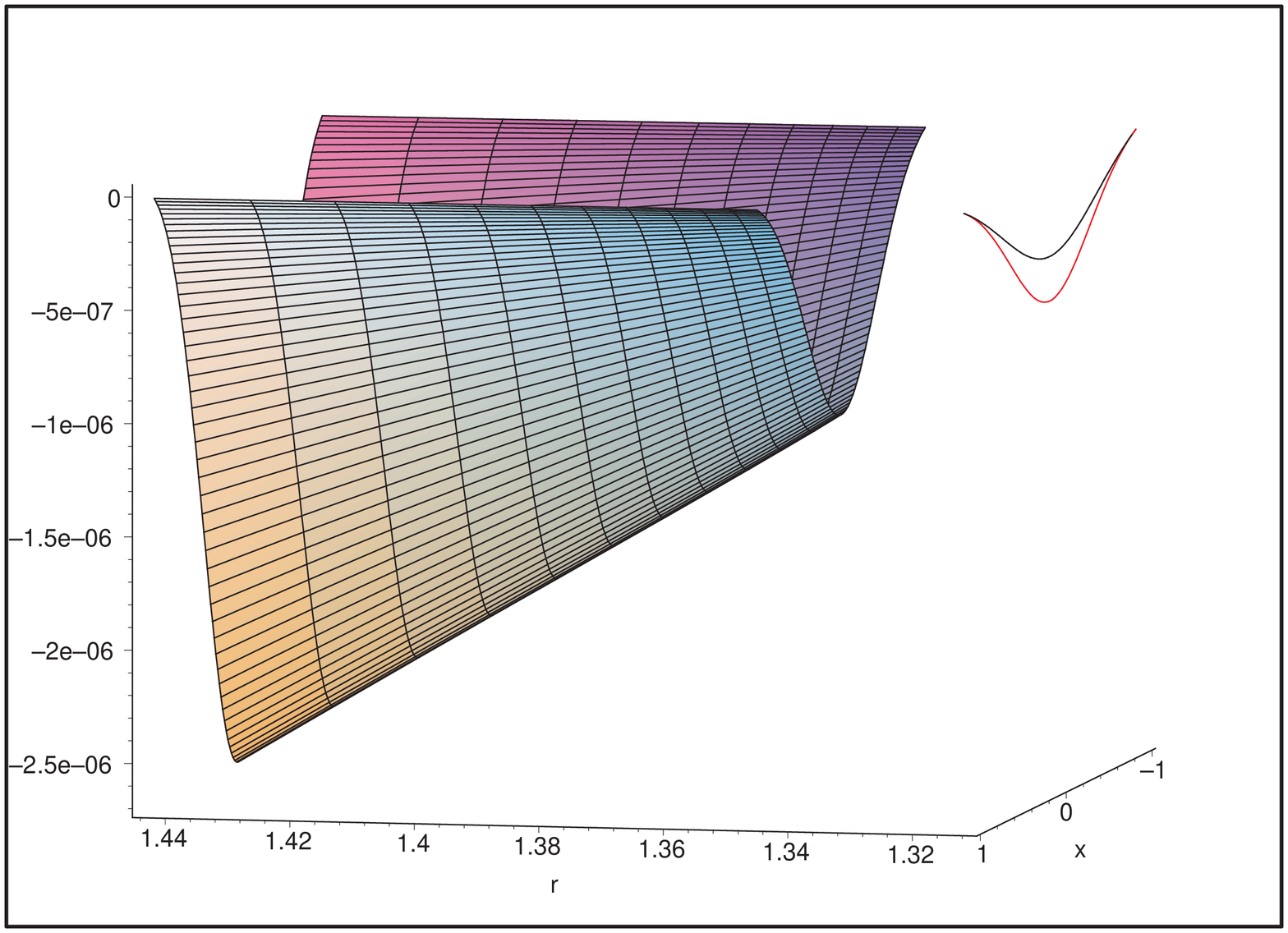}
\caption{$\frac{1}{4\pi}\Delta^2\vac{\hat{T}_{t\phi}}{CCH^--B^-}$}             \label{fig:delta2Ttphi_cch_b_past}
\end{figure}

\begin{figure}[H]
\rotatebox{90}
\centering
\includegraphics*[width=70mm,angle=270]{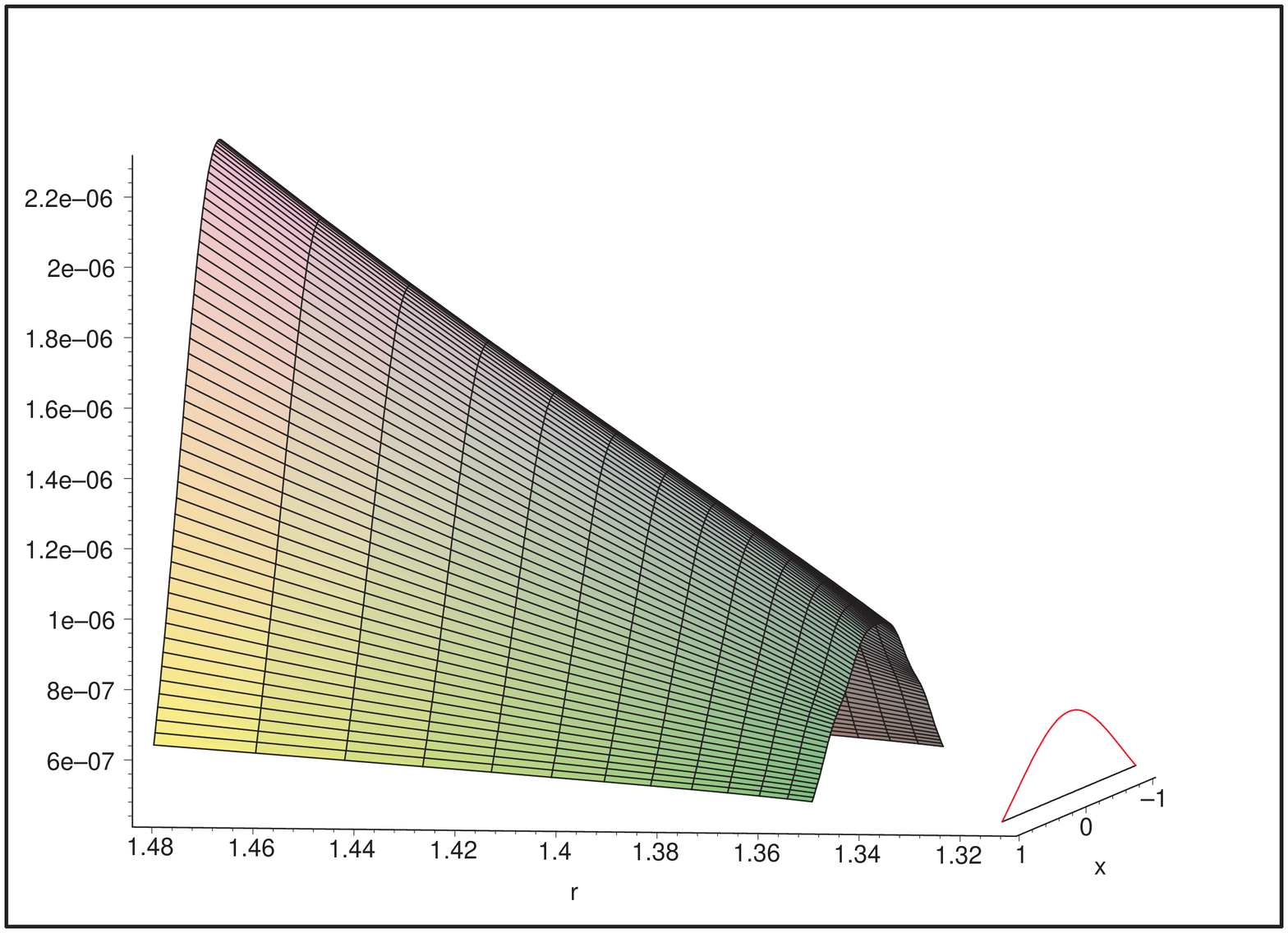}
\caption{$\frac{1}{4\pi}\Delta^2\vac{\hat{T}_{\theta\theta}}{CCH^--B^-}$}      \label{fig:delta2Tthetatheta_cch_b_past_bis}
\includegraphics*[width=70mm,angle=270]{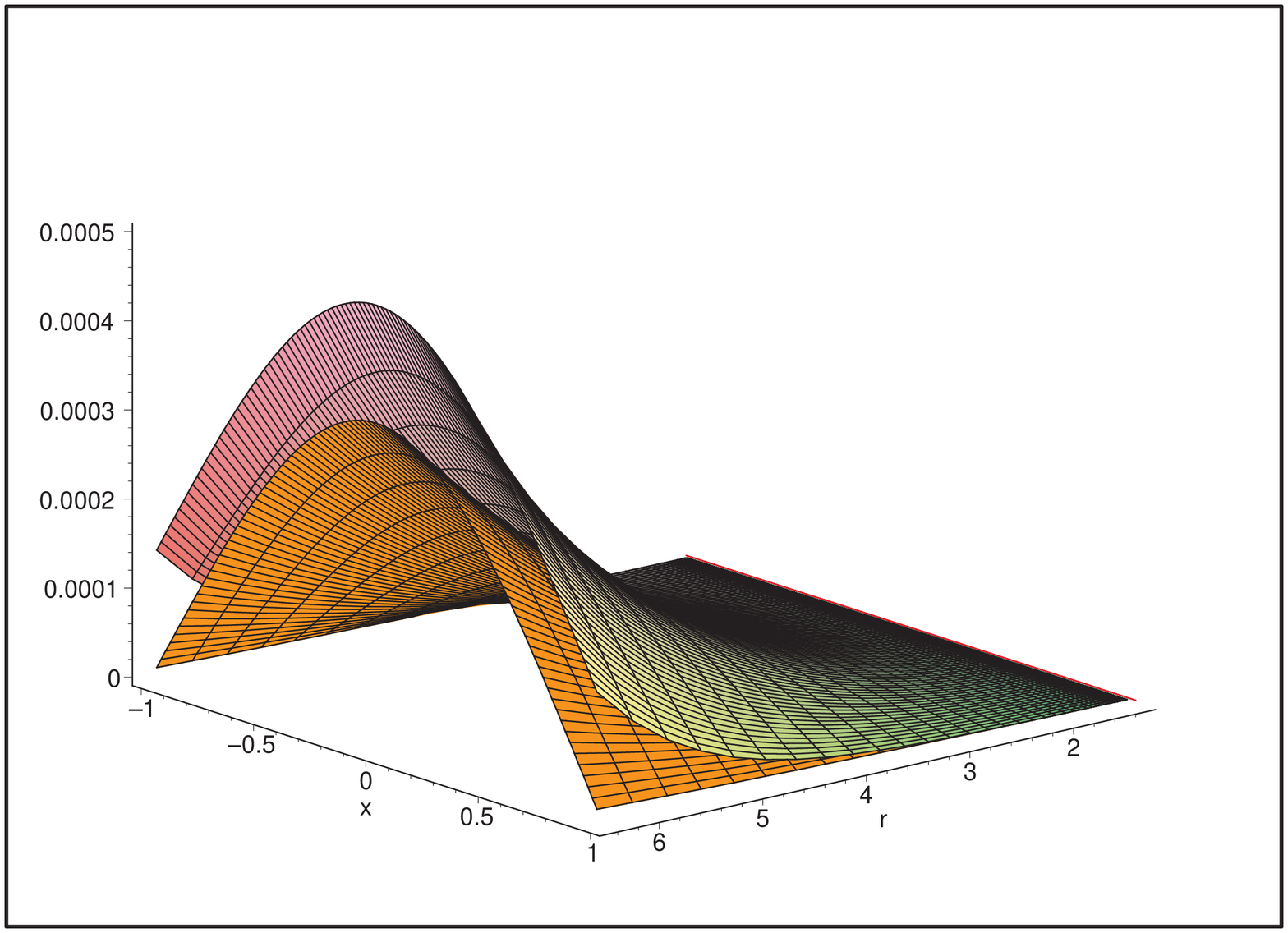} 
\caption{$\frac{1}{4\pi}\Delta^2\vac{\hat{T}_{\theta\theta}}{CCH^--B^-}$ and 
$\frac{1}{4\pi}\Delta^2\vac{\hat{T}_{\theta\theta}}{U^--B^-}$(orange)}    \label{fig:delta2Tthetatheta_u_b_cch_b_past}
\end{figure}

\begin{figure}[H]
\rotatebox{90}
\centering
\includegraphics*[width=70mm,angle=270]{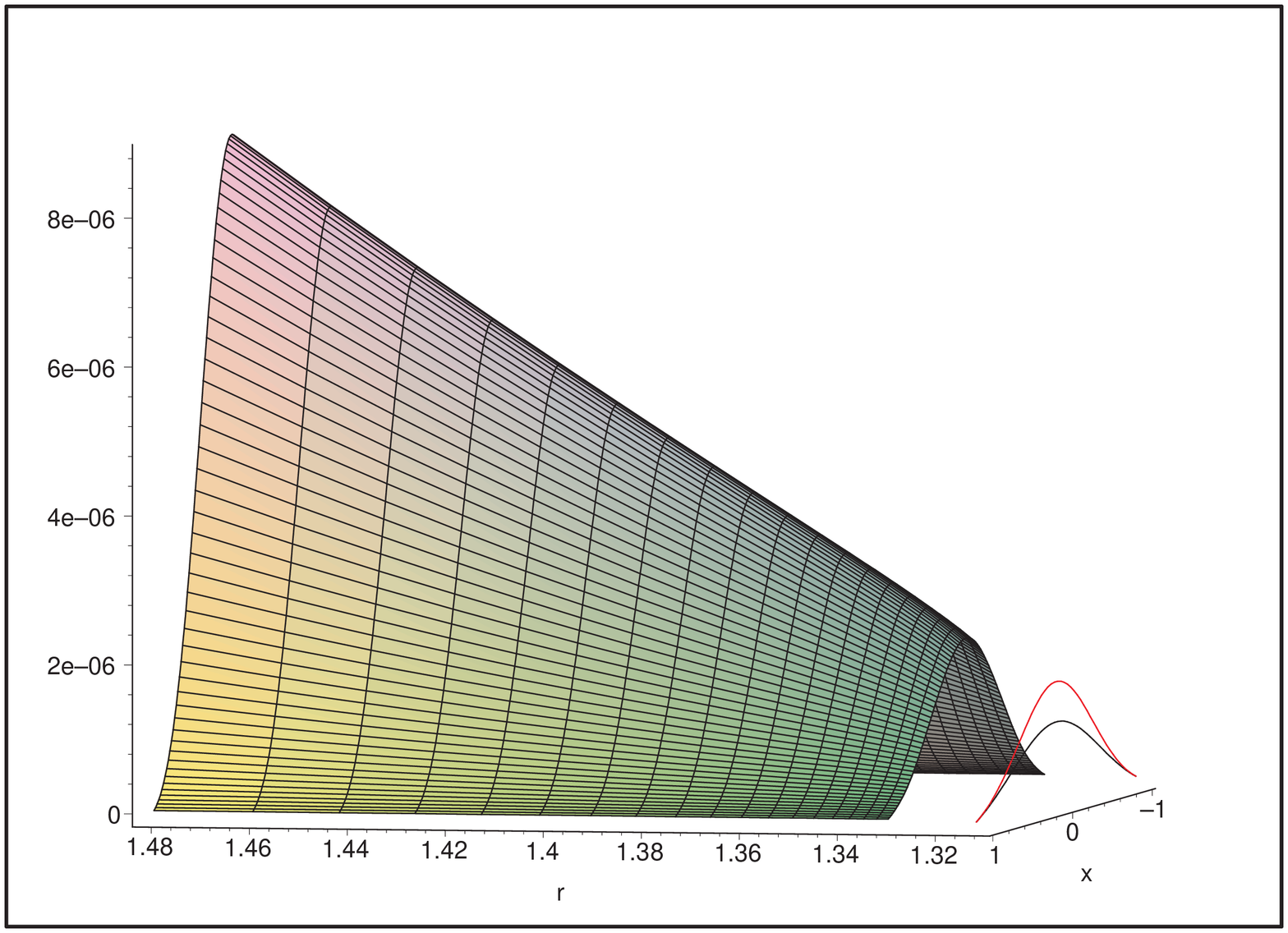} 
\caption{$\frac{1}{4\pi}\Delta^2\vac{\hat{T}_{\phi\phi}}{CCH^--B^-}$}          \label{fig:delta2Tphiphi_cch_b_past}
\end{figure}

Graphs \ref{fig:delta2Ttt_cch_b_past}--\ref{fig:delta2Tphiphi_cch_b_past}
show that the RSET when the field is in the past Boulware vacuum approaches a thermal distribution rotating
with the horizon rather than CCH's result (\ref{eq:eq.3.7CCH;mine}). 
The red lines in the graphs correspond to the thermal stress tensor (\ref{eq:thermal RR stress tensor}) 
rotating with the horizon located at $r=r_+\simeq 1.3122$.
The black lines are also located at $r=r_+$ and correspond to CCH's result (\ref{eq:eq.3.7CCH;mine}).
It can be seen in the graphs that as $r$ becomes closer to the horizon, $\vac[ren]{\hat{T}{}_{\mu\nu}}{CCH^--B^-}$ approaches 
the thermal stress tensor (\ref{eq:thermal RR stress tensor}) (red line) rather than CCH's 
corrected equation (\ref{eq:eq.3.7CCH;mine}) (black line). At the poles, however, it can be straight-forwardly checked
analytically that the two coincide, as expected. Only for the $rr$-component, which is the only component that diverges like
$O(\Delta^{-3})$ close to the horizon, we did not seem to be able to obtain a clear plot, which we do not include. 

Within the range of $r$ considered in Graphs \ref{fig:delta2Ttt_cch_b_past}--\ref{fig:delta2Tphiphi_cch_b_past}
(except \ref{fig:delta2Tthetatheta_u_b_cch_b_past})
for the difference between the states $\ket{CCH^-}$ and $\ket{B^-}$ of the various expectation values, 
the corresponding plots for the difference between the states $\ket{U^-}$ and $\ket{B^-}$ are identical. 
This is the expected behaviour since for small radius $r$ the `up' modes dominate in these RSETs.
Graph \ref{fig:delta2Tthetatheta_u_b_cch_b_past} includes the two differences 
for the $\theta\theta$-component of the stress-energy tensor up to a value of $r$ large enough so that the two differences become clearly distinct.

\begin{figure}[H]
\rotatebox{90}
\centering
\includegraphics*[width=80mm,angle=270]{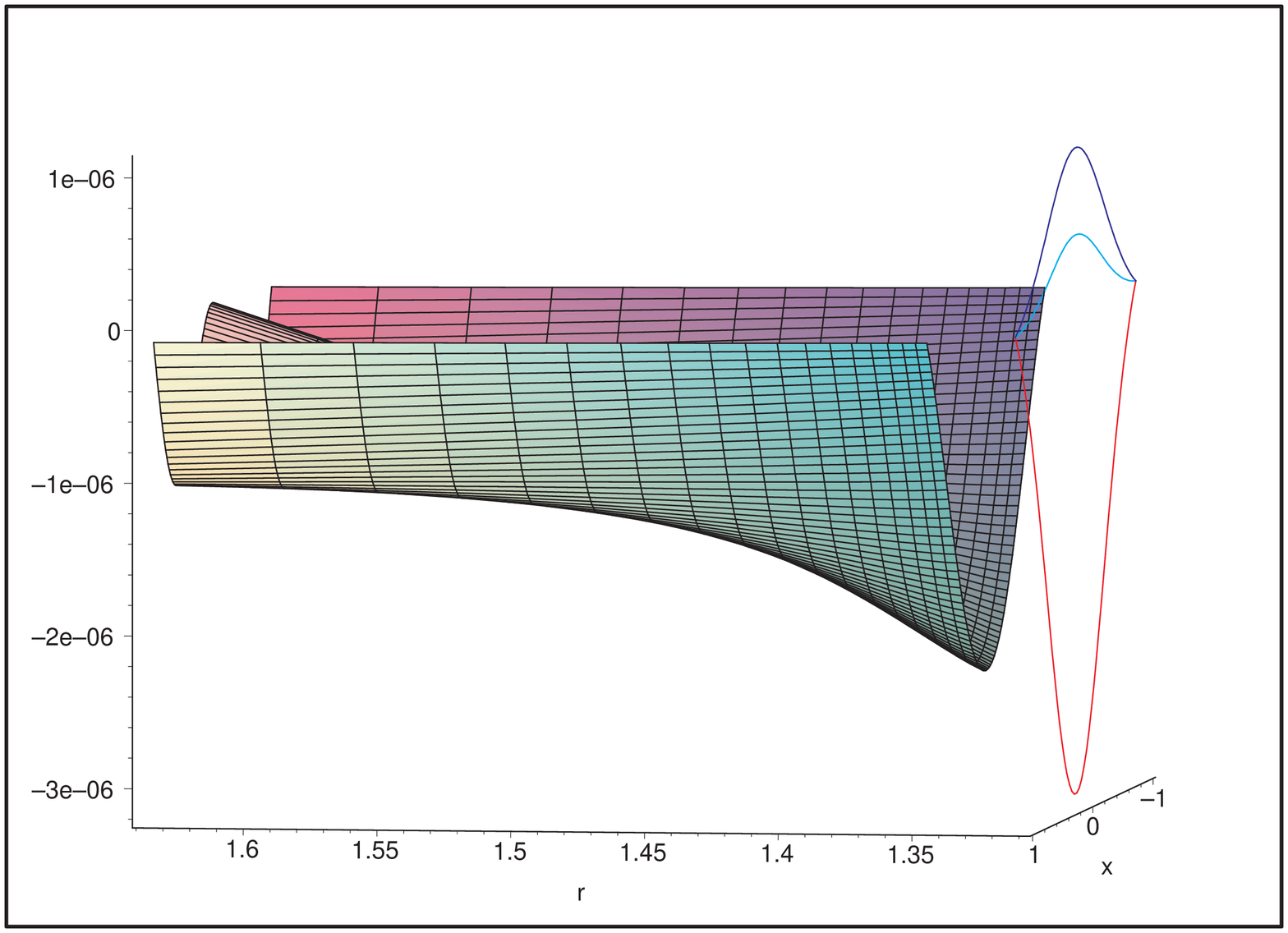} 
\caption{$\frac{1}{4\pi}\Delta\vac{\hat{T}_{t_+\phi_+}}{CCH^--B^-}$, $\frac{1}{4\pi}\Delta T^{\text{(th,RR)}}_{t_+\phi_+}$ (red), 
$\frac{1}{4\pi}\Delta T^{\text{(th,ZAMO)}}_{t_+\phi_+}$ (blue)
and $\frac{1}{4\pi}\Delta T^{\text{(th,Carter)}}_{t_+\phi_+}$ (cyan).}           \label{fig:deltaTtplusphiplus_cch_b_past}
\includegraphics*[width=80mm,,angle=270]{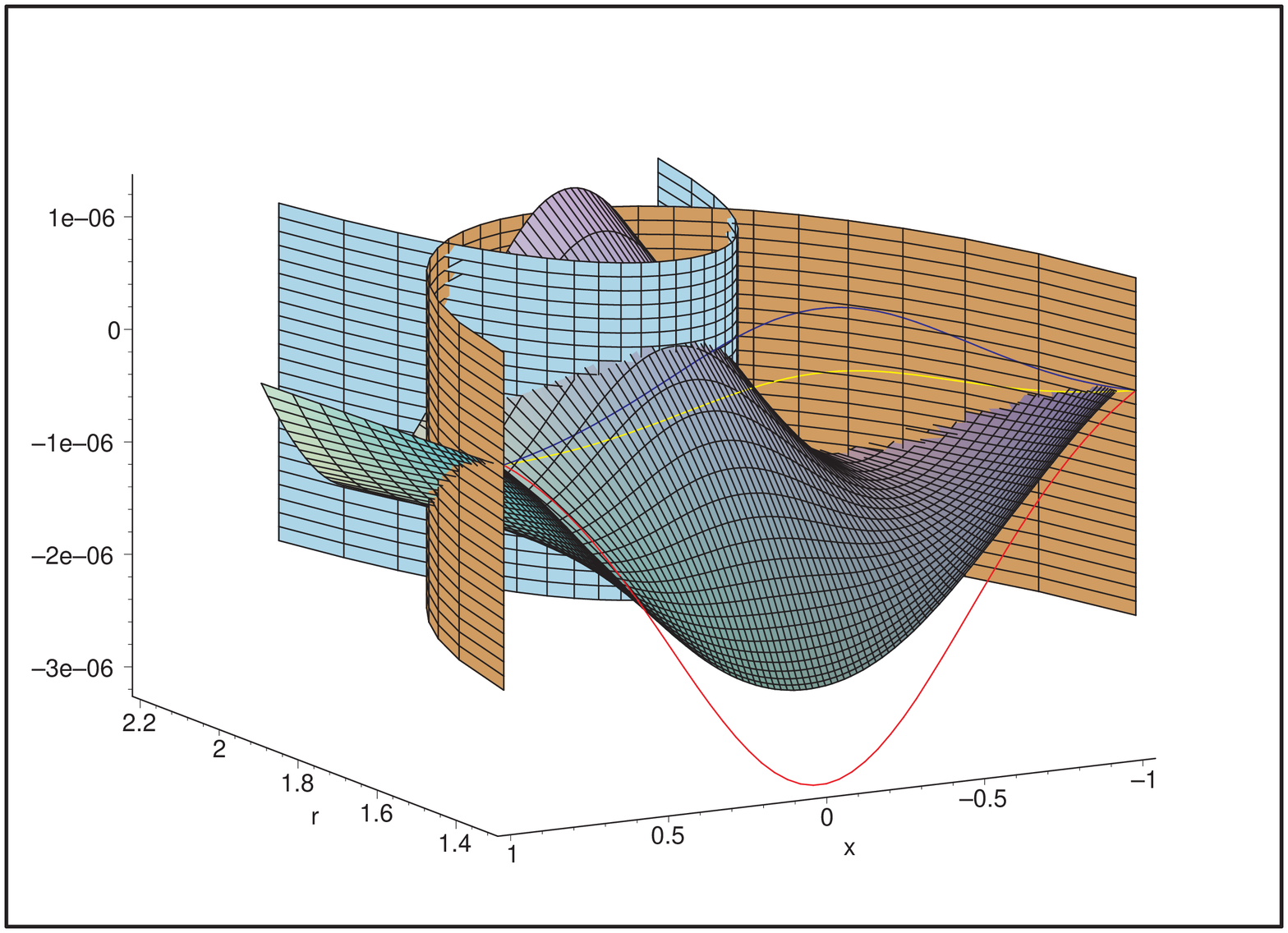} 
\caption{$\frac{1}{4\pi}\Delta\vac{\hat{T}_{t_+\phi_+}}{CCH^--B^-}$, $\frac{1}{4\pi}\Delta T^{\text{(th,RR)}}_{t_+\phi_+}$ (red), 
$\frac{1}{4\pi}\Delta T^{\text{(th,ZAMO)}}_{t_+\phi_+}$ (blue)
and $\frac{1}{4\pi}\Delta T^{\text{(th,Carter)}}_{t_+\phi_+}$ (yellow). The light blue and brown surfaces correspond to the speed-of-light and 
the static limit surfaces respectively.}                     \label{fig:deltaTtplusphiplus_cch_b_past_surfs}
\end{figure}

In following with the notation used in (\ref{eq:tetrad of stationary obs.}) and the one used so far for tensor components 
in Boyer-Lindquist co-ordinates, we use the obvious
notation of `$(\alpha\beta)$-component' to refer to the stress-energy tensor component $T{}_{\mu\nu}e_{(\alpha)}{}^{\mu}e_{(\beta)}{}^{\nu}$
in the tetrad of a stationary observer.
Since the angular velocities of a RRO, ZAMO
and Carter observer all equal $\Omega_+$ at the horizon, each one of the diagonal components of a stress tensor for a thermal
distribution will be the same 
in any of the three tetrads adapted to these observers.
The $(r\theta)$-component will also be the same 
in any of the three tetrads
since the tetrad vectors $\vec{e}_{(r)}$ and $\vec{e}_{(\theta)}$ do not depend
on the rate of rotation. The $(t\phi)$-component,
however, vanishes to leading order for the radial functions as $r\rightarrow r_+$.
To the next leading order for the radial functions this component does depend on the rate of rotation of 
the stationary observer that the tetrad is adapted to.
Graphs \ref{fig:deltaTtplusphiplus_cch_b_past}--\ref{fig:deltaTtplusphiplus_cch_b_past_surfs} for $\vac[ren]{\hat{T}^{}_{t_+\phi_+}}{CCH^--B^-}$ 
show that the rate of rotation of the thermal distribution
approaches, to next order in $\Delta$, that of a RRO, rather than that of a ZAMO or a Carter observer.
This result tallies with Duffy~\cite{th:GavPhD}'s results for the spin-0 case in the Kerr space-time 
modified with a mirror when the field is in the $\ket{H_{\mathcal{M}}}$ state. 
He also numerically shows that $\vac[ren]{\hat{T}{}_{\mu\nu}}{U^--B^-}$ is, close to the horizon and for the scalar field, 
thermal and rotating at the rate of a RRO to $O(\Delta)$ in the angular frequency. 
We calculated and plotted $\vac[ren]{\hat{T}_{t_+\phi_+}}{U^--B^-}$ and 
it fully coincided with $\vac[ren]{\hat{T}_{t_+\phi_+}}{CCH^--B^-}$ in the region of Graphs 
\ref{fig:deltaTtplusphiplus_cch_b_past}--\ref{fig:deltaTtplusphiplus_cch_b_past_surfs}, which is why we do not include them.
We conclude that the rate of rotation close to the horizon for the difference between the states $\ket{U^-}$ and $\ket{B^-}$ is also
that of a RRO, in agreement with Duffy's results.

\begin{figure}[H]
\centering
\includegraphics*[width=70mm,angle=270]{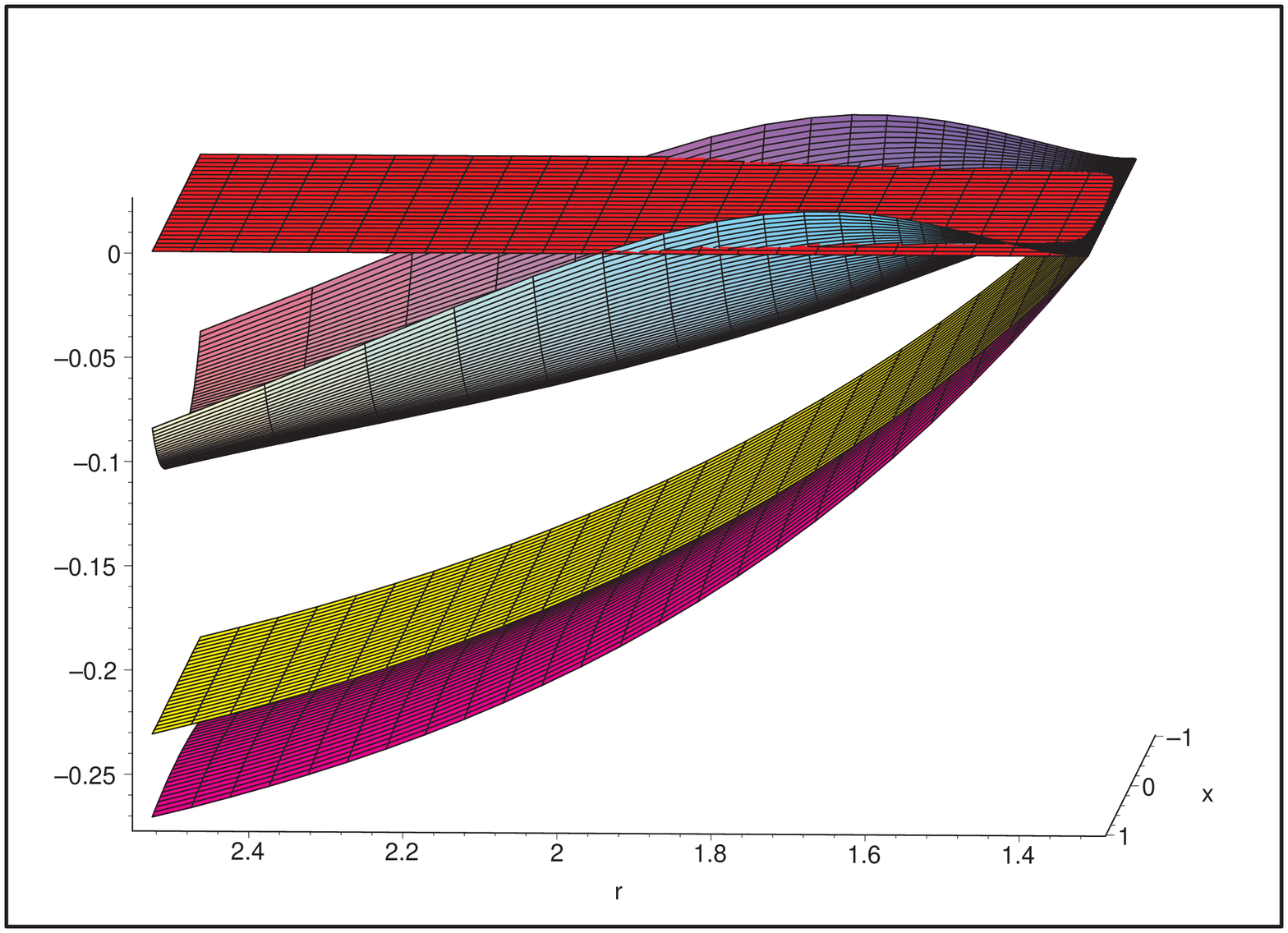} \\
$(\Omega_{\text{ZEFO}}-\Omega_+)$ \\
\includegraphics*[width=70mm,angle=270]{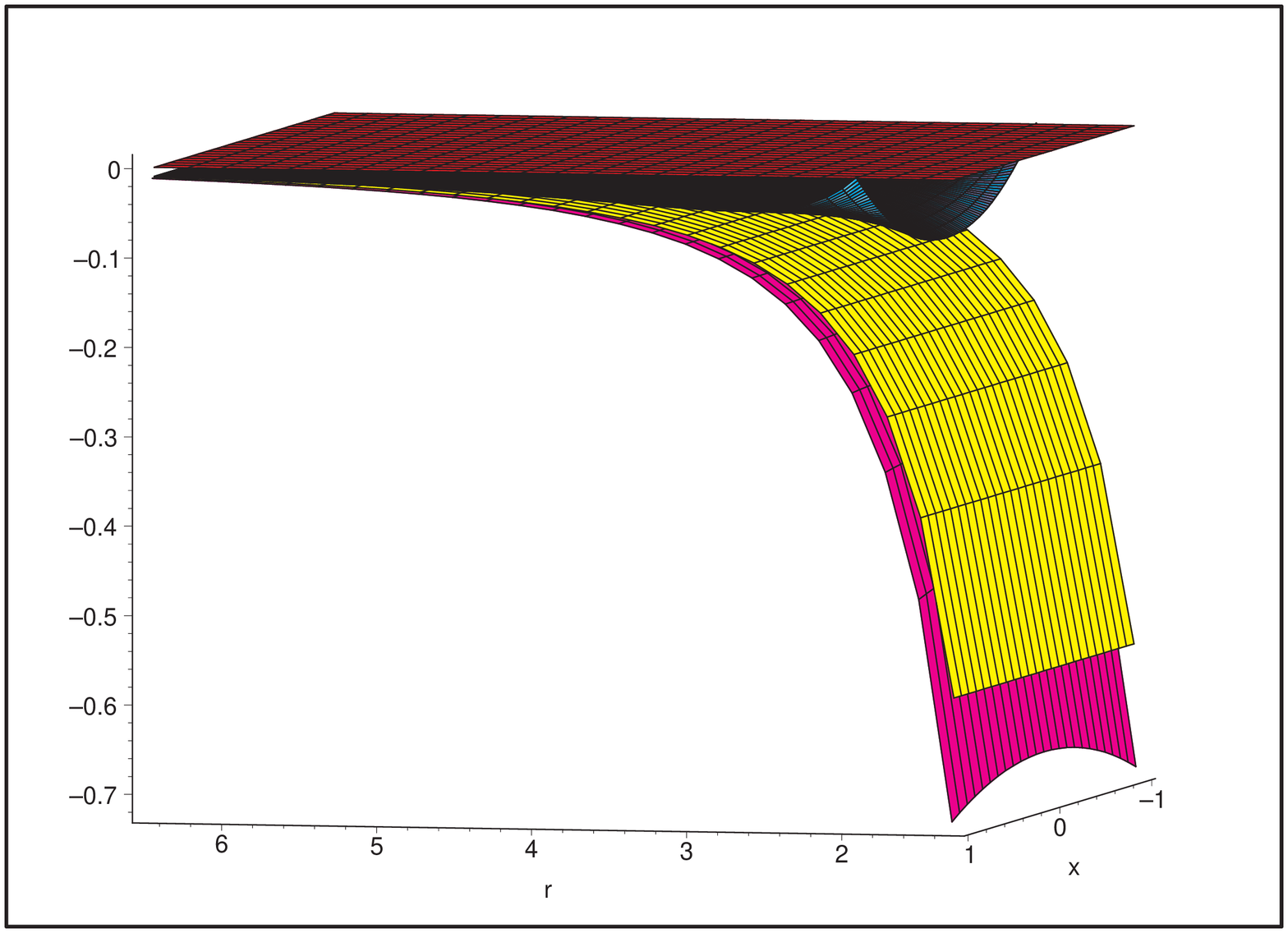} \\
$\left(\Omega_{\text{ZEFO}}-\Omega_+\right)/\Delta$ \\
\caption{Plots of $(\Omega_{\text{ZEFO}}-\Omega_+)$ and  $\left(\Omega_{\text{ZEFO}}-\Omega_+\right)/\Delta$ (dark surfaces) where 
$T_{\mu\nu}$ is replaced by $\vac[ren]{\hat{T}^{}_{\mu\nu}}{CCH^--B^-}$ in (\ref{eq:omega_ZEFO}), together with the corresponding
plots with the angular velocities of a RRO (red), ZAMO (magenta) and Carter observer (yellow).
}
\label{fig:freqCarterZAMO_omega_hor_cch_b_past}
\end{figure}

An alternative technique for investigating what is the rate of rotation of the thermal distribution at the horizon
is as follows. We find what is the frequency $\Omega=\Omega_{\text{ZEFO}}$ of rotation of the tetrad frame 
(\ref{eq:tetrad of stationary obs.}) such that $T_{(t\phi)}=0$, where
the term ZEFO stands for zero energy flux observer. The answer is
\begin{equation} \label{eq:omega_ZEFO}
\Omega_{\text{ZEFO}}=\frac{-2C}{B+\sqrt{B^2-4AC}}
\end{equation}
where
\begin{equation} \label{eq:def. of A,B,C for ZEFO}
\begin{aligned}
A&=g_{\phi\phi}T_{t\phi}-g_{t\phi}T_{\phi\phi} \\
B&=g_{\phi\phi}T_{tt}-g_{tt}T_{\phi\phi}\\
C&=g_{t\phi}T_{tt}-g_{tt}T_{t\phi}
\end{aligned}
\end{equation}
We then plot $\Omega_{\text{ZEFO}}$ where $T_{\mu\nu}$ is replaced by 
$\vac[ren]{\hat{T}_{\mu\nu}}{CCH^--B^-}$ in (\ref{eq:def. of A,B,C for ZEFO}). This plot is compared against that of the angular velocities 
of a RRO, ZAMO and Carter observer in Figure \ref{fig:freqCarterZAMO_omega_hor_cch_b_past}.
We also plotted $\Omega_{\text{ZEFO}}$ where $T_{\mu\nu}$ is replaced by 
$\vac[ren]{\hat{T}_{\mu\nu}}{U^--B^-}$ and it fully coincided with the corresponding one for $\vac[ren]{\hat{T}_{\mu\nu}}{CCH^--B^-}$
in the region of Figure \ref{fig:freqCarterZAMO_omega_hor_cch_b_past}.

\begin{figure}[H]
\centering
\begin{tabular}{cc}
\includegraphics*[height=65mm,width=75mm,angle=270]{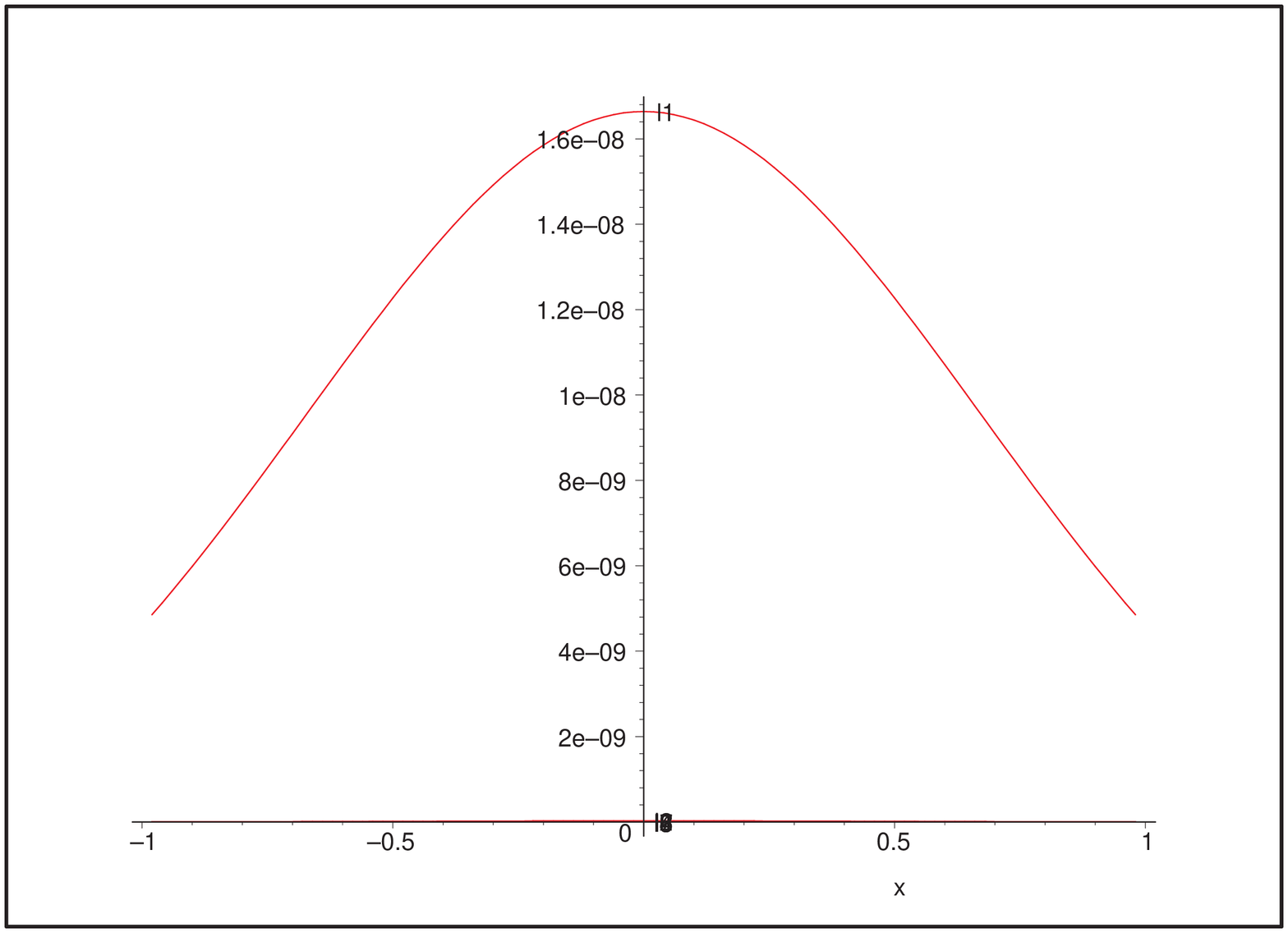} &
\includegraphics*[height=65mm,width=75mm,angle=270]{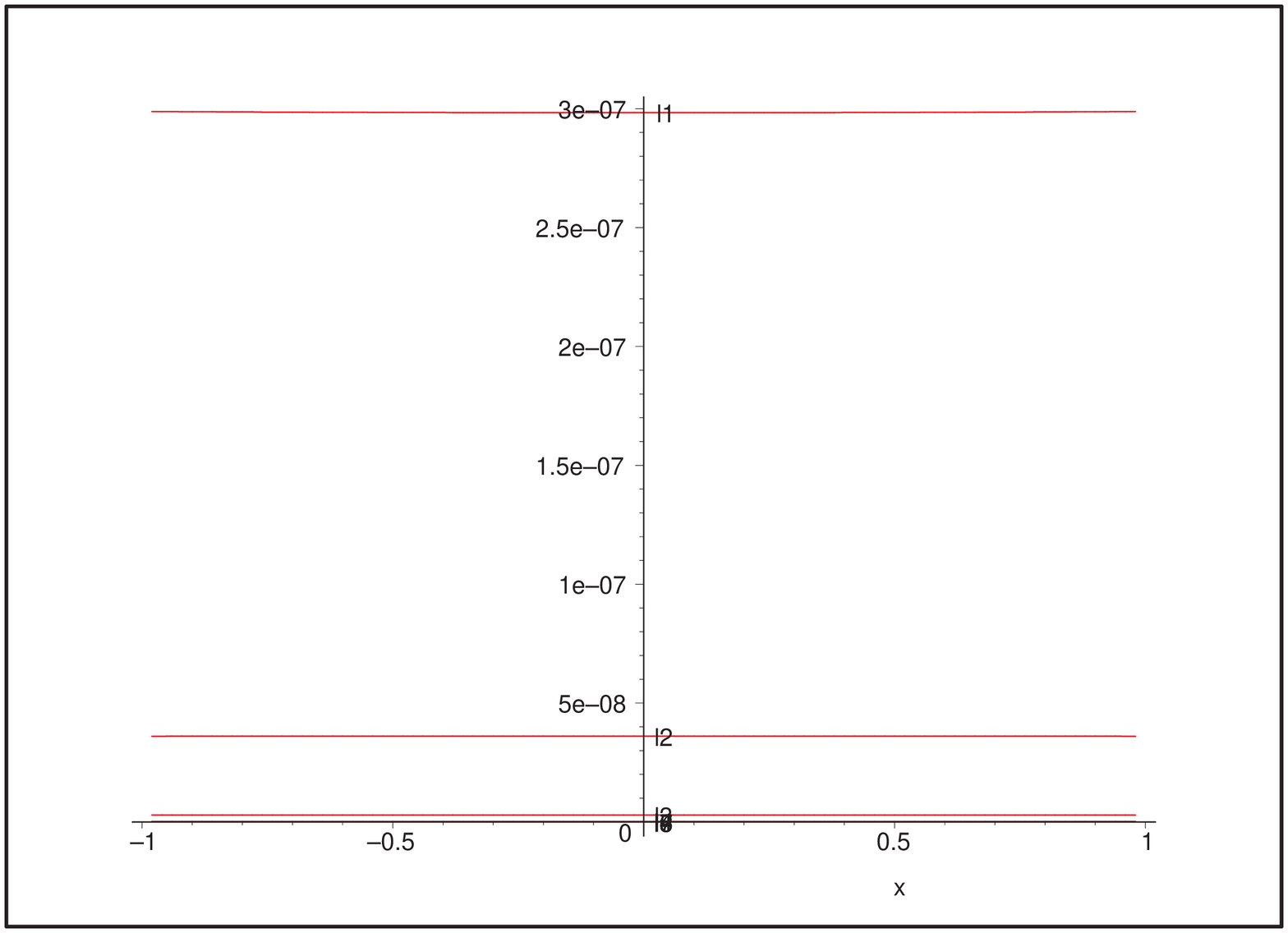} \\
$\frac{1}{4\pi}\vac{\hat{T}_{\theta\theta}(r \simeq 1.36,\theta)}{CCH^--U^-}$ & $\frac{1}{4\pi}\vac{\hat{T}_{\theta\theta}(r \simeq 10.26,\theta)}{CCH^--U^-}$\\
\includegraphics*[height=65mm,width=75mm,angle=270]{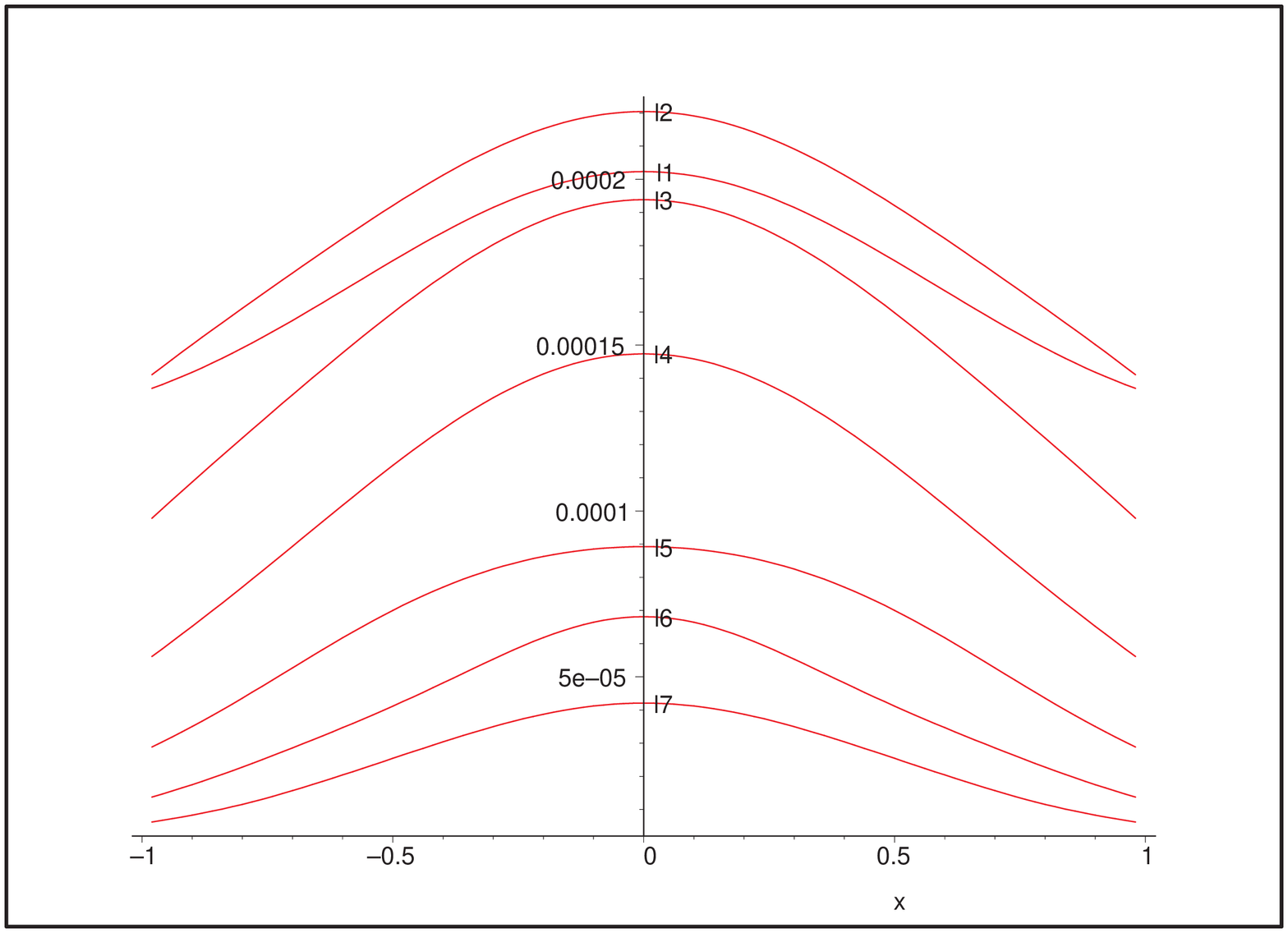} &
\includegraphics*[height=65mm,width=75mm,angle=270]{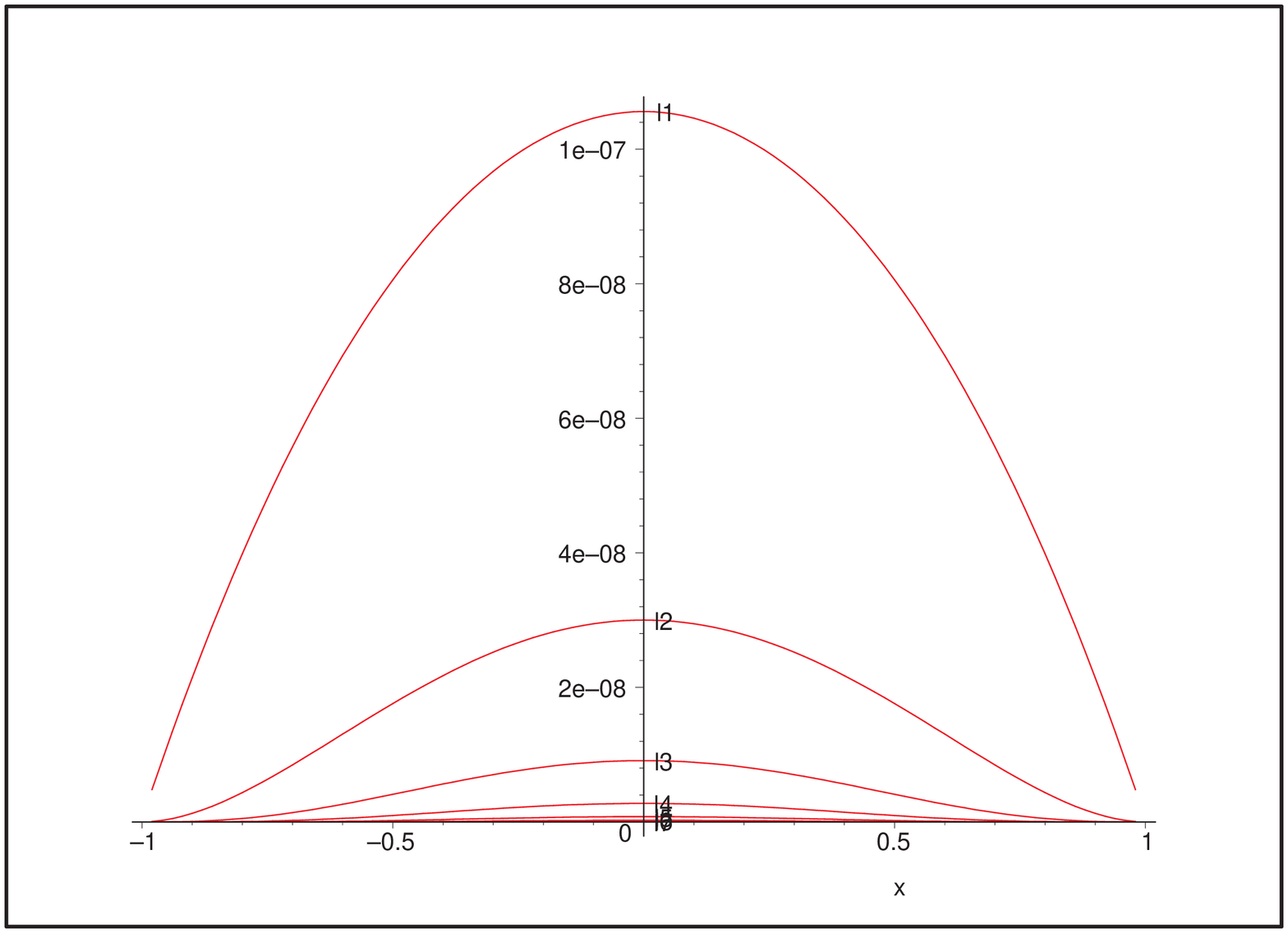} \\
$\frac{1}{4\pi}\vac{\hat{T}_{\theta\theta}(r \simeq 1.36,\theta)}{U^-B^-}$    & $\frac{1}{4\pi}\vac{\hat{T}_{\theta\theta}(r \simeq 10.26,\theta)}{U^-B^-}$\\
\end{tabular}
\caption{
The sum over $l$ has not been performed. For each value of $l$ the sum $\sum_{m=-l}^{l}$ has been performed.
In the case of $\vac{\hat{T}_{\theta\theta}}{CCH^--U^-}$ the low-$l$ modes clearly dominate close to the horizon. 
On the other hand, the high-$l$ modes dominate close to the horizon in the case of $\vac{\hat{T}_{\theta\theta}}{U^--B^-}$.
}
\label{fig:Tthetatheta_cch_u_past_u_b_spher_last_r100r150_n}
\end{figure}

Graphs \ref{fig:Tthetatheta_cch_u_past_u_b_spher_last_r100r150_n}
show the behaviour of the various modes as the horizon is approached. 
The behaviour of the `up' modes close to the horizon are explained by the 
horizon asymptotics developed in Appendix \ref{sec:asympts. close to r_+}.



\newpage


\appendix
\section{Radial function} \label{sec:radial func.}

This Appendix is valid in the Kerr-Newman background with the understanding that the 
background is considered fixed and the NP Maxwell scalars include the electromagnetic field
associated to the background as well as an electromagnetic perturbation. 
The quantity $Q$ represents the charge of the Kerr-Newman black hole as viewed from infinity;
in this Appendix, the quantities $\Delta$ and $r_{\pm}$ take the following values: $\Delta \equiv r^2-2Mr+a^2+Q^2$
and $r_{\pm}=M\pm\sqrt{M^2-a^2-Q^2}$.
The Kerr background is recovered in the limit $Q=0$.

The potential in the radial Teukolsky equation is a long-range potential.
Our numerical calculations followed Detweiler ~\cite{ar:Detw'76}, who derived from the radial Teukolsky equation 
another differential equation with a real, short-range potential.
We numerically solved this equation and obtained its solution $X_{lm\omega}$ and its derivative.
We then find from these the radial functions ${}_{+1}R_{lm\omega}$, ${}_{-1}R_{lm\omega}$ and its derivative with expressions in ~\cite{ar:Detw'76}.
The independent variable in Detweiler's differential equation is the tortoise co-ordinate $r_*$ defined by $\d{r_*}/\d{r}=(r^2+a^2)/\Delta$.
Detweiler's potential tends to $-\omega^2+O(r_*^{-2})$ at infinity ($r_*\to +\infty$) and to 
$-\tilde{\omega}^2$ at the horizon ($r_*\to -\infty$), where $\tilde{\omega}\equiv\omega-m\Omega_+$.
We can therefore define two sets of solutions with the following asymptotic behaviours:

\begin{subequations} \label{eq:X_in/up}
\begin{align}  
X^{\text{in}}_{lm\omega} & \sim
\begin{cases} 
B^{\text{in}}_{lm\omega}e^{-i\tilde{\omega}r_*} & (r\rightarrow r_+) \\
e^{-i\omega r_*}+A^{\text{in}}_{lm\omega}e^{+i\omega r_*} & (r\rightarrow +\infty)
\end{cases} \label{eq:X_in}
\\
X^{\text{up}}_{lm\omega} & \sim
\begin{cases}
e^{+i\tilde{\omega}r_*}+A^{\text{up}}_{lm\omega}e^{-i\tilde{\omega}r_*} & 
(r\rightarrow r_+) \\
B^{\text{up}}_{lm\omega}e^{+i\omega r_*} & (r\rightarrow +\infty)
\end{cases} \label{eq:X_up}
\end{align}
\end{subequations}
When the behaviour of the solution modes in terms of the time $t$ and the angle $\phi$
is included, we can find the asymptotic behaviour of the solution modes in terms
of the advanced ($v\equiv t+r_*$) and retarded ($u\equiv t-r_*$) time co-ordinates:
\begin{subequations} \label{eq:X_in/up as func. of u,v}
\begin{align}  
X^{\text{in}}_{lm\omega}e^{+im\phi-i\omega t} & \sim
\begin{cases} 
e^{-i\omega v+im\phi} & \text{at}\ \mathcal{I}^- \\
A^{\text{in}}_{lm\omega}e^{-i\omega u+im\phi} & \text{at}\ \mathcal{I}^+ \\
0 & \text{at}\ \mathcal{H}^- \\
B^{\text{in}}_{lm\omega}e^{-i\tilde{\omega}v+im\phi_+} & \text{at}\ \mathcal{H}^+
\end{cases} \label{eq:X_in as func. of u,v}
\\
X^{\text{up}}_{lm\omega}e^{+im\phi-i\omega t} & \sim
\begin{cases} 
0 & \text{at}\ \mathcal{I}^- \\
B^{\text{up}}_{lm\omega}e^{-i\omega u+im\phi} & \text{at}\ \mathcal{I}^+ \\
e^{-i\tilde{\omega}u+im\phi_+} & \text{at}\ \mathcal{H}^- \\
A^{\text{up}}_{lm\omega}e^{-i\tilde{\omega}v+im\phi_+} & \text{at}\ \mathcal{H}^+
\end{cases} \label{eq:X_up as func. of u,v}
\end{align}
\end{subequations}
Equation (\ref{eq:X_in as func. of u,v}) represents a wave emerging from $\mathcal{I}^-$, being partially
scattered back to $\mathcal{I}^+$ and partially transmitted through to $\mathcal{H}^+$. 
Similarly, (\ref{eq:X_up as func. of u,v}) represents a wave emerging from $\mathcal{H}^-$, being partially
scattered back to $\mathcal{H}^+$ and partially transmitted through to $\mathcal{I}^+$. 

Both sets of modes are eigenfunctions of the hamiltonians $\hat{H}_{\vec{k}_{\manifold{I}}}\equiv i\vec{k}_{\manifold{I}}$ and $\hat{H}_{\vec{k}_{\manifold{H}}}\equiv i\vec{k}_{\manifold{H}}$ with 
eigenvalues $\omega$ and $\tilde{\omega}$ respectively:
\begin{equation} \label{eq:hamiltonians on in/up modes}
\begin{aligned}
\hat{H}_{\vec{k}_{\manifold{I}}}X^{\bullet}_{lm\omega}e^{+im\phi-i\omega t} &=\omega X^{\bullet}_{lm\omega}e^{+im\phi-i\omega t}\\
\hat{H}_{\vec{k}_{\manifold{H}}}X^{\bullet}_{lm\omega}e^{+im\phi-i\omega t} &=\tilde{\omega}X^{\bullet}_{lm\omega}e^{+im\phi-i\omega t}
\end{aligned}
\end{equation}
where the symbol $\bullet$ indicates either `in' or `up'.
We will restrict the definition of the `in' and `up' modes to those modes with positive $\omega$ and 
positive $\tilde{\omega}$ respectively. 
It then follows that the `in' and `up' modes are positive frequency with respect to 
the Killing vectors $\vec{k}_{\manifold{I}}$ and $\vec{k}_{\manifold{H}}$ respectively.

The asymptotic behaviour of solutions ${}_{\indhel}R^{\text{in/up}}_{lm\omega}$ 
$\forall h=0, \pm 1/2, \pm 1, \pm3/2, \pm2$ of the radial Teukolsky equation is:
\begin{subequations} \label{eq:R_in/up}
\begin{align}
{}_{\indhel}R^{\text{in}}_{lm\omega} & \sim
\begin{cases} \label{eq:R_in}
{}_{\indhel}R^{\text{in,tra}}_{lm\omega}\Delta^{-\indhel }e^{-i\tilde{\omega}r_*} & (r\rightarrow r_+) \\
{}_{\indhel}R^{\text{in,inc}}_{lm\omega}r^{-1}e^{-i\omega r_*}+{}_{\indhel}R^{\text{in,ref}}_{lm\omega}r^{-1-2\indhel}e^{+i\omega r_*} & 
(r\rightarrow +\infty)
\end{cases}
\\
{}_{\indhel}R^{\text{up}}_{lm\omega} & \sim
\begin{cases}\label{eq:R_up}
{}_{\indhel}R^{\text{up,inc}}_{lm\omega}e^{+i\tilde{\omega}r_*}+{}_{\indhel}R^{\text{up,ref}}_{lm\omega}\Delta^{-\indhel }e^{-i\tilde{\omega}r_*} & 
\qquad (r\rightarrow r_+) \\
{}_{\indhel}R^{\text{up,tra}}_{lm\omega}r^{-1-2\indhel}e^{+i\omega r_*} & 
\qquad (r\rightarrow +\infty)
\end{cases}
\end{align}
\end{subequations}

From (\ref{eq:X_in/up}) and expressions in ~\cite{ar:Detw'76}
we can find the asymptotic coefficients of ${}_{-1}R^{\bullet}_{lm\omega}$ from those of $X^{\bullet}_{lm\omega}$:
\begin{equation} \label{eq:R_1 coeffs from X's}
\begin{aligned}
\frac{{}_{-1}R^{\text{in,ref}}_{lm\omega}}{{}_{-1}R^{\text{in,inc}}_{lm\omega}A^{\text{in}}_{lm\omega}}&=\frac{4\omega^2}{{}_1B_{lm\omega}} &\qquad\quad
\frac{{}_{-1}R^{\text{in,tra}}_{lm\omega}}{{}_{-1}R^{\text{in,inc}}_{lm\omega}B^{\text{in}}_{lm\omega}}&=
\frac{-sgn(\tilde{\omega})|\omega|i}{(r_+^2+a^2)^{1/2}\EuFrak{N}^*} \\
\frac{{}_{-1}R^{\text{up,ref}}_{lm\omega}}{{}_{-1}R^{\text{up,inc}}_{lm\omega}A^{\text{up}}_{lm\omega}}&=\frac{-i{}_1B_{lm\omega}}{4K_+\EuFrak{N}^*} &\quad
\frac{{}_{-1}R^{\text{up,tra}}_{lm\omega}}{{}_{-1}R^{\text{up,inc}}_{lm\omega}B^{\text{up}}_{lm\omega}}&=\frac{|\omega|(r_+^2+a^2)^{1/2}}{|K_+|} \\
{}_{-1}R^{\text{in,inc}}_{lm\omega}&=\frac{1}{2^{3/2}|\omega|} &\quad {}_{-1}R^{\text{up,inc}}_{lm\omega}&=\frac{-2^{1/2}(r_+^2+a^2)^{1/2}\tilde{\omega}}{{}_1B_{lm\omega}}
\end{aligned}
\end{equation}

In the calculation of ${}_{-1}R^{\text{in,tra}}_{lm\omega}/{}_{-1}R^{\text{in,inc}}_{lm\omega}$ in (\ref{eq:R_1 coeffs from X's}) 
we needed an extra term in the asymptotic
expansion of the ingoing part (the outgoing part is simply obtained by complex conjugation since Detweiler's potential is real)
of $X_{lm\omega}$ close to the horizon. By introducing the asymptotic expansion
\begin{equation} \label{eq:asympt. for r=r_+ of X_in}
\frac{X^{\text{in}}_{lm\omega}}{B^{\text{in}}_{lm\omega}}=\left[1+\alpha_1(r-r_+)+O((r-r_+)^2)\right]e^{-i\tilde{\omega}r_*}
\end{equation}
in Detweiler's differential equation and performing a Taylor series expansion around $r_+$ of Detweiler's 
potential, we find from the second order term that
\begin{equation} \label{eq:val. of alpha_1}
\begin{aligned}
&\alpha_1=\\
&=\frac{-1}{2\EuFrak{N}^*}\left[{}_{-1}\lambda_{lm\omega}-\frac{4Ma^2r_+-Q^2(a^2-r_+^2)}{(r_+^2+a^2)^2}+\frac{a^2+Q^2}{r_+^2+\nu^2}-
\frac{4amr_+\tilde{\omega}}{r_+-r_-}-\frac{2(r_+-M)^2\eta}{(r_+^2+\nu^2)^2}\right]
\end{aligned}
\end{equation}
where $\nu^2 \equiv a^2-am/\omega$ and $\eta\equiv ({}_{1}B_{lm\omega}-{}_{-1}\lambda_{lm\omega})/(2\omega^2)$.
In the calculation of ${}_{-1}R^{\text{in,inc}}_{lm\omega}$, an extra term is also needed in the asymptotic
expansion of the ingoing part of $X_{lm\omega}$ for large $r$:
\begin{equation}
X_{lm\omega}=\left[1+\frac{\beta_1}{r}+O(r^{-2})\right]e^{-i\omega r_*}
\end{equation}
with
\begin{equation}
\beta_1=-\frac{({}_{-1}\lambda_{lm\omega}+2a\omega m)i}{2\omega}
\end{equation}

After obtaining the asymptotic coefficients of ${}_{-1}R^{\bullet}_{lm\omega}$ from those of $X^{\bullet}_{lm\omega}$,
we just need to derive those of ${}_{+1}R^{\bullet}_{lm\omega}$ to complete the asymptotic picture of the solutions to the
radial Teukolsky equation for spin-1. 
This is achieved by using the asymptotic behaviour in (\ref{eq:R_in/up}) together with 
a transformation in ~\cite{ar:Detw'76} that relates ${}_{+1}R$ to ${}_{-1}R$; the result is:
\begin{equation} \label{eq:R1 coeffs from R_1's}
\begin{aligned}
\frac{{}_{+1}R^{\text{in,inc}}_{lm\omega}}{{}_{-1}R^{\text{in,inc}}}&=-2^3\omega^2; & \qquad
\frac{{}_{+1}R^{\text{in,ref}}_{lm\omega}}{{}_{-1}R^{\text{in,ref}}_{lm\omega}}&=
\frac{{}_{+1}R^{\text{up,tra}}_{lm\omega}}{{}_{-1}R^{\text{up,tra}}_{lm\omega}}=-\frac{{}_1B_{lm\omega}^2}{2\omega^2}     \\
\frac{{}_{+1}R^{\text{in,tra}}_{lm\omega}}{{}_{-1}R^{\text{in,tra}}_{lm\omega}}&=
\frac{{}_{+1}R^{\text{up,ref}}_{lm\omega}}{{}_{-1}R^{\text{up,ref}}_{lm\omega}}=-2^3K_+ \EuFrak{N}^*i; &
\frac{{}_{+1}R^{\text{up,inc}}_{lm\omega}}{{}_{-1}R^{\text{up,inc}}_{lm\omega}}&=\frac{-i{}_1B_{lm\omega}^2}{2K_+\EuFrak{N}} 
\end{aligned}
\end{equation}

It is clear from their asymptotic behaviour that neither ${}_{\indhel}R^{\text{in}}_{lm\omega}$ nor 
${}_{\indhel}R^{\text{up}}_{lm\omega}$ satisfy the symmetry (\ref{eq:R symm.->cc,-s}). 
As a matter of fact, we obtained that, under this symmetry, the functions ${}_{\indhel}R^{\bullet}_{lm\omega}$
transform to the radial funcions that are derived from the solution $X_{lm\omega}^{\bullet *}$ and its derivative.
Indeed, we construct first a new radial function ${}_{-1}\bar{R}_{lm\omega}^{\bullet}$ 
derived from $X_{lm\omega}^{\bullet *}$ in the same manner that ${}_{-1}R_{lm\omega}^{\bullet}$ is derived from
$X_{lm\omega}^{\bullet}$, using an expression in ~\cite{ar:Detw'76}.
It can then be checked using equations (\ref{eq:R_1 coeffs from X's}) and (\ref{eq:R1 coeffs from R_1's}) that the relations
\begin{equation}
\begin{aligned}
{}_{+1}R^{\bullet}_{lm\omega}&=2{}_1B_{lm\omega}\Delta^{-1}{}_{-1}\bar{R}^{\bullet *}_{lm\omega} \\
{}_{+1}\bar{R}^{\bullet}_{lm\omega}&=2{}_1B_{lm\omega}\Delta^{-1}{}_{-1}R^{\bullet *}_{lm\omega} 
\end{aligned}
\end{equation}
are satisfied, where ${}_{+1}\bar{R}^{\bullet}_{lm\omega}$ 
is calculated by applying to ${}_{-1}\bar{R}^{\bullet}_{lm\omega}$ the transformation 
in ~\cite{ar:Detw'76} that relates ${}_{+1}R$ to ${}_{-1}R$ and its derivative.
Renaming ${}_{\pm 1}\bar{R}^{\text{in}}_{lm\omega}$ and ${}_{\pm 1}\bar{R}^{\text{up}}_{lm\omega}$
by ${}_{\pm 1}R^{\text{out}}_{lm\omega}$ and ${}_{\pm 1}R^{\text{down}}_{lm\omega}$
respectively, we have the following two sets of modes:
\begin{subequations} \label{eq:R_out/down}
\begin{align}
{}_{\pm 1}R^{\text{out}}_{lm\omega}&\equiv (2{}_1B_{lm\omega})^{\pm 1}\Delta^{\mp 1}{}_{\mp 1}R^{\text{in} *}_{lm\omega} 
 \sim \\ & \sim
\begin{cases} \label{eq:R_out}
{}_{\pm 1}R^{\text{out,tra}}_{lm\omega}e^{+i\tilde{\omega}r_*} & (r\rightarrow r_+) \\
{}_{\pm 1}R^{\text{out,inc}}_{lm\omega}r^{-1\mp 2}e^{+i\omega r_*}+
{}_{\pm 1}R^{\text{out,ref}}_{lm\omega}r^{-1}e^{-i\omega r_*} & 
(r\rightarrow +\infty)
\end{cases}
\\
{}_{\pm 1}R^{\text{down}}_{lm\omega}&\equiv (2{}_1B_{lm\omega})^{\pm 1}\Delta^{\mp 1}{}_{\mp 1}R^{\text{up} *}_{lm\omega} 
 \sim \\ & \sim
\begin{cases}\label{eq:R_down}
{}_{\pm 1}R^{\text{down,inc}}_{lm\omega}\Delta^{\mp 1}e^{-i\tilde{\omega}r_*}+
{}_{\pm 1}R^{\text{down,ref}}_{lm\omega}e^{+i\tilde{\omega}r_*} & 
(r\rightarrow r_+) \\
{}_{\pm 1}R^{\text{down,tra}}_{lm\omega}r^{-1}e^{-i\omega r_*} & (r\rightarrow +\infty)
\end{cases}
\end{align}
\end{subequations}
where
\begin{equation} \label{eq:def. coeffs. Rs_out/down}
\begin{aligned}
{}_{\pm 1}R^{\text{out,inc/ref/tra}}_{lm\omega}&\equiv (2{}_1B_{lm\omega})^{\pm 1}{}_{\mp 1}R^{\text{in,inc/ref/tra} *}_{lm\omega}, &
{}_{\pm 1}R^{\text{down,inc/ref/tra}}_{lm\omega}&\equiv (2{}_1B_{lm\omega})^{\pm 1}{}_{\mp 1}R^{\text{up,inc/ref/tra} *}_{lm\omega}
\end{aligned}
\end{equation}
Note that the factor $(2{}_1B_{lm\omega})^{\pm 1}$ is needed so that the Teukolsky-Starobinski\u{\i} identities
are satisfied.
Similarly, since the radial modes ${}_{\pm 1}R^{\text{out}}_{lm\omega}$ and ${}_{\pm 1}R^{\text{down}}_{lm\omega}$
are obtained from $X^{\text{in} *}_{lm\omega}$ and $X^{\text{up} *}_{lm\omega}$ respectively,
we rename the latter as
\begin{subequations} \label{eq:X_out/down}
\begin{align}  
X^{\text{out}}_{lm\omega}\equiv X^{\text{in} *}_{lm\omega}   & \sim
\begin{cases} 
B^{\text{out}}_{lm\omega}e^{+i\tilde{\omega}r_*} & (r\rightarrow r_+) \\
e^{+i\omega r_*}+A^{\text{out}}_{lm\omega}e^{-i\omega r_*} & (r\rightarrow +\infty)
\end{cases} \label{eq:X_out}
\\
X^{\text{down}}_{lm\omega}\equiv X^{\text{up} *}_{lm\omega} & \sim
\begin{cases}
e^{-i\tilde{\omega}r_*}+A^{\text{down}}_{lm\omega}e^{+i\tilde{\omega}r_*} & 
(r\rightarrow r_+) \\
B^{\text{down}}_{lm\omega}e^{-i\omega r_*} & (r\rightarrow +\infty)
\end{cases} \label{eq:X_down}
\end{align}
\end{subequations}
with
\begin{equation} \label{eq:def. coeffs. X_out/down}
\begin{aligned}
A^{\text{out}}_{lm\omega}&\equiv A^{\text{in} *}_{lm\omega}, & \qquad \qquad 
A^{\text{down}}_{lm\omega}&\equiv A^{\text{up} *}_{lm\omega} \\
B^{\text{out}}_{lm\omega}&\equiv B^{\text{in} *}_{lm\omega}, &
B^{\text{down}}_{lm\omega}&\equiv B^{\text{up} *}_{lm\omega} &
\end{aligned}
\end{equation}
We follow the same positive-frequency convention for the `out' and `down' modes
as that for the `in' and `up' modes respectively, namely, their definition is restricted
to modes with positive $\omega$ and positive $\tilde{\omega}$ respectively.
The asymptotic behaviour in terms of the advanced and retarded time co-ordinates of these
two new sets of functions is
\begin{subequations} \label{eq:X_out/down as func. of u,v}
\begin{align}  
X^{\text{out}}_{lm\omega}e^{+im\phi-i\omega t} & \sim
\begin{cases} 
A^{\text{out}}_{lm\omega}e^{-i\omega v+im\phi} & \text{at}\ \mathcal{I}^- \\
e^{-i\omega u+im\phi} & \text{at}\ \mathcal{I}^+ \\
B^{\text{out}}_{lm\omega}e^{-i\tilde{\omega}u+im\phi_+} & \text{at}\ \mathcal{H}^- \\
0 & \text{at}\ \mathcal{H}^+
\end{cases} \label{eq:X_out as func. of u,v}
\\
X^{\text{down}}_{lm\omega}e^{+im\phi-i\omega t} & \sim
\begin{cases} 
B^{\text{down}}_{lm\omega}e^{-i\omega v+im\phi} & \text{at}\ \mathcal{I}^- \\
0 & \text{at}\ \mathcal{I}^+ \\
A^{\text{down}}_{lm\omega}e^{-i\tilde{\omega}u+im\phi_+} & \text{at}\ \mathcal{H}^- \\
e^{-i\tilde{\omega}v+im\phi_+} & \text{at}\ \mathcal{H}^+
\end{cases} \label{eq:X_down as func. of u,v}
\end{align}
\end{subequations}
Modes (\ref{eq:X_out as func. of u,v}) describe a wave going out to $\mathcal{I}^+$
whereas modes (\ref{eq:X_down as func. of u,v}) describe a wave going down $\mathcal{H}^+$. 

It is easy to check that the `out'[`down'] NP scalars are related to the `in'[`up'] NP scalars 
under the symmetry transformation $(t,\phi)\to (-t,-\phi)$ in the manner:
\begin{equation} \label{NP scalars in/up->out/down}
\left.
\begin{array}{ll}
{}_{lm\omega}\phi_{-1}^{\text{in/up}}&\to (-1)^{m+1}\Delta^{-1}\rho^{-2}{}_{l-m-\omega}\phi_{+1}^{\text{out/down}} \\
{}_{lm\omega}\phi_{0}^{\text{in/up}}&\to -(-1)^{m+1}{}_{l-m-\omega}\phi_{0}^{\text{out/down}}  \\
{}_{lm\omega}\phi_{+1}^{\text{in/up}}&\to (-1)^{m+1}\Delta\rho^2{}_{l-m-\omega}\phi_{-1}^{\text{out/down}}  
\end{array}
\right\} \text{under} \quad (t,\phi)\to (-t,-\phi)
\end{equation}
where we have used the radial symmetry (\ref{eq:R symm.->cc,-m,-w}) and the angular symmetry (\ref{eq:S symm.->-s,-m,-w}).

Finally, it can be checked that the radial Teukolsky equation admits the following wronskian:
\begin{equation} \label{eq:gral. radial wronsk.,R_+/-1}
W[{}_{+1}R_{lm\omega}^{\bullet},{}_{-1}R_{lm\omega}^{\bullet *}]=
{}_{-1}R^{\bullet *}_{lm\omega}\mathcal{D}^{\dagger}_0\left(\Delta{}_{+1}R_{lm\omega}^{\bullet}\right)-
\Delta{}_{+1}R_{lm\omega}^{\bullet}\mathcal{D}^{\dagger}_0{}_{-1}R^{\bullet *}_{lm\omega}=C^{\bullet}i
\end{equation}
where $C^{\bullet}$ are real constants.
It is useful to note the following two relations satisfied by the wronskians of the `in'
and `up' solutions once the normalization constants (\ref{eq:normalization consts.}) have been included:
\begin{subequations} 
\begin{align}
W[{}_{+1}R_{lm\omega}^{\text{up}},{}_{-1}R_{lm\omega}^{\text{up} *}]&=-W[{}_{+1}R_{lm\omega}^{\text{in}},{}_{-1}R_{lm\omega}^{\text{in} *}]  \label{eq: wronskian in=-up}  \\
W[{}_{+1}R_{lm\omega}^{\bullet},{}_{-1}R_{lm\omega}^{\bullet *}]&=
+W[{}_{+1}R_{l-m-\omega}^{\bullet},{}_{-1}R_{l-m-\omega}^{\bullet *}]
\end{align}
\end{subequations}
As a short-hand notation we will denote $W[{}_{+1}R_{lm\omega}^{\bullet},{}_{-1}R_{lm\omega}^{\bullet *}]$ by
$W[{}_{+1}R,{}_{-1}R^{*}]^{\bullet}_{lm\omega}$.
We give here two important wronskian relations, obtainable in terms of the coefficients of the solution to Detweiler equation by using
relations (\ref{eq:R1 coeffs from R_1's}) and (\ref{eq:R_1 coeffs from X's}):
\begin{subequations}
\begin{align}
\omega \left(1-\abs{A_{lm\omega}^{\text{in}}}^2\right)&=
\tilde{\omega}\abs{B_{lm\omega}^{\text{in}}}^2  \label{eq:wronsk. X in} \\
\tilde{\omega}\left(1-\abs{A_{lm\omega}^\text{up}}^2\right)&=
\omega \abs{B_{lm\omega}^{\text{up}}}^2 \label{eq:wronsk. X up} 
\end{align}
\end{subequations}
Wronskian relation (\ref{eq:wronsk. X in}) shows that for the modes such that $\tilde{\omega}\omega<0$, 
it must be $\abs{A_{lm\omega}^{\text{in}}}^2>1$ , and thus the wave mode is reflected back
with a gain of energy. 
This phenomenon is known as superradiance.
Since the `in' modes are only defined for positive $\omega$, superradiance occurs for these
modes for negative $\tilde{\omega}$ only. 
Similarly, from wronskian relation (\ref{eq:wronsk. X up}), when $\tilde{\omega}\omega<0$ it must be
$|A^{\text{up}}_{lm\omega}|^2>1$. Therefore, the `up' modes, which are defined
for positive $\tilde{\omega}$, that experience superradiance are those for which $\omega<0$.
The condition $\tilde{\omega}\omega<0$ for superradiance, which is the same for scalar and gravitational perturbations, 
clearly shows that this phenomenon is only possible if $a\neq 0$ and therefore it only occurs if the black hole
possesses an ergosphere.

\clearpage


\section{Asymptotics close to the horizon} \label{sec:asympts. close to r_+}

This Appendix is valid in the Kerr-Newman background with the same re-definitions of quantities as in Appendix \ref{sec:radial func.}.

In this Appendix we are interested in finding the asymptotic behaviour close to the horizon
of the solution to the radial Teukolsky equation. 
The main application of this study is in the calculation in Section \ref{sec:RRO}
of the behaviour close to the horizon of the RSET when the field is in the past Boulware state.
We will therefore only consider the `up' modes and, because of the presence of a factor that decreases exponentially with $\tilde{\omega}$,
we will not consider the case of large $\tilde{\omega}$. 

This study is based on one performed by Candelas  ~\cite{ar:Candelas'80}.
Even though Candelas started the calculation for general spin, he soon confined it to the scalar case. It is our
intention to complete his asymptotic calculation for general spin and only specialize to the spin-1 case at the end.

Candelas performs a Taylor series expansion around $r=r_+$ of the coefficients of ${}_{\indhel}R_{lm\omega}$ and its
derivatives appearing in the radial Teukolsky equation. By keeping only the first order terms
in the expansion and also terms that involve parameters which might become very large, the radial Teukolsky equation becomes:
\begin{equation} \label{eq:non-approx eqA4Cand'80}
(r-r_+)\ddiff{{}_{\indhel}R_{lm\omega}}{r}+(\indhel+1)\diff{{}_{\indhel}R_{lm\omega}}{r}- \left[\frac{{}_{\indhel}\lambda_{lm\omega}
-4i\omega \indhel r}{r_+-r_-}-
\frac{q(q-2i\indhel)}{4(r-r_+)}\right]
{}_{\indhel}R_{lm\omega}=0
\end{equation}
where $q \equiv 2K_+/(r_+-r_-)$.
The parameters in (\ref{eq:non-approx eqA4Cand'80}) that might become very large independently of the limit
$r \rightarrow r_+$ are ${}_{\indhel}\lambda_{lm\omega}$, $\omega $ and $\tilde{\omega}$. 
We keep $\tilde{\omega}$ bounded, which means that either both $m$ and $\omega $ are bounded or else that
$\displaystyle \omega \rightarrow \infty$ and $m \sim \omega/\Omega_+$. 
We are going to restrict ourselves to the first possibility (i.e., $m$ and $\omega $ bounded) since it is only 
for this case that we are able to find the behaviour of the angular solutions, needed in the calculation in Section \ref{sec:RRO}. 
We thus have that the only
term in (\ref{eq:non-approx eqA4Cand'80}) that might become very large independently of $r \rightarrow
r_+$ is the one with ${}_{\indhel}\lambda_{lm\omega}$. Since we are keeping $m$ and $\omega $ bounded, ${}_{\indhel}\lambda_{lm\omega}$ can
only become large if we let $l \rightarrow +\infty$.   

When letting $l \rightarrow +\infty$ and keeping $\omega $ and $m$ bounded in the Teukolsky angular 
equation, all the terms in the coefficient of the angular function ${}_{\indhel}S_{lm\omega}(\theta)$ can be ignored
except for ${}_{\indhel}\lambda_{lm\omega}$ and those with a $1/\sin \theta$ in
them. This is equivalent to setting $a\omega=0$ in the angular equation. This means that in the limit $l \rightarrow +\infty$
(with $m$ and $\omega $ bounded) the angular solution reduces to the spin-weighted spheroidal harmonics: 
${}_{\indhel}Z_{lm\omega} \rightarrow {}_{\indhel}Y_{lm}$ and that ${}_{\indhel}\lambda_{lm\omega} \rightarrow (l-\indhel )(l+\indhel+1)\rightarrow l^2$.
The latter expression implies that ${}_1B_{lm\omega}\rightarrow l^2$ in the same limit.
After replacing ${}_{\indhel}\lambda_{lm\omega}$ by $l^2$ and neglecting $4i\omega \indhel r$ in (\ref{eq:non-approx eqA4Cand'80}),
Candelas goes on to prove that the asymptotic behaviour close to the horizon of the `up' radial functions with $l \rightarrow +\infty$
is given by
\begin{equation} \label{eq:eqA6Cand'80}
{}_{\indhel}R_{lm\omega}^{\text{up}} \rightarrow
{}_{\indhel}a_lx^{-\indhel /2}K_{\indhel+iq}(2lx^{1/2})+{}_{\indhel}b_lx^{-\indhel /2}I_{-(\indhel+iq)}(2lx^{1/2})
\qquad (l\rightarrow +\infty,r \rightarrow r_+)
\end{equation}
which is uniformly valid in $l$. The factors ${}_{\indhel}a_l$ and ${}_{\indhel}b_l$ are the coefficients of the two independent solutions.
It is at this point that Candelas' analysis specializes to the scalar case. We pursue it here for general spin.

The asymptotic behaviour of the modified Bessel functions $I_{\nu}(z)$ and $K_{\nu}(z)$ are well known (~\cite{bk:AS}).
If we fix $r$ close to the horizon in (\ref{eq:eqA6Cand'80}) and let $l \rightarrow +\infty$, the value of the radial
potential at that fixed value of $r$ will go to infinity and thus it must be
${}_{\indhel}R^{\text{up}}_{lm\omega} \rightarrow 0$.
From the asymptotic behaviour for large argument of the modified Bessel functions, it is only possible that ${}_{\indhel}R^{\text{up}}_{lm\omega} \rightarrow 0$ 
in the limit $l \rightarrow +\infty$ with $r$ fixed in (\ref{eq:eqA6Cand'80}) 
if the coefficient ${}_{\indhel}b_l$ decreases exponentially with $l$. 
It follows from this result together with the asymptotic behaviour for small argument of the modified Bessel functions
that in the limits $r \rightarrow r_+$ and $l \rightarrow +\infty$, while keeping $lx^{1/2}$ finite,
the second term in (\ref{eq:eqA6Cand'80}) can be neglected with
respect to the first one.   
In this last statement we have made the implicit assumption that if the coefficient ${}_{\indhel}a_l$ 
decreases for large $l$, then it does slower than the coefficient ${}_{\indhel}b_l$. 
This assumption is proved to be correct in what follows. We have from the above discussion that
\begin{equation} \label{eq:approx eqA6Cand'80}
{}_{\indhel}R^{\text{up}}_{lm\omega} \rightarrow {}_{\indhel}a_lx^{-\indhel /2}K_{\indhel+iq}\left(2lx^{1/2}\right)
\qquad (l\rightarrow +\infty,r \rightarrow r_+,lx^{1/2}\ \text{finite})
\end{equation}

We can determine the coefficient ${}_{\indhel}a_l$ by comparison with the WKB
approximation (\ref{eq:R_up}). By taking the limit $lx^{1/2}\rightarrow 0$ on the 
solution (\ref{eq:approx eqA6Cand'80}) we obtain
\begin{equation} \label{eq:approx eqA6Cand'80,r->rplus}
\begin{aligned}
{}_{\indhel}R^{\text{up}}_{lm\omega} \rightarrow 
&
\frac{{}_{\indhel}a_l\pi
(r_+-r_-)^{2\indhel}l^{-\indhel-iq}I_{\tilde{\omega}}^*}{2\sin [(\indhel+iq)\pi]
\Gamma(1-\indhel-iq)}\Delta^{-\indhel }e^{-i\tilde{\omega}r_*}-
\\&-
\frac{{}_{\indhel}a_l\pi
l^{\indhel+iq}I_{\tilde{\omega}}}{2\sin [(\indhel+iq)\pi]
\Gamma(1+\indhel+iq)}e^{+i\tilde{\omega}r_*}  
\qquad 
(l\rightarrow \infty,r \rightarrow r_+,lx^{1/2}\rightarrow 0)
\end{aligned}
\end{equation}
where
\begin{equation}
I_{\tilde{\omega}}\equiv e^{-\tilde{\omega}r_+}\left(4M\kappa_+\right)^{-\frac{i\tilde{\omega}}{2\kappa_+}}\left(-4M\kappa_-\right)^{-\frac{-i\tilde{\omega}}{2\kappa_-}}
\end{equation}
Comparing this asymptotic expression with the WKB approximation (\ref{eq:R_up}) it follows that
\begin{equation} \label{eq:al and Rup,ref}
\begin{aligned}
{}_{\indhel}a_l&=- \frac{2\sin [(\indhel+iq)\pi]\Gamma(1+\indhel+iq)I_{\tilde{\omega}}^*}{\pi}l^{-\indhel-iq}{}_{\indhel}R^{\text{up,inc}}_{lm\omega}   \\
{}_{\indhel}R^{\text{up,ref}}_{lm\omega}&=
-\frac{\Gamma(1+\indhel+iq)}{\Gamma(1-\indhel-iq)}(r_+-r_-)^{2\indhel}I_{\tilde{\omega}}^{* 2}l^{-2(\indhel+iq)}{}_{\indhel}R^{\text{up,inc}}_{lm\omega}
\end{aligned}
\end{equation}

We now specialize to the spin-1 case. Combining equations
(\ref{eq:approx eqA6Cand'80}) and (\ref{eq:al and Rup,ref}), and using
the same normalization as the one used in the numerical results (i.e., setting         
${}_{-1}R^{\text{up,inc}}_{lm\omega}=1$ and  ${}_{+1}R^{\text{up,inc}}_{lm\omega}=-iB^2/\left(2\EuFrak{N}K_+\right)$), we have
\begin{equation} \label{eq:R1 `up' approx l->inf,r->rplus}
\begin{aligned}
{}_{+1}R^{\text{up}}_{lm\omega} &\rightarrow \frac{-2I_{\tilde{\omega}}^*l^{3-iq}}{(r_+-r_-)K_+\Gamma(-iq)}
x^{-1/2}K_{1+iq}(2lx^{1/2}) 
\\ &\qquad \qquad \qquad \qquad\qquad (l\rightarrow +\infty,r \rightarrow r_+,lx^{1/2}\ \text{finite}) \\
{}_{-1}R^{\text{up}}_{lm\omega} &\rightarrow \frac{i(r_+-r_-)I_{\tilde{\omega}}^*l^{+1-iq}}{K_+\Gamma(-iq)}
x^{1/2}K_{-1+iq}(2lx^{1/2}) 
\\ &\qquad \qquad \qquad \qquad\qquad (l\rightarrow +\infty,r \rightarrow r_+,lx^{1/2}\ \text{finite})
\end{aligned}
\end{equation}

It is also useful to give the expressions that the `up' radial functions (\ref{eq:R1 `up' approx l->inf,r->rplus})
adopt in this limit whenever the constants of normalization (\ref{eq:normalization consts.}) 
are included. These expressions are, in compact form:
\begin{equation} \label{eq:R1 `up' approx l->inf,r->rplus;compact version}
|N^{\text{up}}_{-1}|{}_{\indhel}R^{\text{up}}_{lm\omega} \rightarrow 
A_{\indhel}Nx^{-1/2}K_{\indhel+iq}(2lx^{1/2}) \qquad (l\rightarrow +\infty,r \rightarrow r_+,lx^{1/2}\ \text{finite})
\end{equation}
for $\indhel =\pm 1$, where
\begin{equation} \label{eq:def.A_s}
A_{\indhel} \equiv
\left \{
\begin{array}{ll}
\displaystyle -4 &, \indhel=+1 \\
\displaystyle \frac{2i}{l^4} &, \indhel=-1
\end{array}
\right \}
\left[l(r_+-r_-)\right]^{-\indhel }
\end{equation}
and
\begin{equation} \label{eq:def.D}
N \equiv \frac{I_{\tilde{\omega}}^*l^{-iq}}{\sqrt{2^3\pi K_+}\Gamma(-iq)}
\end{equation}
We therefore have finally found the asymptotic behaviour of the `up' radial modes with $l\rightarrow +\infty$ 
close to the horizon with $\tilde{\omega}$, and both $m$ and $\omega$, bounded. We believe that this is the method behind 
the approximation given by CCH in their
TableII. Their result, however, does not exactly coincide with either (\ref{eq:R1 `up' approx l->inf,r->rplus}) or
(\ref{eq:R1 `up' approx l->inf,r->rplus;compact version})
(as a matter of fact, in their table there is a quantity $\rho$ that they have not defined and it cannot be the spin coefficient
as it cannot have a $\theta$-dependency). 

Note that there is no reason why the `up' radial modes (\ref{eq:R1 `up' approx l->inf,r->rplus}) or 
(\ref{eq:R1 `up' approx l->inf,r->rplus;compact version}) should diverge in the stated limits. 
In fact, they clearly do not in the case of helicity $-1$. 
It is only when other factors (as in (\ref{eq:phi_0/2(in/up)})) are included that the resulting expressions we deal with
in Section \ref{sec:RRO} diverge in this limit.

We are also interested in finding the result of applying the operator $\mathcal{D}_0^{\dagger}\Delta$ on the asymptotic
solution (\ref{eq:R1 `up' approx l->inf,r->rplus;compact version}).
It immediately follows from (\ref{eq:R1 `up' approx l->inf,r->rplus;compact version}) and (\ref{eq:def.A_s}) that
\begin{equation} \label{eq:Ddagger Delta R1 `up' approx l->inf,r->rplus}
\begin{aligned}
&|N^{\text{up}}_{-1}|\mathcal{D}_0^{\dagger}\left(\Delta{}_{\indhel}R^{\text{up}}_{lm\omega}\right) \rightarrow \\
&\rightarrow A_{\indhel}N (r_+-r_-)x^{-\indhel /2} \left[\left(-\frac{\indhel}{2}+1+i\frac{q}{2}\right)K_{\indhel+iq}(2lx^{1/2})+
lx^{1/2}K'_{\indhel+iq}(2lx^{1/2})
\right]  \\
& \qquad \qquad \qquad  \qquad \qquad\qquad \qquad\qquad(l\rightarrow +\infty,r \rightarrow r_+,lx^{1/2}\ \text{finite})
\end{aligned}
\end{equation}
When using a recurrence relation (~\cite{bk:AS}) for the modified Bessel function, 
expression (\ref{eq:Ddagger Delta R1 `up' approx l->inf,r->rplus}) for $\indhel =+1$ reduces to
\begin{equation} \label{eq:Ddagger Delta R+1 `up' approx l->inf,r->rplus}
|N^{\text{up}}_{-1}|\mathcal{D}^{\dagger}_0\left(\Delta{}_{+1}R^{\text{up}}_{lm\omega}\right) \rightarrow 
-A_{\indhel}N \frac{(r_+-r_-)}{2x^{1/2}} K_{iq} \qquad (l\rightarrow +\infty,r \rightarrow r_+,lx^{1/2}\ \text{finite})
\end{equation}
In graphs \ref{fig:plot candelas'80 radial approx,w=0.01}--\ref{fig:relat.err. plot candelas'80 radial approx,w=0.01}
we compare the numerical solution with the analytic
asymptotic approximation (\ref{eq:R1 `up' approx l->inf,r->rplus}) we have found.
These graphs show that this approximation indeed tends to the non-approximated (numerical) solution and that as 
$r \rightarrow r_+$ the approximation is better for the higher values of $l$, as predicted.

\begin{figure}[H]
\centering
\includegraphics*[width=70mm]{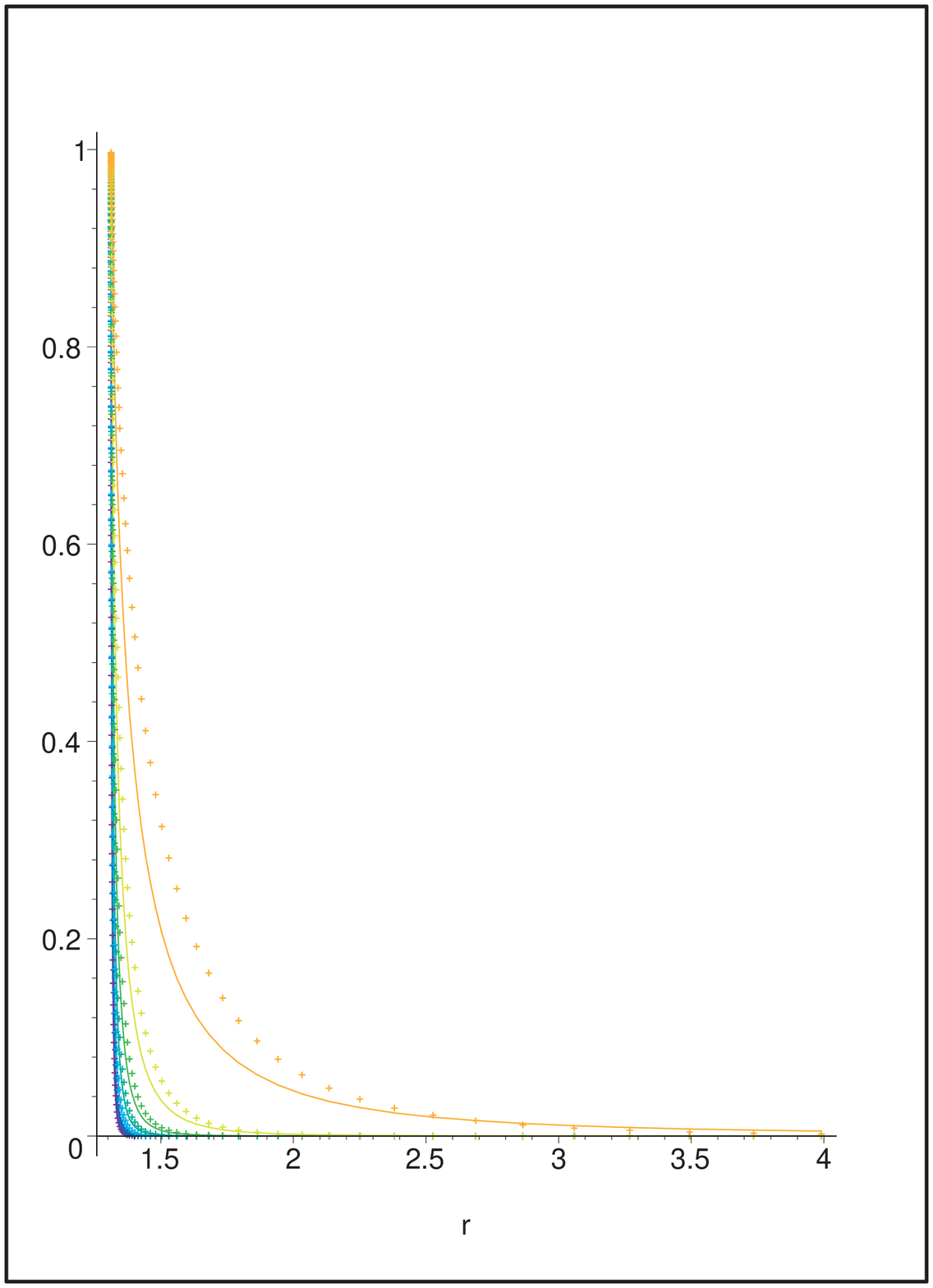}
\caption{$\abs{{}_{-1}R^{\text{up}}_{lm\omega}}^2$, $l=4\dots 7$, $m=0$, $\omega=0.01$.
Correspondence between colours and modes is the same as in Figure \ref{fig:relat.err. plot candelas'80 radial approx,w=0.01}
(dots are the numerical solution and straight lines are the approximation from (\ref{eq:R1 `up' approx l->inf,r->rplus})).}
\label{fig:plot candelas'80 radial approx,w=0.01}
\includegraphics*[width=70mm]{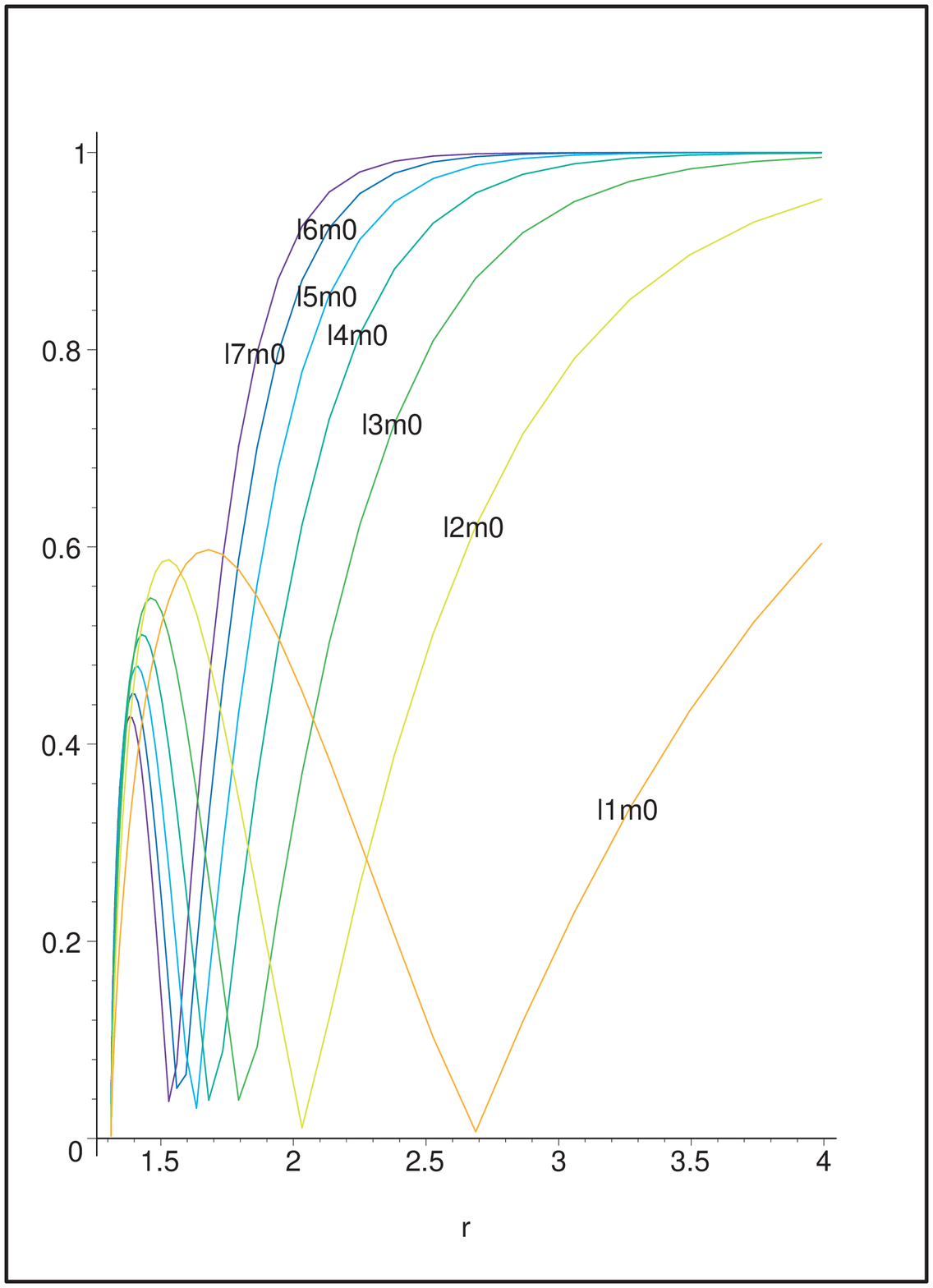} 
\caption{Relative error $\frac{\abs{\abs{{}_{-1}R^{\text{up,num}}_{lm\omega}}^2-\abs{{}_{-1}R^{\text{up,approx}}_{lm\omega}}^2}}
{\abs{{}_{-1}R^{\text{up,num}}_{lm\omega}}^2}$, $l=1\dots 7$, $m=0$, $\omega=0.01$.}
\label{fig:relat.err. plot candelas'80 radial approx,w=0.01}
\end{figure}



\section*{Acknowledgements}

We wish to thank Prof.~V.P.Frolov for helpful comments and for directing us to ~\cite{ar:Bolash&Frolov}.
M.C. wishes to thank Dr.~G.Duffy for many useful discussions.
This research was partly funded by Enterprise Ireland.


\bibliography{./rset.bbl}

\begin{thebibliography}{42}
\expandafter\ifx\csname natexlab\endcsname\relax\def\natexlab#1{#1}\fi
\expandafter\ifx\csname bibnamefont\endcsname\relax
  \def\bibnamefont#1{#1}\fi
\expandafter\ifx\csname bibfnamefont\endcsname\relax
  \def\bibfnamefont#1{#1}\fi
\expandafter\ifx\csname citenamefont\endcsname\relax
  \def\citenamefont#1{#1}\fi
\expandafter\ifx\csname url\endcsname\relax
  \def\url#1{\texttt{#1}}\fi
\expandafter\ifx\csname urlprefix\endcsname\relax\def\urlprefix{URL }\fi
\providecommand{\bibinfo}[2]{#2}
\providecommand{\eprint}[2][]{\url{#2}}

\bibitem[{\citenamefont{Candelas et~al.}(1981)\citenamefont{Candelas,
  Chrzanowski, and Howard}}]{ar:CCH}
\bibinfo{author}{\bibfnamefont{P.}~\bibnamefont{Candelas}},
  \bibinfo{author}{\bibfnamefont{P.}~\bibnamefont{Chrzanowski}},
  \bibnamefont{and} \bibinfo{author}{\bibfnamefont{K.~W.}
  \bibnamefont{Howard}}, \bibinfo{journal}{Phys.\ Rev.\ D}
  \textbf{\bibinfo{volume}{24}}, \bibinfo{pages}{297} (\bibinfo{year}{1981}).

\bibitem[{\citenamefont{Frolov and Zel'nikov}(1985)}]{ar:F&Z'85}
\bibinfo{author}{\bibfnamefont{V.~P.} \bibnamefont{Frolov}} \bibnamefont{and}
  \bibinfo{author}{\bibfnamefont{A.~I.} \bibnamefont{Zel'nikov}},
  \bibinfo{journal}{Phys.\ Rev.\ D} \textbf{\bibinfo{volume}{32}},
  \bibinfo{pages}{3150} (\bibinfo{year}{1985}).

\bibitem[{\citenamefont{Bolashenko and Frolov}(1989)}]{ar:Bolash&Frolov}
\bibinfo{author}{\bibfnamefont{P.}~\bibnamefont{Bolashenko}} \bibnamefont{and}
  \bibinfo{author}{\bibfnamefont{V.~P.} \bibnamefont{Frolov}},
  \bibinfo{journal}{J.\ Math.\ Phys.}  (\bibinfo{year}{1989}).

\bibitem[{\citenamefont{Casals and Ottewill}(2004)}]{ar:Casals&Ott'04}
\bibinfo{author}{\bibfnamefont{M.}~\bibnamefont{Casals}} \bibnamefont{and}
  \bibinfo{author}{\bibfnamefont{A.~C.} \bibnamefont{Ottewill}},
  \bibinfo{journal}{gr-qc/0409012}  (\bibinfo{year}{2004}).

\bibitem[{\citenamefont{Casals}(2004)}]{th:CasalsPhD}
\bibinfo{author}{\bibfnamefont{M.}~\bibnamefont{Casals}}, Ph.D. thesis,
  \bibinfo{school}{University College Dublin} (\bibinfo{year}{2004}).

\bibitem[{\citenamefont{Candelas}(1980)}]{ar:Candelas'80}
\bibinfo{author}{\bibfnamefont{P.}~\bibnamefont{Candelas}},
  \bibinfo{journal}{Phys.\ Rev.\ D} \textbf{\bibinfo{volume}{21}},
  \bibinfo{pages}{2185} (\bibinfo{year}{1980}).

\bibitem[{\citenamefont{Misner et~al.}(1973)\citenamefont{Misner, Thorne, and
  Wheeler}}]{bk:M&T&W}
\bibinfo{author}{\bibfnamefont{C.~W.} \bibnamefont{Misner}},
  \bibinfo{author}{\bibfnamefont{K.~S.} \bibnamefont{Thorne}},
  \bibnamefont{and} \bibinfo{author}{\bibfnamefont{J.~A.}
  \bibnamefont{Wheeler}}, \emph{\bibinfo{title}{Gravitation}}
  (\bibinfo{publisher}{W.H. Freeman}, \bibinfo{address}{San Francisco},
  \bibinfo{year}{1973}).

\bibitem[{\citenamefont{Carter}(1968)}]{ar:Carter'68b}
\bibinfo{author}{\bibfnamefont{B.}~\bibnamefont{Carter}},
  \bibinfo{journal}{Commun.\ Math.\ Phys.} \textbf{\bibinfo{volume}{10}},
  \bibinfo{pages}{280} (\bibinfo{year}{1968}).

\bibitem[{\citenamefont{Newman and Penrose}(1962)}]{ar:N&P'62}
\bibinfo{author}{\bibfnamefont{E.~T.} \bibnamefont{Newman}} \bibnamefont{and}
  \bibinfo{author}{\bibfnamefont{R.}~\bibnamefont{Penrose}},
  \bibinfo{journal}{J.\ Math.\ Phys.} \textbf{\bibinfo{volume}{3}},
  \bibinfo{pages}{566} (\bibinfo{year}{1962}).

\bibitem[{\citenamefont{Chandrasekhar}(1992)}]{bk:Chandr}
\bibinfo{author}{\bibfnamefont{S.}~\bibnamefont{Chandrasekhar}},
  \emph{\bibinfo{title}{The Mathematical Theory of Black Holes}}
  (\bibinfo{publisher}{Oxford University Press}, \bibinfo{address}{Oxford, UK},
  \bibinfo{year}{1992}), \bibinfo{edition}{2nd} ed.

\bibitem[{\citenamefont{Chrzanowski}(1975)}]{ar:Chrzan'75}
\bibinfo{author}{\bibfnamefont{P.~L.} \bibnamefont{Chrzanowski}},
  \bibinfo{journal}{Phys.\ Rev.\ D} \textbf{\bibinfo{volume}{11}},
  \bibinfo{pages}{2042} (\bibinfo{year}{1975}).

\bibitem[{\citenamefont{Teukolsky}(1972)}]{ar:Teuk'72}
\bibinfo{author}{\bibfnamefont{S.~A.} \bibnamefont{Teukolsky}},
  \bibinfo{journal}{Phys.\ Rev.\ Lett.} \textbf{\bibinfo{volume}{29}},
  \bibinfo{pages}{1114} (\bibinfo{year}{1972}).

\bibitem[{\citenamefont{Teukolsky}(1973)}]{ar:Teuk'73}
\bibinfo{author}{\bibfnamefont{S.~A.} \bibnamefont{Teukolsky}},
  \bibinfo{journal}{The Astrophysical Journal} \textbf{\bibinfo{volume}{185}},
  \bibinfo{pages}{635} (\bibinfo{year}{1973}).

\bibitem[{\citenamefont{Chrzanowski et~al.}(1976)\citenamefont{Chrzanowski,
  Matzner, Sandberg, and Ryan~Jr.}}]{ar:CMSR}
\bibinfo{author}{\bibfnamefont{P.~L.} \bibnamefont{Chrzanowski}},
  \bibinfo{author}{\bibfnamefont{R.~A.} \bibnamefont{Matzner}},
  \bibinfo{author}{\bibfnamefont{V.~D.} \bibnamefont{Sandberg}},
  \bibnamefont{and} \bibinfo{author}{\bibfnamefont{M.~P.}
  \bibnamefont{Ryan~Jr.}}, \bibinfo{journal}{Phys.\ Rev.\ D}
  \textbf{\bibinfo{volume}{14}}, \bibinfo{pages}{317} (\bibinfo{year}{1976}).

\bibitem[{\citenamefont{Teukolsky and Press}(1974)}]{ar:Teuk&Press'74}
\bibinfo{author}{\bibfnamefont{S.~A.} \bibnamefont{Teukolsky}}
  \bibnamefont{and} \bibinfo{author}{\bibfnamefont{W.~H.} \bibnamefont{Press}},
  \bibinfo{journal}{The Astrophysical Journal} \textbf{\bibinfo{volume}{193}},
  \bibinfo{pages}{443} (\bibinfo{year}{1974}).

\bibitem[{\citenamefont{Jensen et~al.}(1991)\citenamefont{Jensen, McLaughlin,
  and Ottewill}}]{ar:J&McL&Ott'91}
\bibinfo{author}{\bibfnamefont{B.~P.} \bibnamefont{Jensen}},
  \bibinfo{author}{\bibfnamefont{J.~G.} \bibnamefont{McLaughlin}},
  \bibnamefont{and} \bibinfo{author}{\bibfnamefont{A.~C.}
  \bibnamefont{Ottewill}}, \bibinfo{journal}{Phys.\ Rev.\ D}
  \textbf{\bibinfo{volume}{43}}, \bibinfo{pages}{4142} (\bibinfo{year}{1991}).

\bibitem[{\citenamefont{McLaughlin}(1990)}]{th:McL'90}
\bibinfo{author}{\bibfnamefont{J.~G.} \bibnamefont{McLaughlin}}, Ph.D. thesis,
  \bibinfo{school}{University of Oxford} (\bibinfo{year}{1990}).

\bibitem[{\citenamefont{Jensen et~al.}(1995)\citenamefont{Jensen, McLaughlin,
  and Ottewill}}]{ar:J&McL&Ott'95}
\bibinfo{author}{\bibfnamefont{B.~P.} \bibnamefont{Jensen}},
  \bibinfo{author}{\bibfnamefont{J.~G.} \bibnamefont{McLaughlin}},
  \bibnamefont{and} \bibinfo{author}{\bibfnamefont{A.~C.}
  \bibnamefont{Ottewill}}, \bibinfo{journal}{Phys.\ Rev.\ D}
  \textbf{\bibinfo{volume}{51}}, \bibinfo{pages}{5676} (\bibinfo{year}{1995}).

\bibitem[{\citenamefont{Cohen and Kegeles}(1974)}]{ar:Coh&Keg'74}
\bibinfo{author}{\bibfnamefont{J.~M.} \bibnamefont{Cohen}} \bibnamefont{and}
  \bibinfo{author}{\bibfnamefont{L.~S.} \bibnamefont{Kegeles}},
  \bibinfo{journal}{Phys.\ Rev.\ D} \textbf{\bibinfo{volume}{10}},
  \bibinfo{pages}{1070} (\bibinfo{year}{1974}).

\bibitem[{\citenamefont{Wald}(1978)}]{ar:Wald'78}
\bibinfo{author}{\bibfnamefont{R.~M.} \bibnamefont{Wald}},
  \bibinfo{journal}{Phys.\ Rev.\ Letters} \textbf{\bibinfo{volume}{41}},
  \bibinfo{pages}{203} (\bibinfo{year}{1978}).

\bibitem[{\citenamefont{Christensen}(1978)}]{ar:Christ'78}
\bibinfo{author}{\bibfnamefont{S.~M.} \bibnamefont{Christensen}},
  \bibinfo{journal}{Phys.\ Rev.\ D} \textbf{\bibinfo{volume}{17}},
  \bibinfo{pages}{946} (\bibinfo{year}{1978}).

\bibitem[{\citenamefont{Jensen et~al.}(1999)\citenamefont{Jensen, McLaughlin,
  and Ottewill}}]{ar:J&McL&Ott'88}
\bibinfo{author}{\bibfnamefont{B.~P.} \bibnamefont{Jensen}},
  \bibinfo{author}{\bibfnamefont{J.~G.} \bibnamefont{McLaughlin}},
  \bibnamefont{and} \bibinfo{author}{\bibfnamefont{A.~C.}
  \bibnamefont{Ottewill}}, \bibinfo{journal}{Class.\ Quantum.\ Grav.}
  \textbf{\bibinfo{volume}{5}}, \bibinfo{pages}{L187} (\bibinfo{year}{1999}).

\bibitem[{\citenamefont{Unruh}(1976)}]{ar:Unruh'76}
\bibinfo{author}{\bibfnamefont{W.~G.} \bibnamefont{Unruh}},
  \bibinfo{journal}{Phys.\ Rev.\ D} \textbf{\bibinfo{volume}{14}},
  \bibinfo{pages}{870} (\bibinfo{year}{1976}).

\bibitem[{\citenamefont{Brown and Ottewill}(1983)}]{ar:Brown&Ott'83}
\bibinfo{author}{\bibfnamefont{M.}~\bibnamefont{Brown}} \bibnamefont{and}
  \bibinfo{author}{\bibfnamefont{A.~C.} \bibnamefont{Ottewill}},
  \bibinfo{journal}{Proc.\ R.\ Soc.\ Lond. A} \textbf{\bibinfo{volume}{389}},
  \bibinfo{pages}{379} (\bibinfo{year}{1983}).

\bibitem[{\citenamefont{Grove and Ottewill}(1983)}]{ar:Grove&Ott'83}
\bibinfo{author}{\bibfnamefont{P.}~\bibnamefont{Grove}} \bibnamefont{and}
  \bibinfo{author}{\bibfnamefont{A.~C.} \bibnamefont{Ottewill}},
  \bibinfo{journal}{J.\ Phys.\ A} \textbf{\bibinfo{volume}{16}},
  \bibinfo{pages}{3905} (\bibinfo{year}{1983}).

\bibitem[{\citenamefont{Christensen and Fulling}(1977)}]{ar:Christ&Fulling'77}
\bibinfo{author}{\bibfnamefont{S.~M.} \bibnamefont{Christensen}}
  \bibnamefont{and} \bibinfo{author}{\bibfnamefont{S.~A.}
  \bibnamefont{Fulling}}, \bibinfo{journal}{Phys.\ Rev.\ D}
  \textbf{\bibinfo{volume}{15}}, \bibinfo{pages}{2088} (\bibinfo{year}{1977}).

\bibitem[{\citenamefont{Hartle and Hawking}(1976)}]{ar:Hartle&Hawk'76}
\bibinfo{author}{\bibfnamefont{J.~B.} \bibnamefont{Hartle}} \bibnamefont{and}
  \bibinfo{author}{\bibfnamefont{S.~W.} \bibnamefont{Hawking}},
  \bibinfo{journal}{Phys.\ Rev.\ D} \textbf{\bibinfo{volume}{13}},
  \bibinfo{pages}{2188} (\bibinfo{year}{1976}).

\bibitem[{\citenamefont{Kay and Wald}(1991)}]{ar:Kay&Wald'91}
\bibinfo{author}{\bibfnamefont{B.~S.} \bibnamefont{Kay}} \bibnamefont{and}
  \bibinfo{author}{\bibfnamefont{R.~M.} \bibnamefont{Wald}},
  \bibinfo{journal}{Physics Reports} \textbf{\bibinfo{volume}{207}},
  \bibinfo{pages}{51} (\bibinfo{year}{1991}).

\bibitem[{\citenamefont{Jensen et~al.}(1992)\citenamefont{Jensen, McLaughlin,
  and Ottewill}}]{ar:J&McL&Ott'92}
\bibinfo{author}{\bibfnamefont{B.~P.} \bibnamefont{Jensen}},
  \bibinfo{author}{\bibfnamefont{J.~G.} \bibnamefont{McLaughlin}},
  \bibnamefont{and} \bibinfo{author}{\bibfnamefont{A.~C.}
  \bibnamefont{Ottewill}}, \bibinfo{journal}{Phys.\ Rev.\ D}
  \textbf{\bibinfo{volume}{45}}, \bibinfo{pages}{3002} (\bibinfo{year}{1992}).

\bibitem[{\citenamefont{Frolov and Thorne}(1989)}]{ar:F&T'89}
\bibinfo{author}{\bibfnamefont{V.~P.} \bibnamefont{Frolov}} \bibnamefont{and}
  \bibinfo{author}{\bibfnamefont{K.~S.} \bibnamefont{Thorne}},
  \bibinfo{journal}{Phys.\ Rev.\ D} \textbf{\bibinfo{volume}{39}},
  \bibinfo{pages}{2125} (\bibinfo{year}{1989}).

\bibitem[{\citenamefont{Ottewill and
  Winstanley}(2000{\natexlab{a}})}]{ar:Ott&Winst'00}
\bibinfo{author}{\bibfnamefont{A.~C.} \bibnamefont{Ottewill}} \bibnamefont{and}
  \bibinfo{author}{\bibfnamefont{E.}~\bibnamefont{Winstanley}},
  \bibinfo{journal}{Phys.\ Rev.\ D} \textbf{\bibinfo{volume}{62}},
  \bibinfo{pages}{084018} (\bibinfo{year}{2000}{\natexlab{a}}).

\bibitem[{\citenamefont{Duffy}(2002)}]{th:GavPhD}
\bibinfo{author}{\bibfnamefont{G.}~\bibnamefont{Duffy}}, Ph.D. thesis,
  \bibinfo{school}{University College Dublin} (\bibinfo{year}{2002}).

\bibitem[{\citenamefont{Schumaker and Caves}(1985)}]{ar:Schum&Caves'85}
\bibinfo{author}{\bibfnamefont{B.~L.} \bibnamefont{Schumaker}}
  \bibnamefont{and} \bibinfo{author}{\bibfnamefont{C.~M.} \bibnamefont{Caves}},
  \bibinfo{journal}{Phys.\ Rev.\ A} \textbf{\bibinfo{volume}{31}},
  \bibinfo{pages}{3093} (\bibinfo{year}{1985}).

\bibitem[{\citenamefont{Louisell}(1990)}]{bk:Louisell}
\bibinfo{author}{\bibfnamefont{W.~H.} \bibnamefont{Louisell}},
  \emph{\bibinfo{title}{Quantum Statistical Properties of Radiation}}
  (\bibinfo{publisher}{Wiley-Interscience}, \bibinfo{year}{1990}).

\bibitem[{\citenamefont{Weinberg}(1972)}]{bk:Weinberg}
\bibinfo{author}{\bibfnamefont{S.}~\bibnamefont{Weinberg}},
  \emph{\bibinfo{title}{Gravitation and Cosmology: Principles and Applications
  of the General Theory of Relativity}} (\bibinfo{publisher}{John Wiley \&
  Sons, Inc.}, \bibinfo{address}{New York, London, Sydney and Toronto},
  \bibinfo{year}{1972}).

\bibitem[{\citenamefont{Dirac}(1975)}]{bk:Dirac}
\bibinfo{author}{\bibfnamefont{P.}~\bibnamefont{Dirac}},
  \emph{\bibinfo{title}{General Theory of Relativity}}
  (\bibinfo{publisher}{Wiley}, \bibinfo{address}{New York},
  \bibinfo{year}{1975}).

\bibitem[{\citenamefont{Page}(1976)}]{ar:PageII'76}
\bibinfo{author}{\bibfnamefont{D.~N.} \bibnamefont{Page}},
  \bibinfo{journal}{Phys.\ Rev.\ D} \textbf{\bibinfo{volume}{14}},
  \bibinfo{pages}{3260} (\bibinfo{year}{1976}).

\bibitem[{\citenamefont{Candelas and Deutsch}(1977)}]{ar:Cand&Deutsch'77}
\bibinfo{author}{\bibfnamefont{P.}~\bibnamefont{Candelas}} \bibnamefont{and}
  \bibinfo{author}{\bibfnamefont{D.}~\bibnamefont{Deutsch}},
  \bibinfo{journal}{Proc.\ R.\ Soc.\ Lond. A} \textbf{\bibinfo{volume}{354}},
  \bibinfo{pages}{79} (\bibinfo{year}{1977}).

\bibitem[{\citenamefont{Ottewill and
  Winstanley}(2000{\natexlab{b}})}]{ar:Ott&Winst'00Lett}
\bibinfo{author}{\bibfnamefont{A.~C.} \bibnamefont{Ottewill}} \bibnamefont{and}
  \bibinfo{author}{\bibfnamefont{E.}~\bibnamefont{Winstanley}},
  \bibinfo{journal}{Phys.\ Lett.\ A} \textbf{\bibinfo{volume}{273}},
  \bibinfo{pages}{149} (\bibinfo{year}{2000}{\natexlab{b}}).

\bibitem[{\citenamefont{Gradshteyn and Ryzhik}(1995)}]{bk:GR}
\bibinfo{author}{\bibfnamefont{I.~S.} \bibnamefont{Gradshteyn}}
  \bibnamefont{and} \bibinfo{author}{\bibfnamefont{I.~M.}
  \bibnamefont{Ryzhik}}, \emph{\bibinfo{title}{Table of Integrals, Series and
  Products}} (\bibinfo{publisher}{Academic Press}, \bibinfo{address}{San
  Diego}, \bibinfo{year}{1995}), \bibinfo{edition}{5th} ed.

\bibitem[{\citenamefont{Detweiler}(1976)}]{ar:Detw'76}
\bibinfo{author}{\bibfnamefont{S.~L.} \bibnamefont{Detweiler}},
  \bibinfo{journal}{Proc.\ R.\ Soc.\ Lond. A} \textbf{\bibinfo{volume}{349}},
  \bibinfo{pages}{217} (\bibinfo{year}{1976}).

\bibitem[{\citenamefont{Abramowitz and Stegun}(1965)}]{bk:AS}
\bibinfo{author}{\bibfnamefont{M.}~\bibnamefont{Abramowitz}} \bibnamefont{and}
  \bibinfo{author}{\bibfnamefont{I.~A.} \bibnamefont{Stegun}},
  \emph{\bibinfo{title}{Handbook of Mathematical Functions}}
  (\bibinfo{publisher}{Dover Publications,Inc.}, \bibinfo{address}{New York,
  USA}, \bibinfo{year}{1965}), \bibinfo{edition}{ninth} ed.

\end{thebibliography}

\end{document}